\documentclass[fleqn,usenatbib]{mnras}
\usepackage{newtxtext,newtxmath}
\usepackage[T1]{fontenc}

\DeclareRobustCommand{\VAN}[3]{#2}
\let\VANthebibliography\thebibliography
\def\thebibliography{\DeclareRobustCommand{\VAN}[3]{##3}\VANthebibliography}

\usepackage{graphicx}	
\usepackage{amsmath}	

\usepackage{multicol}

\usepackage{xspace} 
\renewcommand{\AA}{\normalfont\r{A}\xspace} 

\usepackage{color,soul}
\definecolor{lightblue}{rgb}{.70,.95,1}
\sethlcolor{lightblue}

\newcommand{\teff}{\ensuremath{T_{\mathrm{eff}}}\xspace}

\newcommand{\kms}{\ensuremath{\rm{km}\,s^{-1}}\xspace}
\newcommand{\logg}{\ensuremath{\log g}\xspace}
\newcommand{\feh}{\rm{[Fe/H]}\xspace}
\newcommand{\cfe}{\rm{[C/Fe]}\xspace}

\newcommand{\alphafe}{\rm{[\ensuremath{\alpha}/Fe]}\xspace}
\newcommand{\Gaia}{\textit{Gaia}\xspace}
\newcommand{\CaHK}{\emph{CaHK}\xspace}
\newcommand{\Pristine}{\emph{Pristine}\xspace}
\newcommand{\FERRE}{{\tt FERRE}\xspace}
\newcommand{\ULySS}{{\tt ULySS}\xspace}

\newcommand{\RGC}{\ensuremath{R_\mathrm{GC}}\xspace}

\title[PIGS orbits]{The Pristine Inner Galaxy Survey (PIGS) VIII: Characterising the orbital properties of the ancient, very metal-poor inner Milky Way}

\author[Ardern-Arentsen et al.]{
Anke Ardern-Arentsen,$^{1}$\thanks{Email: \url{anke.arentsen@ast.cam.ac.uk}},
Giacomo Monari,$^{2}$ Anna B. A. Queiroz,$^{3,4,5}$ Else Starkenburg,$^{6}$ 
\newauthor 
Nicolas F. Martin,$^{2,7}$  Cristina Chiappini,$^{5}$ David S. Aguado,$^{3,4}$ Vasily Belokurov,$^{1}$ Ray Carlberg,$^{8}$ 
\newauthor 
Stephanie Monty,$^{1}$ GyuChul Myeong,$^{1}$ Mathias Schultheis,$^{9}$ Federico Sestito,$^{10}$ Kim A. Venn,$^{10}$  
\newauthor 
Sara Vitali,$^{11}$ Zhen Yuan,$^{2}$ Hanyuan Zhang,$^{1}$ Sven Buder, $^{12, 13}$ Geraint F. Lewis,$^{14}$  
\newauthor 
William H. Oliver,$^{15, 16, 14}$ Zhen Wan,$^{17}$ Daniel B. Zucker$^{18,19}$
\newauthor
\\
$^{1}$ Institute of Astronomy, University of Cambridge, Madingley Road, Cambridge CB3 0HA, UK \\
$^{2}$ Universit\'e de Strasbourg, CNRS, Observatoire astronomique de Strasbourg, UMR 7550, F-67000 Strasbourg, France\\
$^{3}$ Instituto de Astrof{\'\i}sica de Canarias, E-38205 La Laguna, Tenerife, Spain  \\
$^{4}$ Universidad de La Laguna, Departamento de Astrof\'{\i}sica, 38206 La Laguna, Tenerife, Spain \\
$^{5}$ Leibniz-Institut f\"ur Astrophysik Potsdam (AIP), An der Sternwarte 16, D-14482 Potsdam, Germany \\
$^{6}$ Kapteyn Astronomical Institute, University of Groningen, Postbus 800, 9700 AV, Groningen, the Netherlands\\
$^{7}$ Max-Planck-Institut f\"ur Astronomie, K\"onigstuhl 17, D-69117 Heidelberg, Germany\\
$^{8}$ Department of Astronomy \& Astrophysics, University of Toronto, Toronto, ON M5S 3H4, Canada \\
$^{9}$ Universit\'e C\^ote d'Azur, Observatoire de la C\^ote d'Azur, CNRS, Laboratoire Lagrange, Blvd de l'Observatoire, F-06304 Nice, France\\
$^{10}$ Department of Physics \& Astronomy, University of Victoria, Victoria, BC,
V8W 3P2, Canada \\
$^{11}$ Instituto de Estudios Astrof\'isicos, Facultad de Ingenier\'ia y Ciencias, Universidad Diego Portales, Av. Ej\'ercito Libertador 441, Santiago, Chile \\
$^{12}$ Research School of Astronomy and Astrophysics, Australian National University, Canberra, ACT 2611, Australia\\
$^{13}$ ARC Centre of Excellence for All Sky Astrophysics in 3 Dimensions (ASTRO 3D), Australia\\
$^{14}$ Sydney Institute for Astronomy, School of Physics A28, The University of Sydney, NSW 2006, Australia \\
$^{15}$ Universit\"{a}t Heidelberg, Interdisziplin\"{a}res Zentrum f\"{u}r Wissenschaftliches Rechnen, Im Neuenheimer Feld 205, D-69120 Heidelberg, Germany\\
$^{16}$ Universit\"{a}t Heidelberg, Zentrum f\"{u}r Astronomie, Institut f\"{u}r Theoretische Astrophysik, Albert-Ueberle-Stra{\ss}e 2, D-69120 Heidelberg, Germany\\
$^{17}$ School of Astronomy and Space Science, University of Science and Technology of China, Hefei, 230026, China \\
$^{18}$ School of Mathematical and Physical Sciences, Macquarie University, Sydney, NSW 2109, Australia \\
$^{19}$ Macquarie University Astrophysics and Space Technologies Research Centre, Sydney, NSW 2109, Australia \\
}

\date{Accepted 2024 April 15. Received 2024 April 12; in original form 2023 December 7}

\pubyear{2024}

\begin{document}
\label{firstpage}
\pagerange{\pageref{firstpage}--\pageref{lastpage}}
\maketitle

\begin{abstract}
The oldest stars in the Milky Way (born in the first few billion years) are expected to have a high density in the inner few kpc, spatially overlapping with the Galactic bulge. 
We use spectroscopic data from the Pristine Inner Galaxy Survey (PIGS) to study the dynamical properties of ancient, metal-poor inner Galaxy stars. We compute distances using \texttt{StarHorse}, and orbital properties in a barred Galactic potential. With this paper, we release the spectroscopic AAT/PIGS catalogue (13\,235 stars).
We find that most PIGS stars have orbits typical for a pressure-supported population. The fraction of stars confined to the inner Galaxy decreases with decreasing metallicity, but many very metal-poor stars (VMP, [Fe/H] $< -2.0$) stay confined ($\sim60\%$ stay within 5~kpc). The azimuthal velocity v$_\phi$ also decreases between [Fe/H]~$=-1.0$ and $-2.0$, but is constant for VMP stars (at $\sim+40$ km\,s$^{-1}$). 
The carbon-enhanced metal-poor (CEMP) stars in PIGS appear to have similar orbital properties compared to normal VMP stars. 
Our results suggest a possible transition between two spheroidal components -- a more metal-rich, more concentrated, faster rotating component, and a more metal-poor, more extended and slower/non-rotating component. We propose that the former may be connected to pre-disc in-situ stars (or those born in large building blocks), whereas the latter may be dominated by contributions from smaller galaxies. 
This is an exciting era where large metal-poor samples, such as in this work (as well as upcoming surveys, e.g., 4MOST), shed light on the earliest evolution of our Galaxy. \\
\end{abstract}



\section{Introduction}

Low-mass stars born in the first few billion years after the Big Bang allow us to probe the early Universe through detailed local observations. The field of Galactic Archaeology makes use of these ancient, metal-poor stars to learn about the first stellar generations and to decipher the earliest phases of galaxy formation -- in particular they allow us to disentangle the history of the (early) Milky Way \citep[see e.g. the reviews by][]{freebland02, frebelnorris15}. The \textit{oldest} metal-poor stars in the Milky Way are expected to be very centrally concentrated -- overlapping with the Galactic bulge, in the inner $\sim 5$~kpc of the Milky Way. This expectation comes from the hierarchical build-up scenario of galaxy formation as well as inside-out growth of the main galaxy progenitor, and is supported by a variety of simulations investigating the spatial distributions of the oldest metal-poor stars in our Galaxy \citep[e.g.][]{white00, tumlinson10, starkenburg17a, elbadry18, belokurovkravtsov22}. This ancient, concentrated stellar population is expected to be the result of mergers of many building blocks, small and larger, together assembling the primordial Milky Way. If one of the building blocks clearly dominates, it can be considered the main progenitor and stars formed inside it could be labelled as formed ``\textit{in-situ}''. Up until that point, the distinction between stars born in-situ or accreted is not as clear as later on in the life of our Galaxy. 

The inner $\sim 5$~kpc of the Milky Way host a complex mixture of stellar populations. What is typically called ``the bulge'' is a central over-density extending outside of the Galactic plane, comprised of relatively metal-rich stars ($-1.0 <$ \feh\footnote{[X/Y] $ = \log(N_\mathrm{X}/N_\mathrm{Y})_* - \log(N_\mathrm{X}/N_\mathrm{Y})_{\odot}$, where the asterisk subscript refers to the considered star, and N is the number density. Throughout this work, we use [Fe/H] to refer to ``metallicity''.} $< +0.5$, with multiple peaks in the metallicity distribution), showing cylindrical rotation around the Galactic centre, and is thought to originate (predominantly) from instabilities in the Galactic disc, namely a buckling bar \citep[for an overview, further references and discussions around the presence of a pressure-supported component, see][]{barbuy18}. The term ``bulge'' is also sometimes used to simply refer to the spatial location of stars, e.g. within $3.5$ or $5$~kpc. Metal-poor stars are rare in the bulge region compared to the Galactic halo -- for example only 4\% and 0.2\% of stars in the relatively metallicity-blind ARGOS bulge survey had \feh~$<-1.0$ and $<-2.0$, respectively \citep{ness13a}, whereas the metallicity distribution function of the halo peaks below $\feh = -1.0$ \citep[e.g.][and references therein]{youakim20}. The properties of the metal-poor inner Galaxy have therefore remained elusive for many years. 

Early studies show that metal-poor\footnote{For this summary, our definition of ``metal-poor'' is $\feh < -1.0$, which is more metal-poor than what typical bulge studies would use the term for -- some might even refer to $\feh < 0.0$ as metal-poor. For example, \citet{zoccali17} find that metal-poor stars with $\feh < 0.0$ have a more centrally concentrated/spheroidal distribution than super-solar metallicity stars.} (MP, $\feh < -1.0$) stars in the inner Galaxy have different properties compared to the metal-rich bulge stars. For example the ARGOS survey showed that they rotate slower around the Galactic centre and have a higher velocity dispersion \citep{ness13b, wylie21} -- they connected these stars with the Galactic halo. Works based on inner Galaxy RR Lyrae stars (which are expected to be old and mostly metal-poor, e.g. \citealt{savino20} find that the inner Galaxy RR Lyrae spectroscopic metallicities peak around $\feh = -1.4$) show that their 3D distribution does not closely trace that of metal-rich stars, although it might be slightly bar and/or peanut-shaped \citep{dekany13, pietrukowicz15, semczuk22}, and that they have a high velocity dispersion and rotate slowly, if at all \citep{kunder16, wegg19, kunder20}. The latter authors also suggest that the RR Lyrae might trace multiple overlapping Galactic components. 

The APOGEE spectroscopic survey \citep{apogee} has been very important in improving our understanding of the metal-poor inner Galaxy, as it is the only large spectroscopic survey covering both the inner Galaxy (very close to the Galactic mid-plane) and other regions of the Milky Way. The sample of low-metallicity stars is very sparse for $\feh < -1.5$, but the survey has still revealed interesting properties above this metallicity through detailed stellar chemical compositions. It was found that the fraction of stars with hints of globular-cluster-like chemistry increases towards the inner Galaxy, compared to the rest of the halo \citep{schiavon17, horta21b}. These authors, as well as \citet{belokurovkravtsov23} have interpreted this as an increased contribution from disrupted (ancient) globular clusters in the central regions of the Milky Way. The metal-poor APOGEE data, combined with \Gaia DR2 astrometry \citet{gaia16, gaiadr2}, also revealed leftovers of a large inner Galaxy building block/accreted galaxy (\citealt{horta21a, horta23a}, in agreement with the inference from globular clusters by \citealt{kruijssen20}). \citet{queiroz21} used APOGEE plus \Gaia EDR3 \citep{gaiaedr3} astrometry to study the orbital properties of Milky Way bulge stars and found signatures of a pressure-supported inner Galaxy population (alongside other co-existing stellar populations -- inner thick and thin disc, stars in bar-shaped orbits), which is more prominent among metal-poor stars. Follow-up of some of those metal-poor ($-2.0 < \feh < -1.0$) stars showed high alpha abundances \citep[][]{razera22}, consistent with a population born inside the Milky Way rather than accreted later on (the authors refer to it as a ``spheroidal bulge''). In the Solar neighbourhood, \citet{belokurovkravtsov22} use APOGEE data combined with \Gaia DR3 \citep{gaiadr3} to identify metal-poor ($\feh \lesssim -1.3$) stars belonging to a chemically distinct, isotropic population, which they infer to be the tail of an ancient centrally concentrated in-situ component that formed in the main Milky Way progenitor before the onset of the Galactic disc -- named ``Aurora'' by the authors. 

Observations of samples of \textit{very} metal-poor (VMP, $\feh < -2.0$) stars in the inner Galaxy have mostly been made possible thanks to efficient photometric pre-selection methods, e.g. using infrared photometry \citep{schlaufmancasey14}, or narrow-band optical photometry around the Ca H\&K lines \citep[e.g. from the SkyMapper and Pristine surveys,][]{wolf18, starkenburg17b}. The spectroscopic follow-up observations of inner Galaxy VMP stars show that globally, they look chemically similar to ``normal'' halo stars observed locally and/or further out into the halo \citep{garciaperez13, caseyschlaufman15, koch16, howes14, howes15, howes16, lucey19, lucey22, reggiani20, arentsen21, sestito23}, although there are some hints from the population that they may have been born in different/larger building blocks. For example, \cite{caseyschlaufman15} find that their three inner Galaxy stars have low scandium (although this was not found in other works, e.g. \citealt{koch16}), \cite{lucey19} and \cite{koch16} find a low dispersion in alpha abundances, \cite{howes15, howes16} and \cite{arentsen21} find a low frequency of carbon-enhanced metal-poor (CEMP) stars, and \cite{lucey22} uncover different correlations between chemical abundances for stars with different orbital properties (although this is mostly for stars with $\feh > -2.0$). Some of these signatures (low scandium, low carbon) could be connected to a larger contribution from pair instability supernovae, which are expected to occur more often in larger systems \citep[see e.g.][]{pagnini23}. 
Dedicated high-resolution spectroscopic follow-up of PIGS so far finds many VMP stars with ``typical'' halo chemistry, as well as individual stars with peculiar abundance patterns, such as those typical for globular cluster stars or ultra-faint dwarf galaxies \citep{sestito23, sestito23_ghost}, or the CEMP-r/s star presented by \citet{mashonkina23} which has undergone binary interaction. 

Dynamically, the most metal-poor inner Galaxy stars extend the trends already observed for normal metal-poor stars in the region: low (or even non-existent) rotation around the Galactic centre, a high velocity dispersion and a large fraction of stars not confined to the inner $\sim$few kpc (\citealt{arentsen20_I, lucey21, rix22, sestito23}, and this work). For example, in the high-resolution spectroscopic PIGS follow-up sample of \citet{sestito23}, less than half of the VMP stars are confined to within 5~kpc. The observations to date are consistent with recent simulation results looking at the rotation and velocity dispersion \citep[e.g.][]{fragkoudi20} and halo interlopers \citep{orkney23} for the metal-poor inner Galaxy. 

Large samples of metal-poor inner Galaxy stars are now also becoming available thanks to the release of the (very low resolution) XP spectra in Gaia DR3 \citep{gaiadr3}. These are spatially more homogeneous than previous samples, and can uncover the spatial distribution of the most metal-poor stars. For example, \citet{rix22} use XP metallicities and \Gaia radial velocities to show that there is a centrally concentrated, barely rotating population of stars at low metallicity ([M/H]~$< -1.5$), with most stars within $|l|, |b| < 15^{\circ}$, having a Gaussian extent of $\sigma_R \sim2.7$~kpc. They interpret this population as being the result of chaotic early Galaxy assembly, where in-situ and accreted become less strictly separable, and refer to this as the ``proto-Galaxy'' \citep[see also][]{chandra23}.

There still are several open questions regarding the nature of the ancient, metal-poor inner Galaxy (see also the discussion in \citealt{rix22}). For example: what are the building blocks contributing to the ancient inner Milky Way? Can we clearly distinguish ``in-situ'' stars from ``accreted'' stars, and/or are these terms not meaningful anymore when discussing the early Galaxy? How much mass is there in the pressure-supported inner Galaxy? Where (and when) do the most metal-poor stars in the inner Galaxy come from? What contribution do disrupted globular clusters have in the central VMP population? What new information can we learn about the first stars and small galaxies from VMP stars in this different Galactic environment?

In this article, we extend the Pristine Inner Galaxy Survey (PIGS) study of the dynamical properties of (V)MP inner Milky Way stars started in \citet{arentsen20_I} -- now adding more stars and deriving detailed orbital properties rather than using only projected radial velocities. It also extends the work by \citet{rix22}, with a larger sample of VMP stars with reliable distances, radial velocities and metallicities, reaching closer to the Galactic centre due to the fainter magnitude limit in PIGS compared to \Gaia, and adopting a more realistic Galactic potential with a bar. 

The overview of this article is as follows. Section~\ref{sec:pigs} describes the PIGS observations and Section~\ref{sec:dyn} describes how we derive distances with {\tt StarHorse} \citep[][]{Santiago16, Queiroz18} and orbital properties by integrating in a barred Galactic potential \citep{portail2017,sormani22}. We discuss the results in Section~\ref{sec:results}, focusing on the confinement of metal-poor stars to the inner Galaxy, their rotation around the Galactic centre, a comparison to the dynamics of metal-rich inner Galaxy stars from \citet{queiroz21}, and the kinematics of carbon-enhanced metal-poor (CEMP) stars. We then discuss the results in the context of simulations and previously observations in Section~\ref{sec:discussion}, and point forward to possible improvements for future work. We summarise our findings in Section~\ref{sec:summary}.

\section{The Pristine Inner Galaxy Survey (PIGS)}\label{sec:pigs}

PIGS is a survey targeting the most metal-poor stars in the inner Galaxy \citep{arentsen20_II}, pre-selecting metal-poor candidates using metallicity-sensitive narrow-band \CaHK photometry from MegaCam \citep{boulade03} at the Canada-France-Hawaii Telescope (CFHT). PIGS is an extension of the main \Pristine survey \citep{starkenburg17b, martin23}, which is mostly targeting the Galactic halo, and faces unique challenges in the dusty and crowded inner Galaxy. Details about the PIGS target selection are described in \citet{arentsen20_II}, and the sub-survey targeting the Sagittarius dwarf galaxy is described in \citet{vitali22}. In short, the \CaHK photometry was combined with PanSTARRS-1 \citep[PS1,][]{panstarrs} or \Gaia~DR2 \citep{gaiadr2} broad-band photometry to create colour-colour diagrams, from which the follow-up targets were selected, with the most metal-poor candidates given the highest priority. A cut on the \Gaia DR2 parallax and its uncertainty was applied to remove foreground stars (which are mostly metal-rich). The choice of broad-band photometry evolved while the survey was progressing. When using PS1 we selected targets with $14 < g < 17$ and, when using \Gaia, we selected stars in the range $13.5 < G < 16.5$, with the goal of reaching giant stars in the Galactic bulge region. PIGS focuses on the region with absolute longitudes and latitudes $< 12$~degrees, E(B-V)~$< 0.8$ and declination $>-30$~degrees. 

Of the metal-poor candidates in PIGS, $13\,000$ were followed up from 2017 to 2020 with the Anglo Australian Telescope (AAT) using AAOmega+2dF \citep{saunders04, lewis02, sharp06}, obtaining simultaneous low-resolution optical spectra ($R \sim 1300$, $3700-5500$~\AA) and medium-resolution calcium triplet spectra ($R \sim 11\,000$, $8400-8800$~\AA). The spectra were analysed with two independent pipelines to derive stellar parameters, \FERRE\footnote{FERRE \citep{allende06} is available from \url{http://github.com/callendeprieto/ferre}} and \ULySS\footnote{ULySS \citep{koleva09} is available from \url{http://ulyss.univ-lyon1.fr/}}, as described in \citet{arentsen20_II}. Both are full-spectrum fitting codes, but \ULySS uses an empirical spectral library and \FERRE a synthetic library. In this work we adopt the \FERRE stellar parameters, which were found to be better at very low metallicity in our previous work and also include an estimate of the carbon abundance. We apply the \FERRE parameter quality cuts as described in \citet{arentsen20_II}, after which the sample has median uncertainties of 149~K, 0.41~dex, 0.16~dex and 0.23~dex for \teff, \logg, \feh and \cfe, respectively. 

\begin{figure}
\centering
\includegraphics[width=1.0\hsize,trim={0.0cm 0.0cm 0.0cm 0.0cm}]{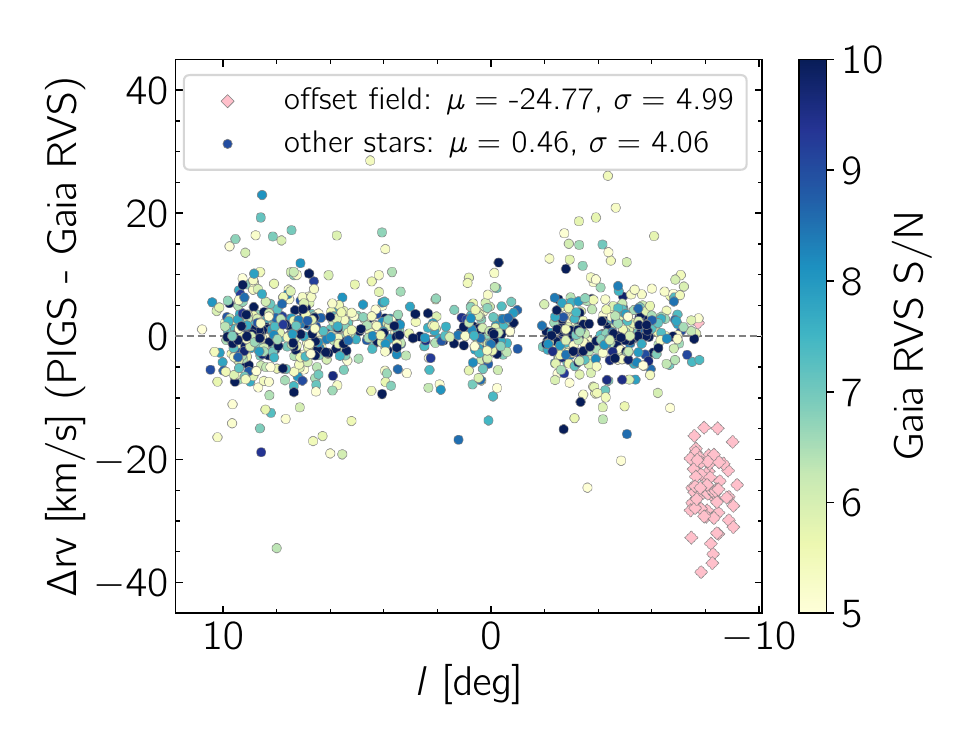}
\caption{Comparison of radial velocities between PIGS and \Gaia DR3 colour-coded by \Gaia RVS S/N, for stars with S/N $> 5$ in \Gaia. One field (Field251.2-29.7) is offset from the rest and has been highlighted in pink. The maximum of the colour bar is fixed to 10 but there are stars with higher S/N.}
\label{fig:rvcomp} 
\end{figure}

The radial velocities in PIGS were derived from the calcium triplet AAT spectra using the {\tt FXCOR} package in {\tt IRAF}\footnote{IRAF (Image Reduction and Analysis Facility) is distributed by the National Optical Astronomy Observatories, which are operated by the Association of Universities for Research in Astronomy, Inc., under contract with the National Science Foundation.}, with the statistical uncertainties estimated to be on the order of $2~\kms$ \citep[for details see][]{arentsen20_II}. Previously there was not enough overlap between PIGS and other spectroscopic surveys to do an external test of the velocities. However, there are now $\sim 1000$ stars in common with the \Gaia DR3 RVS sample (with \Gaia S/N $>5$). We compare the PIGS and \Gaia DR3 velocities to test for any systematic issues (see Figure~\ref{fig:rvcomp}). The general agreement is excellent, with a small offset ($\sim0.5~\kms$) and a dispersion of $\sim 4~\kms$. The dispersion goes down to $\sim 2~\kms$ for stars with \Gaia S/N $> 10$. We found that one AAT field (Field251.2-29.7) has a systematically different radial velocity compared to \Gaia than the rest of the fields, with a mean offset of $-24.77~\kms$ and a slightly larger dispersion ($5~\kms$). The night log reveals that there were some technical issues with the fibre plate just before the observations of this field, which could have caused a shift in the wavelength. We decided to apply a correction of $24.77~\kms$ to this field for the remainder of this work.

Some PIGS follow-up specifically targeted the Sagittarius dwarf galaxy \citep[see][]{vitali22} -- for the current work we removed these Sagittarius stars from our analysis using their \Gaia EDR3 proper motions and PIGS radial velocities (removing stars with $\sqrt{(\mu_\alpha + 2.69)^2 + (\mu_\delta + 1.35)^2} < 0.65$ and RV $> 100$~\kms). 

In the first PIGS paper we explored the kinematics of metal-poor stars in the inner Galaxy as a function of metallicity \citep{arentsen20_I} using radial velocities and \ULySS metallicities for the spectroscopic PIGS sample observed from 2017 to 2019. In this paper, we extend the study of the kinematics of stars in PIGS to the full footprint (which includes more fields in the Northern bulge), adopting the \FERRE metallicities, and adding more orbital information beyond the radial velocities. Throughout this work we will often divide our sample in four different metallicity groups, see Table~\ref{table:mpgroups}. They are split in horizontal branch (HB) and not HB (because we found systematic differences between those two groups, see the next section). 

\begin{table}
\centering
\caption{\label{table:mpgroups}Metallicity groups used in this work. The $f_{3.5}$ is the fraction of (not HB) stars that has $r < 3.5$~kpc.}
\begin{tabular}{lcccc}
 \hline
 Name & metallicity range & N$_\mathrm{not HB}$ & N$_\mathrm{HB}$ & $f_{3.5}$\\
 \hline
 Metal-rich (MR) & $[-1.0, -0.5]$ & 699 & 20 & 0.52 \\
 Metal-poor (MP) & $[-1.5, -1.0]$ & 1327  &  1587 & 0.87 \\ 
 Intermediate MP (IMP) & $[-2.0, -1.5]$ & 3858 &  752 & 0.93 \\
 Very MP (VMP) & $< -2.0$ & 1704 & 108 & 0.89 \\
 \hline
\end{tabular}
\end{table}

With this paper we also release the full spectroscopic AAT/PIGS catalogue (13\,235 stars, among which are $\sim800$ Sagittarius stars). See Tables~\ref{tab:log} and \ref{tab:pigscat} for an overview of the observations and the contents of the catalogue, which can be downloaded in full from the CDS\footnote{Before publication, the PIGS data release can be found \href{https://drive.google.com/drive/folders/1roMPX7p-TkePx-guqPFr8H5vqaih3sga?usp=sharing}{here}.}.

\subsection{Comparison with APOGEE and \Gaia XP metallicities}

\begin{figure}
\centering
\includegraphics[width=1.0\hsize,trim={0.0cm 0.0cm 0.0cm 0.0cm}]{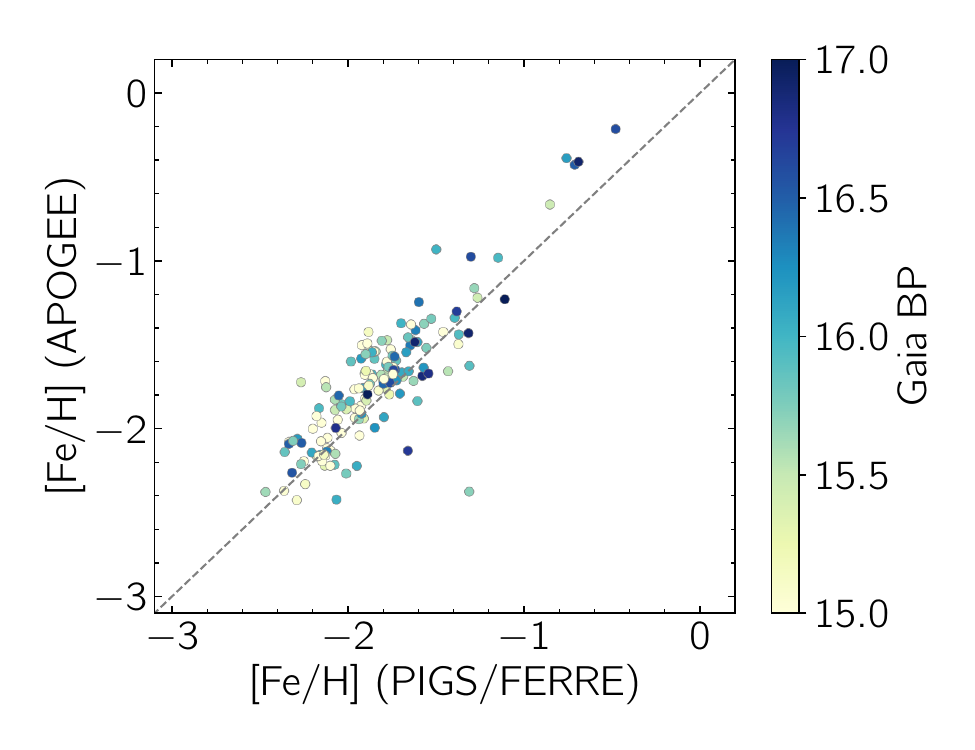}
\includegraphics[width=1.0\hsize,trim={0.0cm 0.7cm 0.0cm 0.7cm}]{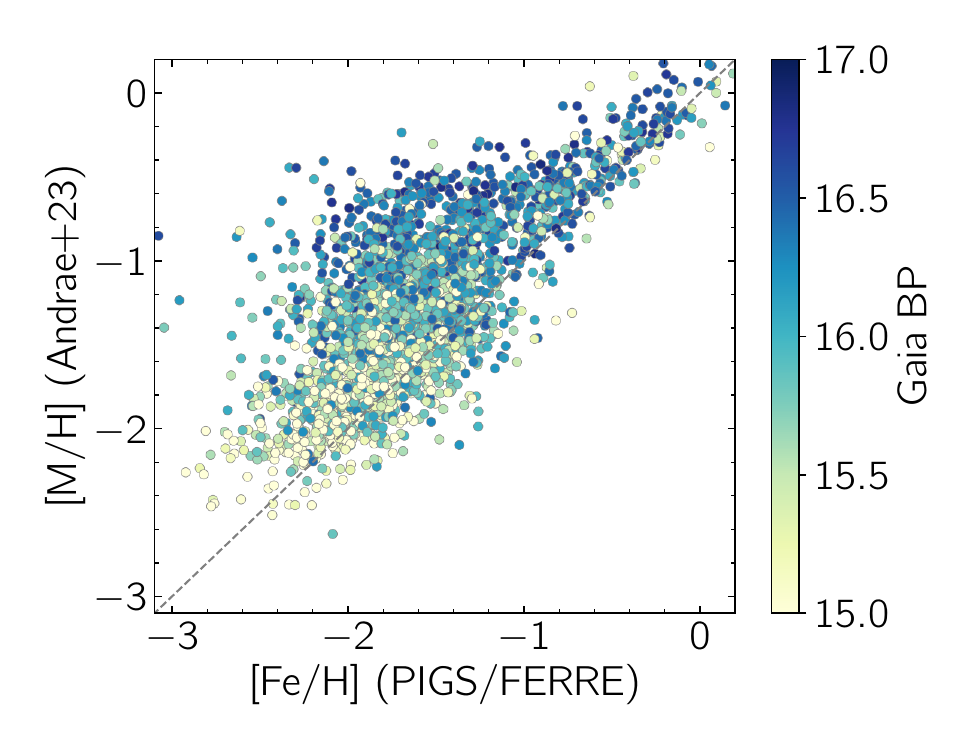}
\caption{Top: Comparison of spectroscopic PIGS metallicities to APOGEE DR17 metallicities, colour-coded by the \Gaia BP magnitude. The grey dashed line is the 1-1 line. Bottom: same, but compared to \Gaia XP metallicities from \citet{andrae23} for their ``vetted giant sample''. There are $\sim 2800$ stars in this comparison. }
\label{fig:andrae} 
\end{figure}

We present a comparison between the PIGS (FERRE) spectroscopic metallicities and the metallicities from APOGEE DR17 \citep{apogeedr17} in the top panel of Figure~\ref{fig:andrae}, for stars with APOGEE SNR~$>30$ (the results do not change when using a stricter cut of e.g. SNR~$>70$). This is an updated comparison with respect to \citet{arentsen20_II}, now with a larger sample of stars (154). The overlap between the two surveys is mostly thanks to dedicated APOGEE follow-up observations of PIGS within the bulge Cluster APOgee Survey (CAPOS) project \citep{geisler21} -- ``randomly'' there would not have been much overlap at low metallicities. There is a small offset between the metallicities from both surveys (median of APOGEE$-$PIGS = $+0.13$~dex), which is not surprising given the different methodologies, resolutions and wavelength coverage used. Overall the agreement is good, with a scatter of 0.17~dex. 

The largest publicly available set of metal-poor inner Galaxy candidates comes from \cite{andrae23}, who derive stellar parameters from the \Gaia XP spectra combined with broadband photometry. They train the XGBoost algorithm on APOGEE DR17 stellar parameters, also adding in a small set of very/extremely metal-poor stars from \citet{li22}, and find very good consistency in their comparisons with the literature. The analysis of the metal-poor inner Galaxy by \citet{rix22} was based on an earlier version of these metallicities. We present a comparison between the \cite[][A23]{andrae23} ``vetted giant sample'' and our PIGS spectroscopic metallicities in the bottom panel of Figure~\ref{fig:andrae}, colour-coded by \Gaia BP magnitude. The agreement is generally good, especially for brighter and/or more metal-rich stars. 
There is a small systematic offset between the two (median of A23$-$PIGS metallicity $= +0.19$~dex for bright metal-poor stars, with a scatter of 0.22~dex) in the same direction as for the PIGS -- APOGEE comparison, which is not surprising since \cite{andrae23} trained on APOGEE metallicities. There is a larger offset and scatter (median of A23$-$PIGS metallicity $= +0.31, \sigma=0.32$~dex) between $-2.0 <$~[Fe/H]~$<-1.0$, and the offset appears to be correlated with the BP magnitude. \cite{andrae23} applied a cut in \Gaia $G<16$ to their vetted giant sample, but the G band is very broad and in highly extincted regions the blue part of the spectrum will be significantly fainter than the red. It is less problematic for metal-rich stars, which have strong features, but affects metal-poor stars more strongly because their features are weaker and some of the main information (e.g. Ca H\&K) is in the blue part of the spectrum. Some metallicity biases may therefore be introduced when making magnitude cuts in these kind of samples. 

\section{Dynamical analysis}\label{sec:dyn}

\subsection{Distance determination with {\tt StarHorse}}\label{sec:dist}

We derive distances using {\tt StarHorse} \citep[][]{Santiago16, Queiroz18}, a Bayesian isochrone matching method capable of deriving distances, extinctions and ages based on a set of observables and priors. All this extra information is necessary because the \Gaia parallaxes alone are not good enough to derive distances to stars in the inner Galaxy (distances from the Sun of $\sim 4-12$ kpc). The {\tt StarHorse} method has been extensively validated with simulations and external samples, and has previously also been used in the Galactic bulge and the whole disc \citep[e.g.][]{queiroz20, queiroz21}. Details of the method and its assumptions can be found in \citet{Queiroz18} and \citet{Anders19, Anders22}. The resulting distance distributions for any given star are not necessarily well-represented by a single Gaussian -- instead, we use a 3-component Gaussian mixture model representation of the probability distribution functions (see the next section). 

For the distances derived in this work, we give as input to the code the \Gaia EDR3 parallaxes \citep{gaiaedr3}, photometry from PanSTARRS \citep{panstarrs}, 2MASS \citep{2mass} and WISE \citep{wise}, and the spectroscopic parameters (\FERRE \teff, \logg, \feh) from PIGS -- similar to what \citet{queiroz23} did for other spectroscopic surveys. The main reference magnitude used is PanSTARRS $g_\mathrm{PS}$. As in previous {\tt StarHorse} papers, we used the PARSEC isochrones \citep{bressan12, marigo17}. These are based on [M/H], and since no \alphafe estimates are available for PIGS, we converted our \feh estimates into [M/H] following \citet{salaris93}, with a fixed \alphafe of 0.4 -- appropriate for metal-poor stars (although some accreted stars could have lower \alphafe). The lowest [M/H] in the isochrones is $-2.2$, corresponding to $\feh \approx -2.5$. There are stars with lower metallicities in PIGS, but the lowest metallicity isochrone should be appropriate for these since for giant stars at this metallicity the isochrones do not change much anymore. We find that {\tt StarHorse} converged for 92\% of the input sample described in the previous section. Roughly half of the non-converged stars do not have a good $g_\mathrm{PS}$ magnitude, and among the other half there is a relatively high fraction with $\feh > -1.0$, for which our stellar parameters are less reliable, or low fidelity from \citet{rybizki22}, indicating spurious astrometry.  

\begin{figure}
\centering
\includegraphics[width=0.9\hsize,trim={0cm 0.0cm 0cm 0cm}]{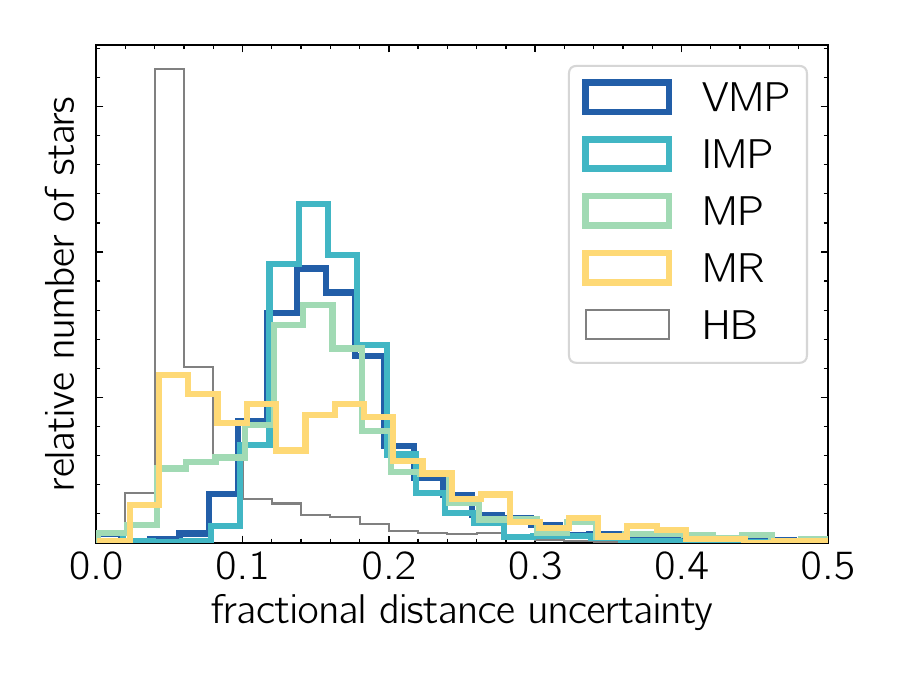}
\includegraphics[width=0.9\hsize,trim={0cm 0.0cm 0cm 0cm}]{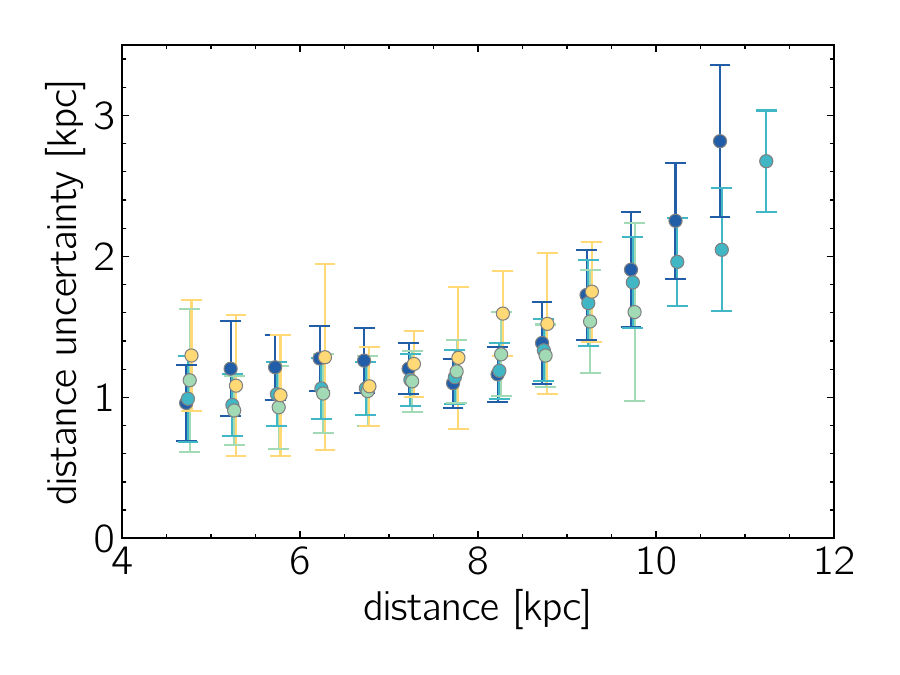}
\caption{Distance uncertainties in PIGS. Top: Distributions of fractional distance uncertainties (dist84-dist16)/(dist50 $\times$ 2) for the four main metallicity groups in this work. HB stars were removed from the coloured bins, and are shown separately in the thin grey histogram (although there still appear to be some helium-burning stars left in the MR and MP samples). The metallicity group labels are defined in Table~\ref{table:mpgroups}. Bottom: median distance uncertainty as a function of the median distance, with the same colour-coding. Error bars indicate the standard deviation (after removing $3\sigma$ outliers). }
\label{fig:disterrhist} 
\end{figure}

The uncertainties on the PIGS/{\tt StarHorse} distances are between $10-20\%$ for the bulk of the sample, peaking at $15\%$ -- see Figure~\ref{fig:disterrhist}. There is also a population with much smaller uncertainties ($< 10\%$, peaking at $5\%$), which turns out to be stars that {\tt StarHorse} puts on the horizontal branch (HB). Such stars should indeed be good distance indicators and would be expected to have smaller uncertainties. However, that requires being able to trust that the spectroscopic parameters for these stars do not have any particular biases (and our assumed alpha over iron ratios are appropriate), as well as knowing the absolute magnitudes and therefore having accurate HBs in the isochrones. Given that there are enough RGB stars present in all metallicity ranges for our purposes, and they are simpler to deal with, we decide to not use the HB stars for the bulk of this work unless explicitly stated (although we checked that the main results do not change if they are included). Some additional discussion on HB stars in PIGS can be found in Appendix~\ref{sec:hb}, including how we removed them.

\subsubsection{Distance comparison with APOGEE}
\begin{figure}
\centering
\includegraphics[width=0.9\hsize,trim={0.7cm 0.0cm 0.7cm 0cm}]{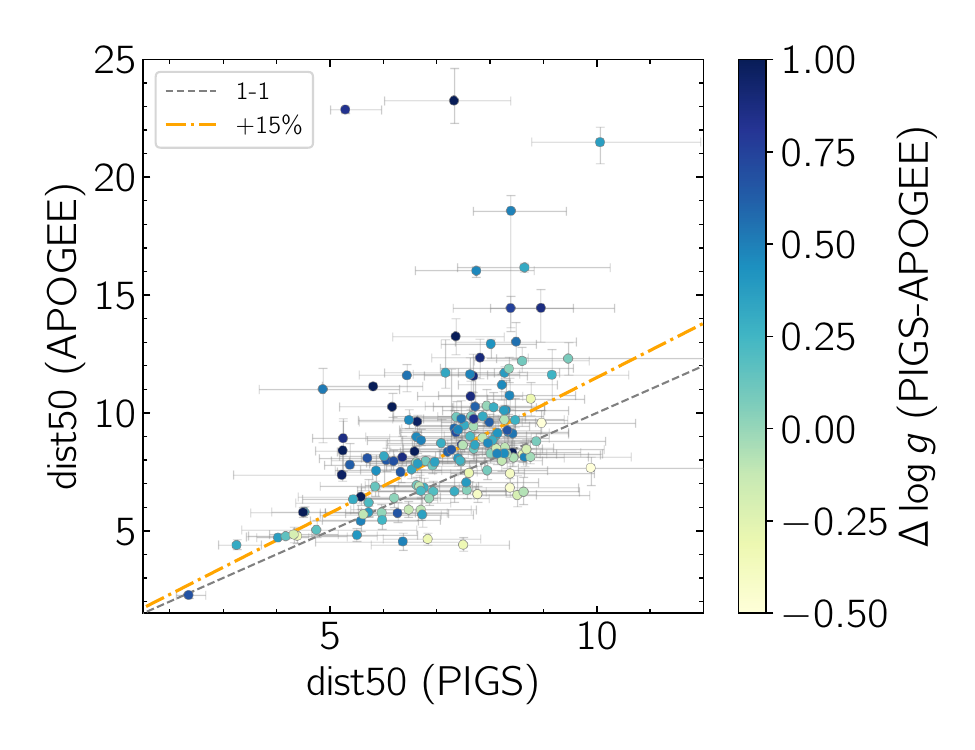}
\caption{Comparison of spectro-photometric {\tt StarHorse} distances for PIGS (this work) and APOGEE \citep{queiroz23}, colour-coded by the difference in spectroscopic \logg. In grey is the 1-1 line, in orange this relation is increased by $15\%$ of the PIGS distance.}
\label{fig:distcomp} 
\end{figure}

For stars in common between PIGS and APOGEE DR17, we compare our {\tt StarHorse} distances with the spectro-photometric APOGEE-{\tt StarHorse} distances from \citet{queiroz23}, see Figure~\ref{fig:distcomp}. The APOGEE distances appear to be larger by $\sim 15\%$ on average compared to PIGS. They have been derived with the same methodology and assumptions as our distances, so any differences should be due to differences in the stellar parameters/metallicities and their uncertainties. Indeed there appears to be a correlation with the difference in spectroscopic \logg, as shown by the colour-coding of the figure. This is not unexpected, given that \logg is the main driver for the brightness, and hence the distance. 
\citet{Queiroz18} investigated the effect of systematic stellar parameter differences on {\tt StarHorse} distances, and found that for giants, an offset in \logg of 0.2~dex results in a $\sim 10\%$ distance bias (their Figure~6). 
The median difference (PIGS$-$APOGEE) in \logg for this sample is $+0.33$~dex (and $+110$~K for \teff) -- extrapolating the \citet{Queiroz18} results for \logg predicts a distance bias of $\sim 15\%$, which is indeed what we find. 

Some of the more severe outliers are distant stars according to APOGEE ($>15$~kpc), with low APOGEE \logg. This could indicate that some of the more distant stars in PIGS are not recognised as such and are placed closer to us than they are. It might be the result of biases in \logg for stars (especially towards the tip of the RGB, apparently), and/or the relatively large \logg uncertainties in PIGS which allow the {\tt StarHorse} inner Galaxy density prior to play a larger role.

How could this impact our results? For our main analyses, we have performed tests with distances increased (or decreased) by $15\%$ to account for possible systematics, and our main conclusions are not affected (see Appendix~\ref{sec:biastest}). Furthermore, the overlap sample with APOGEE is somewhat biased towards more evolved RGB stars compared to the full PIGS sample, because it is on average a magnitude brighter. The spectroscopic \logg differences appear to be larger for such stars. Therefore the distance bias, as well as the number of outliers, might be overestimated from this specific comparison.

A final note: the \texttt{StarHorse} distances for stars in Sagittarius (Sgr) are not as reliable as those for the inner Galaxy -- a number of the Sgr stars (especially those further away from M54, where \texttt{StarHorse} no longer includes a Sgr distance prior) have distances that put them close to the inner Galaxy instead of at the distance of Sgr. This is consistent with what was seen above for distant stars in the APOGEE comparison. We briefly explore this in Appendix~\ref{sec:sgrappendix}. This is a warning to use the distances for Sgr stars with caution.

\subsubsection{Spatial coordinates and metallicity samples}
For the analysis of the orbits of the PIGS stars, we describe the Galaxy in Cartesian and cylindrical coordinates, $(x,y,z)$ and  $(\RGC,\phi,z)$ respectively. In these coordinates, we position the Sun at $(x,y,z)=(-R_\odot,0,0)$, where $R_\odot = 8.2~$kpc \citep{blandhawthorngerhard16,portail2017} is the cylindrical distance of the Sun from the Galactic centre\footnote{We neglect the distance of the Sun from the Galactic plane $z_\odot$, setting it to zero, since its value is very small and does not have a significant influence in our analysis.}. The cylindrical radius $\RGC$ is $\RGC\equiv\sqrt{x^2+y^2}$, while the azimuthal angle $\phi$ is set to be 0 at the Sun and increases in the direction of Galactic rotation (clockwise). 

\begin{figure}
\centering
\includegraphics[width=1.0\hsize,trim={0.5cm 2.0cm 0.5cm 1.0cm}]{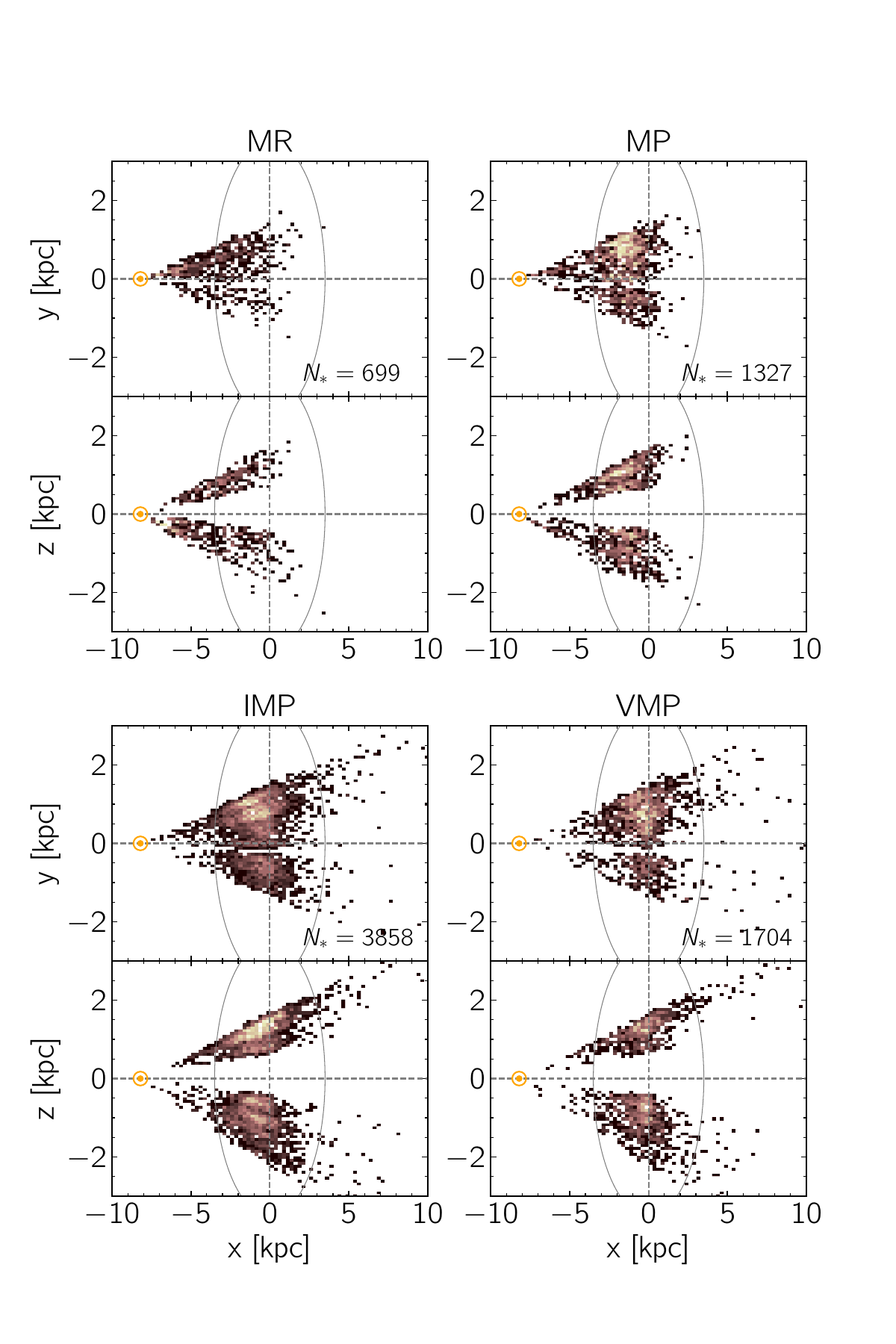}
\caption{Distribution of median $x$, $y$ and $z$ for different metallicity bins, colour-coded by relative density in each panel. The grey lines indicate a circle in $x-y$/$x-z$ with a radius of 3.5~kpc. The metallicity group labels are defined in Table~\ref{table:mpgroups}. The location of the Sun is indicated with the orange symbol. }
\label{fig:xyz-feh} 
\end{figure}

Figure~\ref{fig:xyz-feh} shows the median $x$-$y$ and $x$-$z$ distributions of stars in our different metallicity bins (assuming the position of the Sun and using the 50 Monte Carlo draws as described in the next sub-section). HB stars are excluded. The grey lines indicate a radius of 3.5~kpc from the Galactic centre. Of the stars in the MR, MP, IMP and VMP metallicity ranges (Table~\ref{table:mpgroups}), 52\%, 87\%, 93\% and 89\% have $r < 3.5$~kpc, respectively, where the spherical radius $r = \sqrt{x^2 + y^2 + z^2}$. This is the sample we will use for the remainder of this work, unless otherwise mentioned.  

\subsection{Orbit integration in the \citet{portail2017}/\citet{sormani22} potential}\label{sec:orbits}

To take into account the significant uncertainty in the measurements of the PIGS stars position and kinematics (radial velocities from PIGS and proper motions from \Gaia DR3 data, \citealt{gaiadr3}), we decide to draw samples for each star in the 6-dimensional space $(s,\alpha,\delta,\mu_\alpha,\mu_\delta,v_\text{los})$, where $s$ is a star's heliocentric distance, $\alpha$ and $\delta$ the right ascension and declination, $\mu_\alpha$ and $\mu_\delta$ the corresponding proper motions, and $v_\text{los}$ the line-of-sight velocity. These samples allow us to explore the probability distribution function for the various orbital parameters of the stars in a Monte Carlo fashion. The samples are drawn from the probability distribution function (p.d.f.)
\begin{equation}\label{eq:pdf}
    p(\mathbf{o}|\hat{\mathbf{o}},\sigma_{\mathbf{o}})=f(s;\hat{s},\sigma_s)\delta_\text{D}(\alpha-\hat{\alpha})\delta_\text{D}(\delta-\hat{\delta})\prod_{k=\mu_\alpha,\mu_\delta,v_\text{los}}G(k;\hat{k},\sigma_k),
\end{equation}
where $\mathbf{o}=(s,\alpha,\delta,\mu_\alpha,\mu_\delta,v_\text{los})$, $\hat{\mathbf{o}}=(\hat{s},\hat{\alpha},\hat{\delta},\hat{\mu}_\alpha,\hat{\mu}_\delta,\hat{v}_\text{los})$ represent the measured parameters, and $\sigma_\mathbf{o}$ their uncertainty (except in the case of the distance $s$, see below). We neglect the uncertainty in the sky position $(\alpha,\delta)$, hence their contribution to the p.d.f. are Dirac deltas $\delta_D$. $G(k;\hat{k},\sigma_k)$ is a Gaussian p.d.f. centred at $\hat{k}$ with standard deviation $\sigma_k$, and is a good description of the p.d.f.s of the proper motions and $v_\text{los}$. In the case of the distance $s$, from the {\tt StarHorse} output, we make use of an approximation of the true distance p.d.f. given by a mixture model of three components \citep{Anders22}: 
\begin{equation}
  f(s;\hat{s},\sigma_s)=\frac{\sum_{i=1}^{3}w_iG(s;\hat{s}_i,\sigma_{s,i})}{\int_0^\infty\sum_{i=1}^{3}w_iG(s';\hat{s}_i,\sigma_{s,i})~\text{d}s'}.
\end{equation}
Here, $w_i$ are weights, $\hat{s}=(\hat{s}_1,\hat{s}_2,\hat{s}_3)$, and $\sigma_s=(\sigma_{s,1},\sigma_{s,2},\sigma_{s,3})$. 

Using the p.d.f. of Eq.~\ref{eq:pdf}, we draw 50 samples for each PIGS star, which is expected to be a sufficient representation of the underlying distribution. The values drawn for each star are transformed to Cartesian positions $(x,y,z)$ and velocities $(v_x,v_y,v_z)$, using the distance of the Sun $R_\odot$ and assuming that the velocity of the Sun in Cartesian coordinates is $\mathbf{v}_\odot=(U_\odot,v_0+V_\odot,W_\odot)=(11.10,246.6,7.24)~$kms$^{-1}$, where $v_0$ is the value of the circular speed of the Galaxy at the Sun. The values for $U_\odot$ and $W_\odot$ come from \cite{schonrich2010}, while $v_0+V_\odot$ is in agreement with the proper motion of Sgr A* measured by \cite{reidbrunthaler2004}. The distribution of the median position of the samples in the $(x,y,z)$ space for different metallicity bins is shown in Fig.~\ref{fig:xyz-feh}. The Cartesian $v_x$ and $v_y$ velocities can be transformed to Galactocentric radial and tangential velocities, $v_R$ and $v_\phi$, by
\begin{equation}\label{eq:vRvphi}
    v_R=\frac{xv_x+yv_y}{\RGC},\quad v_\phi=-\frac{xv_y-yv_x}{\RGC},
\end{equation}
where the minus sign in front of the r.h.s. of Eq.~\ref{eq:vRvphi} is used to have $v_\phi$ positive in the sense of the rotation of the Galaxy (i.e. clockwise). The distribution of the samples in the $(v_R,v_\phi)$ space is shown in Figure~\ref{fig:vphivr_draws} for VMP stars.

We integrate the orbits in the \citet[][hereafter S22]{sormani22} barred potential $\Phi(x,y,z,t)$, in its {\tt AGAMA} implementation \citep{vasiliev2019}. This potential is an analytical approximation of the \cite{portail2017} potential, which is a numerical M2M $N$-body model of the Galactic centre, but adding a dark halo to respect constraints on the rotation curve of the Milky Way between $6-8$~kpc \citep{sofue09}. The S22 potential includes an X-shaped inner bar, two long bars, an axisymmetric disc (covering the region outside the bar), a central mass concentration (represented by a triaxial disk), and a flattened axisymmetric dark halo. In this model, the whole potential figure rotates at a fixed pattern speed $\Omega_\text{bar}=39~$\kms~kpc$^{-1}$, in the same sense as the Galactic rotation. The bar is initially inclined at 28~deg from the line connecting the Sun and the Galactic centre, leading with respect to the sense of Galactic rotation. From {\tt AGAMA}, we can also build the `background' axisymmetric and static part of the potential $\Phi_0(x,y,z)$, averaging $\Phi$ along the azimuthal angle. $\Phi_0$, allows us to calculate the circular velocity curve of the Galaxy, which at $R_\odot$ is $v_0=239.17~$\kms (hence, the peculiar velocity of the Sun in the $y$ direction in the Local Standard of Rest of our model is $V_\odot=7.43~$\kms).

From $\Phi_0$, using the `St\"ackel fudge' implemented in {\tt AGAMA}, we can determine the values of the radial, vertical, and azimuthal actions \citep[$J_R$, $J_z$, and $J_\phi$, respectively, see][]{BT08}, for the various samplings of the PIGS stars. $J_\phi$ is just the angular momentum projected along the $z$-axis, $L_z=\RGC v_\phi$. Neither the actions nor the energy,
\begin{equation}\label{eq:energy}
    E = \frac{v_x^2+v_y^2+v_z^2}{2}+\Phi(x,y,z,t),
\end{equation}
are conserved quantities in this barred, time-dependent potential -- they just give a present-day snapshot. We will not use these quantities for any main conclusions. 
From {\tt AGAMA} and $\Phi_0$ we also derive the (axisymmetric) orbital frequencies of the different samples $\Omega_R$, $\Omega_z$, and $\Omega_\phi$. These can vary a lot between the different stars and samples, and provide an estimate of the time that we need to integrate orbits to obtain various orbital parameters. After some tests, we find that integrating orbits (forward in time) in the S22 potential for a time $t_\text{int}=5T_R=5\times 2\pi/\Omega_R$, is good enough to have realistic estimates. We repeat the integration for each sample of each star in PIGS, and compute the positions and velocities on the orbit at 100 equispaced time steps $t_i$, from $t=0$ (now), to $t_\text{int}$. 

For each star sample we compute the pericentre of the orbit as the minimum $r$ between the ones computed at the various time steps $t_i$, i.e. $r_\text{min}=\text{min}\{r(t_i)\}$. Similarly, we define the apocentre of the orbit as $r_\text{max}=\text{max}\{r(t_i)\}$, the maximum height from the plane as $z_\text{max}=\text{max}\{z(t_i)\}$, and the eccentricity as $e=(r_\text{max}-r_\text{min})/(r_\text{max}+r_\text{min})$.

We also derived the orbital properties in three other potentials: the axisymmetric version of the S22 potential, the \citet[][hereafter MM17]{mcmillan17} potential and an adapted {\tt MWPotential2014} \citep[][hereafter aMW14]{bovy15} following \citet[][]{belokurovkravtsov22}, who make it slightly more massive to reproduce the circular velocity at the Sun's radius \citep{blandhawthorngerhard16}. We will use these in later sections to make some sanity checks and/or comparisons with previous work.

It is a challenge to get precise and accurate orbital properties for our metal-poor inner Galaxy stars. The uncertainties in the radial velocities, proper motions and distances result in large uncertainties on the orbital properties of stars -- most of this is driven by the uncertainty in distance. For the main analyses in this work we will use our Monte Carlo samples rather than the median/mean/dispersions, for a better representation of the resulting probability density distributions. Some further discussion and visualisation of the uncertainties are presented in Appendix~\ref{sec:biastest}.

\subsection{Dynamics catalogue}

The distances, velocities and orbital properties for the PIGS sample can be found in Table~\ref{tab:pigscat_orbits}, which is available from the CDS\footnote{Before publication, the PIGS data release can be found \href{https://drive.google.com/drive/folders/1roMPX7p-TkePx-guqPFr8H5vqaih3sga?usp=sharing}{here}.}. For each star we provide the 16th, 50th and 84th percentiles of the distribution. The number of entries (11\,797) is lower than in Table~\ref{tab:pigscat} due to photometric (present and not saturated in $g_\mathrm{PS}$) and spectroscopic quality cuts ({\tt good\_ferre}~=~yes and {\tt rv\_err}~$<$~5~\kms), and/or due to non-convergence in {\tt StarHorse}. 

As discussed previously, the distances for stars in Sagittarius may suffer from biases and should not be used blindly. However, even for Sgr stars with good distances, our derived orbital properties may not be appropriate for these distant stars, given that we have chosen to use a potential that focuses on reproducing the inner Galaxy. 

\section{Results}\label{sec:results}

In this section, we first compare the kinematics of the metal-poor PIGS stars with a different, metal-rich sample of more typical bulge stars. We then discuss how many of the metal-poor stars in PIGS are confined to the inner Galaxy and study the Galactic rotation among these stars. We end with a discussion on the kinematics of CEMP stars. 

\subsection{Comparison with high-metallicity stars from APOGEE}\label{sec:apogee}

Since PIGS is dedicated to observations of the lowest metallicity stars in the inner Galaxy, an internal comparison with metal-rich stars, which should mostly be ``true bulge''/bar stars, is not possible. In this section we instead compare to an external sample of metal-rich stars from APOGEE DR16 \citep{apogeedr16} combined with \Gaia EDR3 \citep{gaiaedr3}, for which \citet[][hereafter Q21]{queiroz21} derived orbital properties for stars towards the bulge, also using spectro-photometric distances from {\tt StarHorse} and employing a very similar Galactic potential (also based on \citealt{portail2017}, but slightly different from the \citealt{sormani22} version) to derive orbital properties as in this work -- this is therefore the closest comparison we can make. The distances (and therefore the derived velocities) for the Q21 sample are of higher quality than those in this work, mostly due to more precise spectroscopic parameters in APOGEE than in PIGS. The authors used a cut in reduced proper motions (RPM) to remove most of the disc stars from their orbital analysis sample and limited their sample to stars with $|z| < 1$~kpc. This RPM sample is the one we compare to. 

\begin{figure}
\centering
\includegraphics[width=0.95\hsize,trim={0.0cm 4.0cm 0.0cm 5.0cm}]{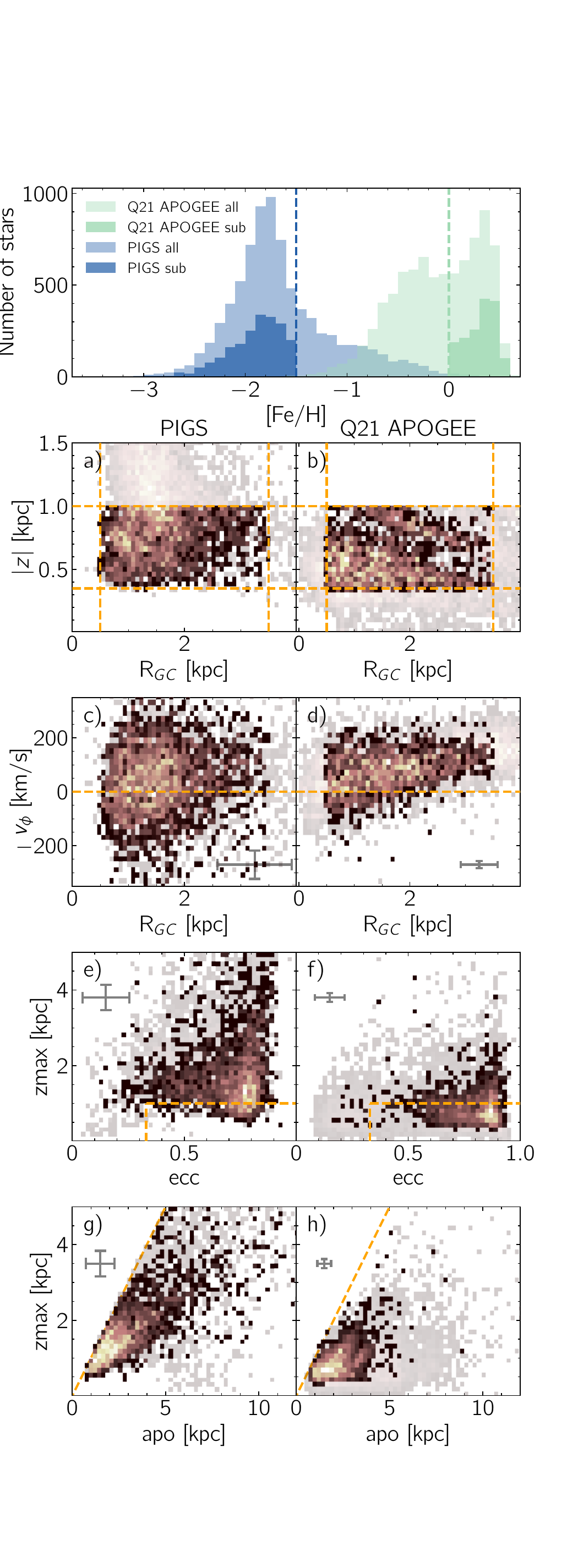}
\caption{Comparisons of metal-poor stars from PIGS and metal-rich stars from APOGEE \citep[the RPM sample from][]{queiroz21}. \textbf{The top panel} shows the metallicity distribution functions for the full samples (light-shaded) and sub-samples highlighted in panels a)-h) (dark-shaded). The sub-samples result from a combination of metallicity cuts ($\feh < -1.5$ and $> 0$ for PIGS and APOGEE, respectively, see vertical lines) and spatial cuts ($0.5 < \RGC < 3.5$~kpc, $0.35 < |z| < 1.0$~kpc [see panels a) and b)] and $|l| < 15^{\circ}$). \textbf{Panels a)-h)} show the spatial and dynamical properties of the full samples (light-shaded) and the sub-samples (dark-shaded), colour-coded by the density of stars, for PIGS (left) and APOGEE (right). The line in panels c) and d) is $v_{\phi} = 0$, the lines in panels e) and f) indicate the bar orbit-dominated region in Q21, and the line in panels g) and h) is the 1-1 line (above which stars cannot lie). Error bars indicate the median uncertainties in the sub-samples (for APOGEE these are standard deviations, for PIGS the 84th - 16th percentile divided by 2). 
}
\label{fig:apogeecomp} 
\end{figure}

\subsubsection{Creating similar spatial sub-samples}

The metallicity distributions for PIGS and the Q21 APOGEE sample (light-shaded histograms) are shown in the upper panel of Figure~\ref{fig:apogeecomp}. They are clearly probing two entirely different (and complementary) metallicity regimes. To make a comparison in the extreme, we further select only PIGS stars with $\feh < -1.5$ and APOGEE stars with $\feh > 0.0$. And although both surveys are targeting the inner Galaxy, the APOGEE and PIGS footprints are quite different, with APOGEE typically looking closer to the Galactic plane. To get a more homogeneous comparison, we further limit both samples to $0.5 < \RGC < 3.5$~kpc, $0.35 < |z| < 1.0$~kpc (see panels $a)$ and $b)$ of Figure~\ref{fig:apogeecomp} for PIGS and Q21 APOGEE, respectively) and $|l| < 15^{\circ}$ for the Q21 APOGEE sample (PIGS is already within that range). The dark-shaded (2D) histograms throughout Figure~\ref{fig:apogeecomp} indicate these sub-samples with our chosen metallicity and spatial cuts (1605 stars in the APOGEE sub-sample, 2053 for PIGS), whereas the light-shaded (2D) histograms represent the full samples (except that for PIGS a cut of $\feh < -1.0$ has been applied to the light-shaded sample to plot the 2D histograms).

\subsubsection{Comparison of the kinematics}
Panels $c)$ and $d)$ in Figure~\ref{fig:apogeecomp} show $v_\phi$ as a function of galactocentric radius, again with PIGS on the left and Q21 APOGEE on the right. Q21 APOGEE has a higher net rotation, and both samples show a decreasing azimuthal velocity and increasing velocity dispersion for stars closer to the Galactic centre. The velocity dispersion in PIGS is much larger than in the Q21 APOGEE sample. This is partly due to the larger distance/velocity uncertainties in PIGS (for this sub-sample, the median uncertainty in $v_\phi$ is 12~\kms for APOGEE and 53~\kms for PIGS), but that is not the only factor. In the Q21 APOGEE sub-sample, the $v_\phi$ dispersion for $\RGC > 1.5$ is 55~\kms (without correcting for uncertainties), whereas in PIGS it is over 90~\kms and increasing for $\feh < -1.5$ (see Figure~\ref{fig:vsig}, corrected for uncertainties). For further details and maps of velocities and velocity dispersions in the Q21 APOGEE sample, see \citet{queiroz21}. Overall, between metal-poor (PIGS) and metal-rich (APOGEE) stars, we see a similar trend of decreasing velocity and increasing velocity dispersions closer to the Galactic centre, but the magnitudes are different for the metal-poor and the metal-rich stars. 

We present the distributions of eccentricity and maximum height above the plane during the orbit in panels $e)$ and $f)$ of Figure~\ref{fig:apogeecomp} for PIGS and Q21 APOGEE stars, respectively. The metal-rich Q21 stars are strongly concentrated in the bottom-right corner of the diagram\footnote{The authors have recently recomputed the orbits with the full \citet{sormani22} potential and find that the $z_\text{max}$ becomes slightly larger, looking a bit more like the PIGS distribution, however, their previous results were robust and the changes do not impact any conclusions (private communication).}, with very high eccentricities and low $z_\mathrm{max}$. \citet{queiroz21} find that stars on bar-shaped orbits mostly lie in this region (eccentricity $>0.66$ and $z_\mathrm{max} < 1$~kpc), using frequency analysis. They also find a significant number of bar-shaped orbits among stars with lower eccentricity ($0.33 - 0.66$) and $z_\mathrm{max} <1$~kpc, but not many in other regions of this space. The region with most bar-dominated orbits has been indicated by orange lines. Most of the stars in PIGS lie outside this bar-dominated region, although stars are still concentrated towards high eccentricity and relatively low $z_\mathrm{max}$. As expected, we find that the distribution of PIGS stars in this space is quite different when we rerun the orbits in the axi-symmetric \citet{sormani22} potential. In this case, the eccentricities are more homogeneously spread between eccentricities of $0.3-0.85$ instead of having such a strong clump at eccentricities $>0.7$, and the distribution of stars extends to slightly lower $z_\mathrm{max}$ for eccentricities $<0.7$. The bar appears to strongly influence the orbital properties of these metal-poor inner Galaxy stars (although it does not strongly affect the apocentres, see the discussion in Section~\ref{sec:confined}).

The final row of Figure~\ref{fig:apogeecomp} presents the distributions of apocentres and maximum height above the plane in PIGS and Q21 APOGEE. The different distributions between the two could partly be due to the different spatial distributions, with the PIGS stars having higher $|z|$ on average, but this is unlikely the full explanation as the differences are larger than 0.5~kpc. The metal-rich Q21 APOGEE stars mostly lie away from the $1-1$ line, indicating that the distribution of stars is flattened, as expected for bar stars. The metal-poor PIGS distribution has more stars closer to the $1-1$ line than the metal-rich Q21 APOGEE stars, indicating a more spheroidal distribution, again as expected. The PIGS distribution looks somewhat different in the axi-symmetric S22 potential. The apocentres remain very similar, but the $z_\mathrm{max}$ changes, which changes the distribution to less of a tight ``cone'' in this plane and more of a ``blob'', centred on (apo,$z_\mathrm{max}$) = (2,1). 

\begin{figure}
\centering
\includegraphics[width=0.9\hsize,trim={0.0cm 0.8cm 0.0cm 0.0cm}]{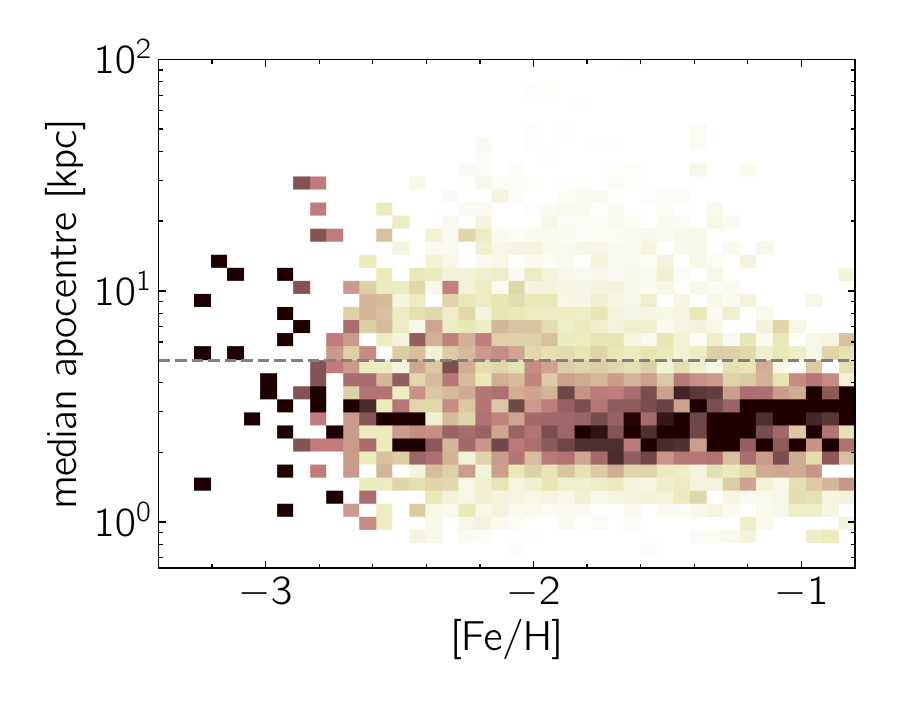}
\includegraphics[width=0.9\hsize,trim={0.0cm 0.8cm 0.0cm 0.0cm}]{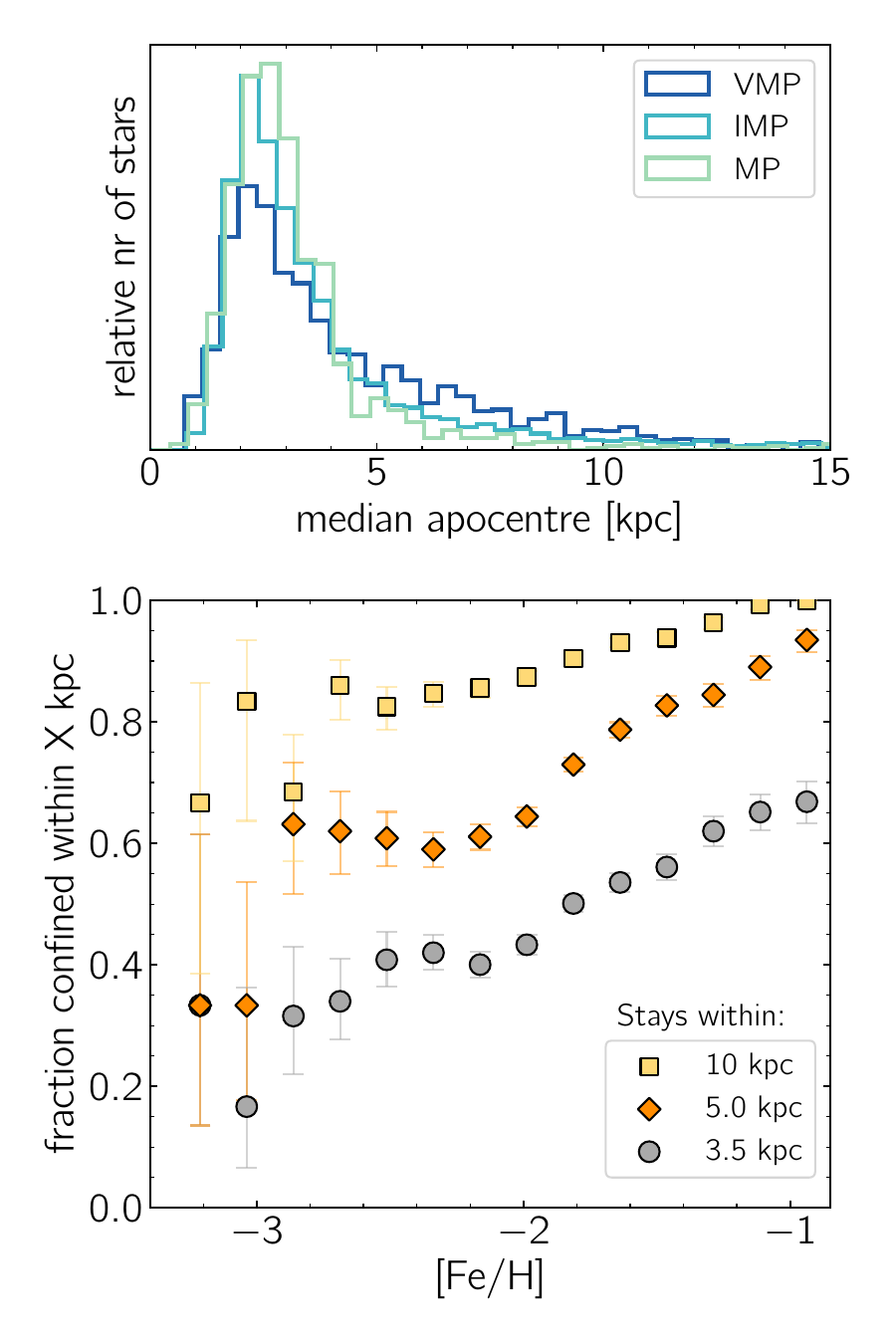}
\caption{Top: column-normalised 2D histogram of metallicity versus median apocentre, for stars with $r < 3.5$~kpc. The darkest pixels have the highest density for a given metallicity. The grey dashed horizontal line indicates 5~kpc. Middle: 1D normalised slices of the distribution above, for the MP, IMP and VMP samples. Bottom: the fraction of stars in the same sample that is confined within a given distance (see legend) from the Galactic centre for $\geq 75\%$ of their orbit samples, as a function of metallicity. Error bars represent $1\sigma$ uncertainties derived from binomial statistics.
} 
\label{fig:interlopers} 
\end{figure}

\begin{figure}
\centering
\includegraphics[width=0.9\hsize,trim={0.0cm 0.8cm 0.0cm 0.0cm}]{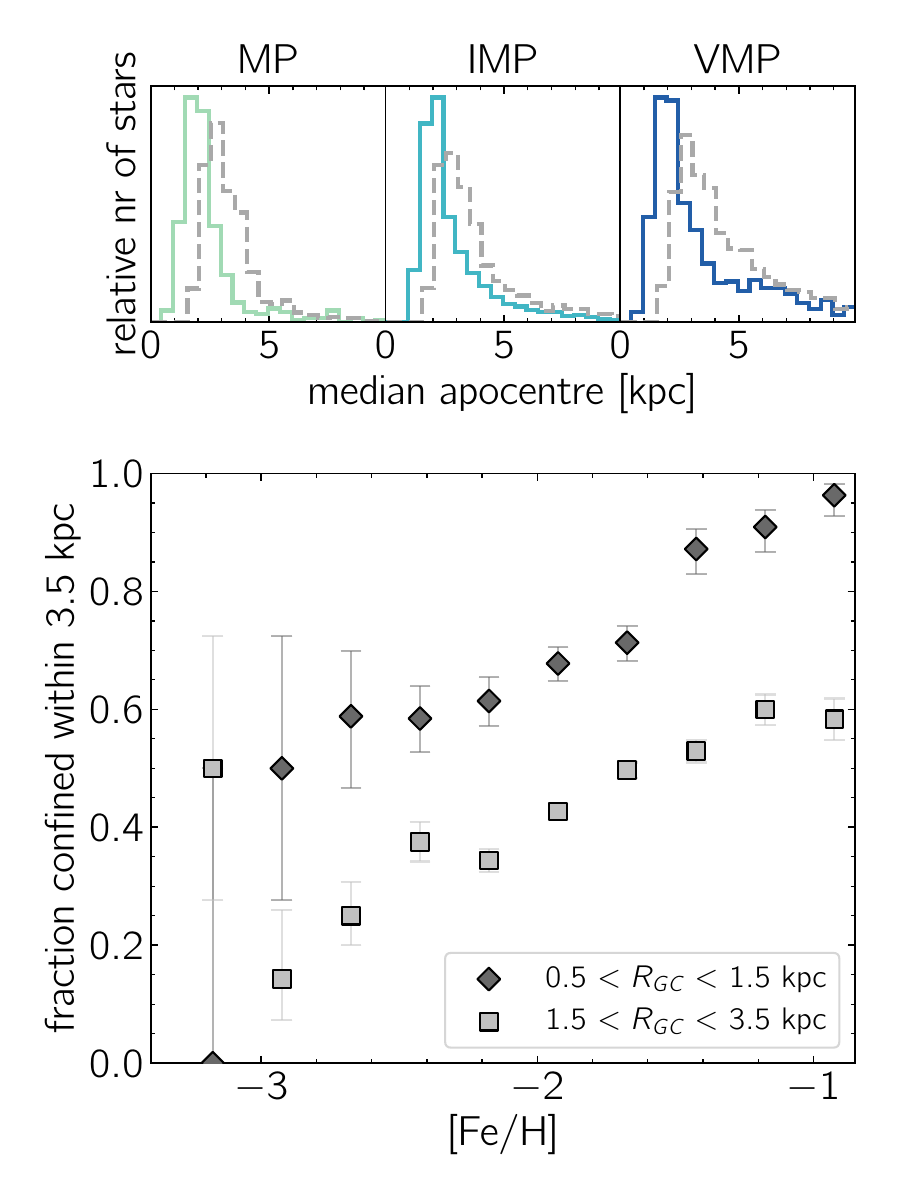}
\caption{Top: distributions of median apocentre for the three metallicity slices for stars with $0.5 < R_{GC} < 1.5$~kpc (coloured solid lines) and for stars with $1.5 < R_{GC} < 3.5$~kpc (dashed grey lines). Bottom: similar to the bottom panel in Figure~\ref{fig:interlopers}, now only for stars staying within 3.5~kpc and split for the same two $R_{GC}$ bins as above. The changes for confinement within 5 or 10~kpc are smaller and not shown.
} 
\label{fig:interlopers_sub} 
\end{figure}

\subsection{Metal-poor stars are less confined than metal-rich stars}\label{sec:confined}

Not all stars that are currently in the inner Galaxy, stay within the inner Galaxy \citep[e.g.][]{kunder20,lucey21, sestito23}. We present the distribution of median apocentres as a function of metallicity for PIGS stars with $r < 3.5$~kpc in the top and middle panels of Figure~\ref{fig:interlopers}. The peak of the apocentre distribution is between 2 and 3~kpc, with lower values for more metal-poor stars -- this is likely partly due to the difference in spatial distributions for the different metallicity bins. The width of the distribution increases with decreasing metallicity, and a prominent tail with apocentres $>5$~kpc becomes visible for $\feh < -2.0$. We perform a Kolmogorov-Smirnov test on the cumulative apocentre distributions for the MP, IMP and VMP metallicity bins and find they are statistically significantly different from each other (p-values $<10^{-9}$). 

The bottom panel of Figure~\ref{fig:interlopers} presents the fraction of stars in a given metallicity bin that is confined to within 3.5~kpc, 5~kpc and 10~kpc in grey circles, orange diamonds and yellow squares, respectively. For this analysis we do not use the median apocentre but require 75\% of the orbit draws of each star to be below the apocentre limit. We find that the fraction of stars confined to each of the three radii drops with decreasing metallicity, although there appears to be a break around $\feh = -2.0$, below which the fraction of confined stars stays relatively constant (except at the lowest metallicities). For a Galactocentric distance of 3.5~kpc, the confined fraction decreases from $65\%$ at $\feh = -1.0$ to $40\%$ for VMP stars. These fractions go up to $95\%$ and $60\%$ for a Galactocentric distance of 5~kpc, and to $100\%$ and $\sim 85\%$ for a Galactocentric distance of 10~kpc. 

We also check how these results change when adopting a different Galactic potential. The differences between the barred and axisymmetric S22 potentials are within the uncertainties. There are differences to the orbits, but they affect the pericentres more than the apocentres. For the MM17 and aMW14 potentials (which are somewhat more massive than the S22 potential), the trend with metallicity becomes steeper and the fractions of confined stars within 3.5 or 5.0 kpc slightly increase (those for 10~kpc remain similar). For example, the confined fraction within 3.5~kpc at $\feh = -1.0$ rises from $\sim 65\%$ to $\sim 70\%$ to $\sim 80\%$ for the S22, MM17 and aMW14 potentials, respectively. A less strong rise is seen at $\feh = -2.0$ (from just under $45\%$ to just over $45\%$ to $\sim 50\%$). The overall trends remain the same and, if anything, we might be underestimating the fraction of confined stars. 

The median apocentre and confined fraction are likely to depend on the distance to the Galactic centre ($R_{GC}$). We confirm this in Figure~\ref{fig:interlopers_sub}, which shows the distributions of the median apocentres (top) and the confined fraction within 3.5~kpc (bottom) for two rings of $R_{GC}$. The top panel shows that the median apocentre distribution peaks strongly at lower apocentres for the inner $R_{GC}$ ring. The bottom panel shows the same trend as seen in the bottom panel of Figure~\ref{fig:interlopers} for both $R_{GC}$ rings, but the overall fraction of confined stars is larger in the inner Galactic radius ring -- as expected.

\subsection{VMP stars still show a coherent rotational signature}\label{sec:rotation}

Previous work has suggested that the rotational signature of stars around the Galactic centre disappears for very metal-poor stars \citep{arentsen20_I, lucey21, rix22}. These papers were limited in various ways, but we can now re-address this question with larger samples of VMP stars that have full orbital properties available. In this section we derive how the azimuthal velocity around the Galactic Centre behaves with metallicity. 

For each of the 50 realisations of the PIGS sample (see Section~\ref{sec:orbits}), we use an Extreme Deconvolution Gaussian Mixture Model (XDGMM, see \citealt{bovy11}) to fit the mean $v_\phi$, $v_R$ and $v_z$ and their velocity dispersions. Velocity uncertainties were assigned per star taking from the 50 draws the 84th percentile minus the 16th percentile divided by two (``1$\sigma$''). 

\begin{figure*}
\centering
\includegraphics[width=0.9\hsize,trim={0.0cm 0.0cm 0.0cm 0.0cm}]{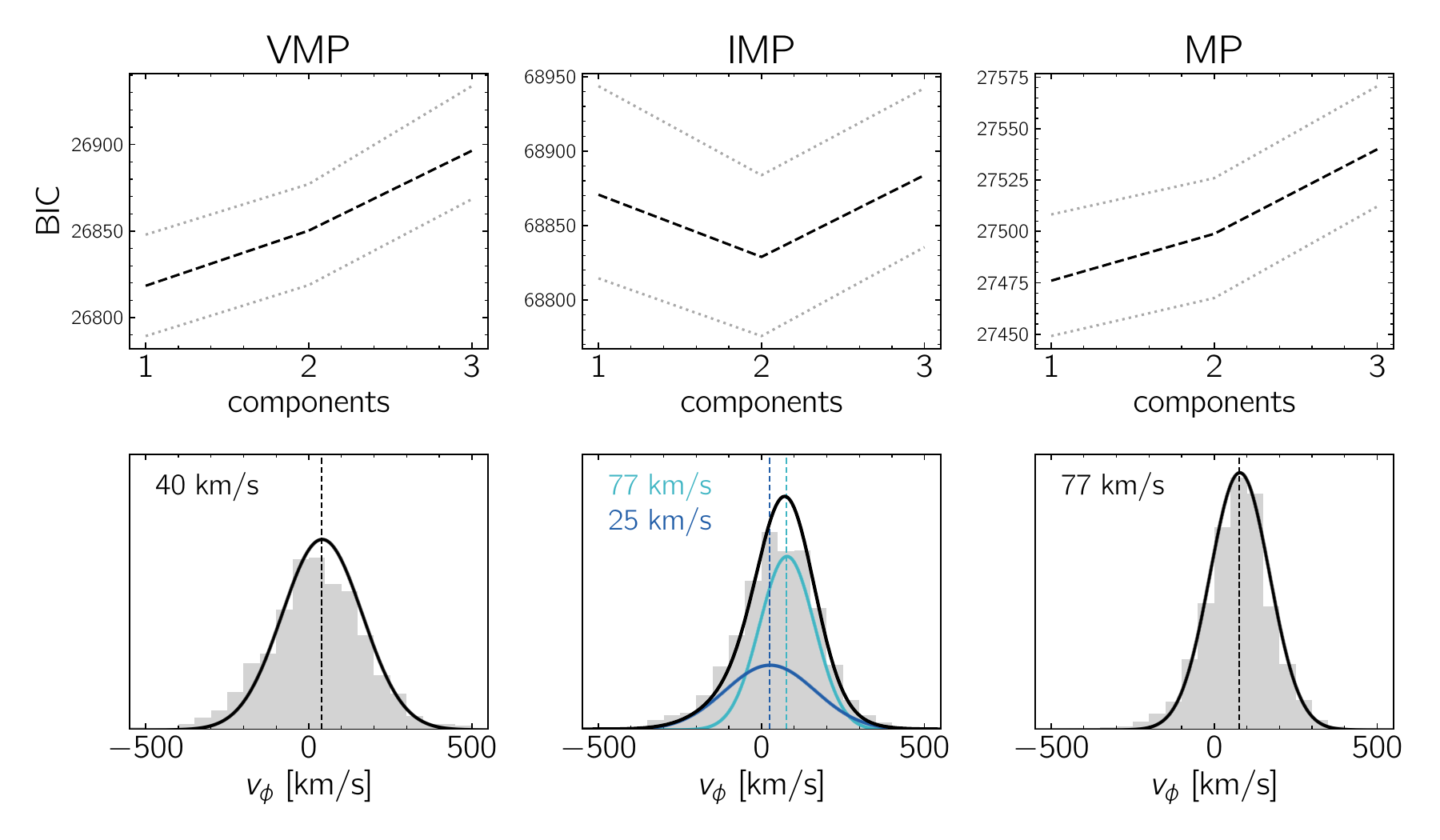}
\caption{Top: Bayesian Information Criterion (BIC) for the 3D XDGMM fit to the 50 VMP, IMP and MP samples, with the black line showing the median and the grey lines $\pm 1\sigma$. Bottom: median $v_\phi$ distributions (grey histograms) with the 1D preferred components from the XDGMM (lines). The middle panel shows the two separate components in colour and their sum in black. The means of the XDGMM Gaussian fits are indicated by dashed lines and in text in the top left corners. Note that the histograms are not corrected for the uncertainties, while the Gaussian curves come from the extreme deconvolution, which takes into account the uncertainties.
} 
\label{fig:gaussians} 
\end{figure*}

We attempt fitting multiple components for the MP, IMP and VMP samples. We employ the Bayesian Information Criterion (BIC) to identify the preferred number of components -- the lowest BIC indicates the preferred number of components. We limit the fit to stars within $1.5 < R_{GC} < 3.5$~kpc (as the uncertainties are more problematic for stars closer to the Galactic centre, see Section~\ref{sec:disp}). The results are summarised in Table~\ref{table:gmm} and Figure~\ref{fig:gaussians}. The figure shows the BIC for the fits in the top row and the 1D $v_\phi$ distributions with preferred Gaussian fits in the bottom row. The BIC for the VMP and MP samples prefers one component, while the BIC for the IMP sample prefers a two-component fit. The two components in the IMP regime are a slowly rotating, hotter component ($\overline{v}_\phi = 18$~km/s, $\sigma = 133$~km/s, amplitude $=0.38$) and a faster rotating, slightly cooler component ($\overline{v}_\phi = 78$~km/s, $\sigma = 86$~km/s, amplitude $=0.62$). 

It is still possible that there are multiple components in the VMP and MP regimes, there do seem to be more stars at negative $v_\phi$ than the 1D model represents in both the VMP and MP distributions. One reason that an extra component would remain undetected might be that for all PIGS stars, the velocity uncertainties are large, and specifically in these two metallicity ranges there are $\sim 2.5$ times fewer stars than the IMP range -- this makes it difficult to detect multiple components with significance. To test what the contribution of the two halo components might be in the MP and VMP regimes, assuming the same components are present as those identified in the IMP regime, we re-ran the XDGMM with two frozen GMM components, only varying the relative amplitudes. In this case, we find the mean amplitudes for the hotter component to be $0.60 \pm 0.02$ in the VMP regime and $0.17 \pm 0.01$ in the MP regime, thus for the cooler component $0.40 \pm 0.02$ (VMP) and $0.83 \pm 0.01$ (MP).


\begin{table}
\centering
\caption{\label{table:gmm} Results from the XDGMM fitting for mean velocity and velocity dispersion for stars in the MP, IMP and VMP samples with $1.5 < R_{GC} < 3.5$ kpc. The values reported are the mean and standard deviation of the fits to the 50 MC samples.}
\begin{tabular}{lcccc}
 \hline
    & $v_R$ [\kms] & $v_\phi$ [\kms] & $v_z$ [\kms] & amplitude\\
 \hline 
 \multicolumn{2}{|l|}{\textbf{MP} \textbf{($N=754)$}} & & & \\
 \hline
 $\mu$ & $9 \pm 2$ & $78 \pm 2 $ & $-6 \pm 1$ & \\
 $\sigma$ & $103 \pm 1$ & $91 \pm 2 $ & $92 \pm 1$ & \\ 
 \hline
 \multicolumn{2}{|l|}{\textbf{IMP} \textbf{($N=1829)$}}  & & & \\
 \hline
 $\mu_1$ & $-4 \pm 3$ & $78 \pm 4 $ & $0 \pm 2$ & $0.62 \pm 0.03$\\
 $\sigma_1$ & $87 \pm 5$ & $84 \pm 2 $ & $88 \pm 2$ & \\ 
 $\mu_2$ & $8 \pm 6$ & $26 \pm 6 $ & $1 \pm 4$ & $0.38 \pm 0.03$\\
 $\sigma_2$ & $152 \pm 7$ & $139 \pm 6 $ & $150 \pm 4$ & \\ 
 \hline
 \multicolumn{2}{|l|}{\textbf{VMP} \textbf{($N=695)$}}  & & & \\
 \hline
 $\mu$ & $6 \pm 3$ & $20 \pm 3 $ & $3 \pm 1$ & \\
 $\sigma$ & $134 \pm 2$ & $123 \pm 3 $ & $127 \pm 2$ & \\ 
 \hline
\end{tabular}
\end{table}

\subsubsection{Mean velocities}

\begin{figure}
\centering
\includegraphics[width=0.9\hsize,trim={0.0cm 0.0cm 0.0cm 0.0cm}]{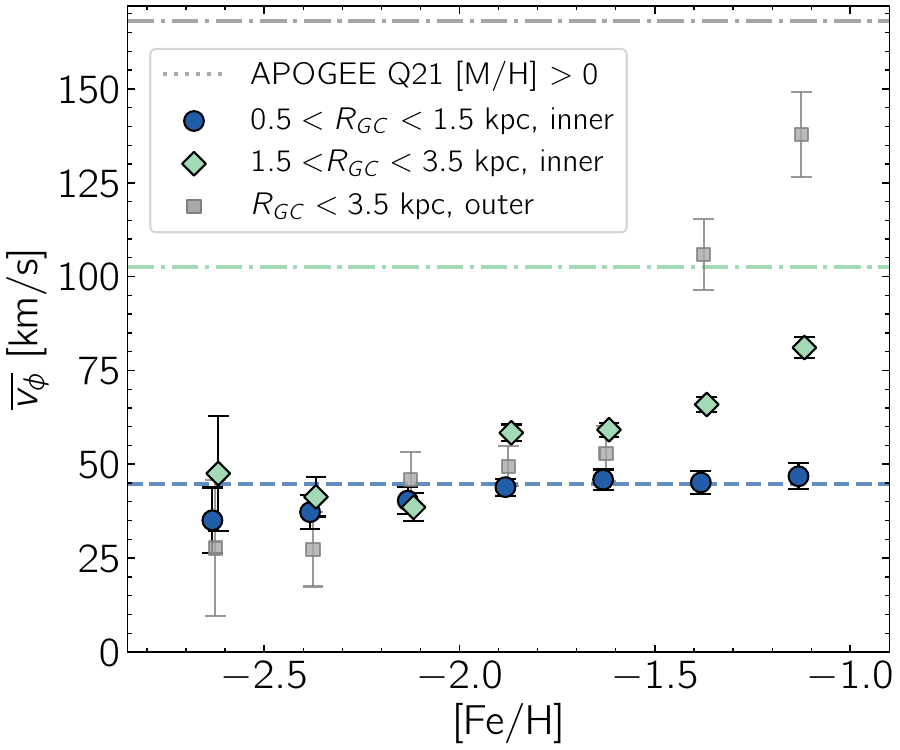}
\includegraphics[width=0.9\hsize,trim={0.0cm 0.0cm 0.0cm 0.0cm}]{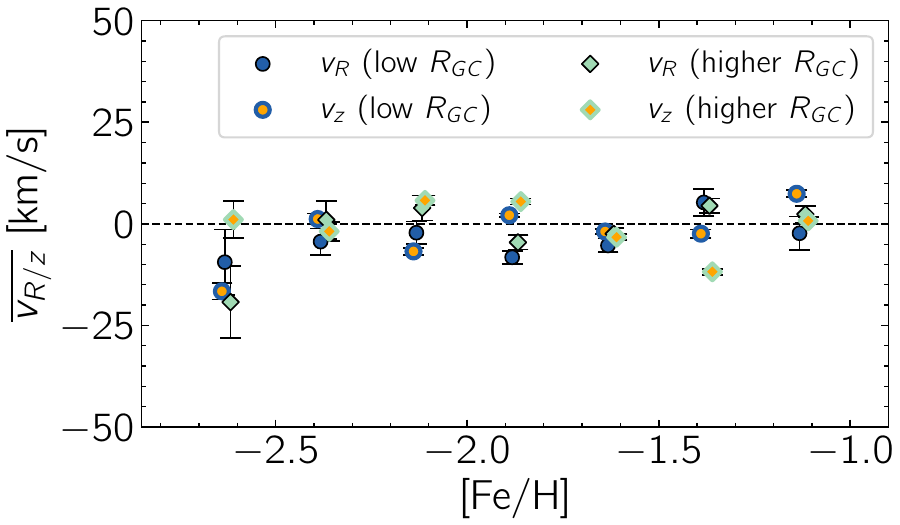}
\caption{Top: Azimuthal velocity $v_\phi$ as a function of metallicity \feh for PIGS stars. The sample is split into three subsets: $0.5 < \RGC < 1.5$~kpc (low $R_{GC}$) ``inner'' and $1.5 < \RGC < 3.5$~kpc (higher $R_{GC}$) ``inner'' (both with median apocentres less than 5~kpc) and $\RGC < 3.5$~kpc ``outer'' (with median apocentres between $5-15$~kpc). The markers and their error bars indicate the mean and the standard deviation of each of the 50 PIGS realisations, using the XDGMM. Small offsets in \feh have been applied for visual clarity. The horizontal lines indicate the median $v_{\phi}$ for the \citet{queiroz21} reduced proper-motion APOGEE sample with [M/H]~$> 0$ (plus spatial cuts, see Figure~\ref{fig:apogeecomp}), in the same $\RGC$ samples. 
Bottom: Vertical velocities $v_z$ and radial velocities $v_R$ as a function of metallicity, for the same apocentre~$< 5$~kpc samples as above. 
} 
\label{fig:vphitrend} 
\end{figure}

For the following analysis, we fit one Gaussian component, in metallicity slices of 0.25~dex wide. The results for $v_\phi$ as a function of metallicity are shown in the top panel of Figure~\ref{fig:vphitrend}, for stars with apocentres $<5$~kpc (this predominantly removes stars in the hotter halo component) and two ranges of median cylindrical Galactocentric radius $\RGC$ with roughly equal number of stars -- blue circles are for stars closer to the Galactic centre, green diamonds are for stars further away. Each of the metallicity bins contains at least 150 stars, except for $\feh < -2.5$ for the blue circles and $\feh < -2.25$ for the green diamonds (those have between 30-110 stars). We also show the results for stars with median apocentres of $5-15$~kpc in grey squares. We find that for stars with $\RGC > 1.5$~kpc and apocentres less than 5~kpc, the azimuthal velocity drops as a function of metallicity. However, it does not drop to zero, even for VMP stars, which still have a $v_\phi \sim 40$~\kms, independent of metallicity. We find a similar trend for the sample with larger apocentres (grey symbols), although the average $v_\phi$ is systematically lower in this sample, except for stars with $\feh > -1.5$, where it is much higher. This is possibly due to a starting contribution from the disc/bar. For PIGS stars with $\RGC < 1.5$~kpc, the azimuthal velocity appears to be constant with metallicity at a level of $\sim 45$~\kms, with a slight drop to $\sim 35$~\kms between $-2.7 < \feh < -2.0$. We find that the average $v_R$ and $v_z$ are consistent with zero, see the bottom panel of Figure~\ref{fig:vphitrend}, as expected.

\begin{figure}
\centering
\includegraphics[width=0.9\hsize,trim={0.5cm 0.5cm 0.5cm 0.0cm}]{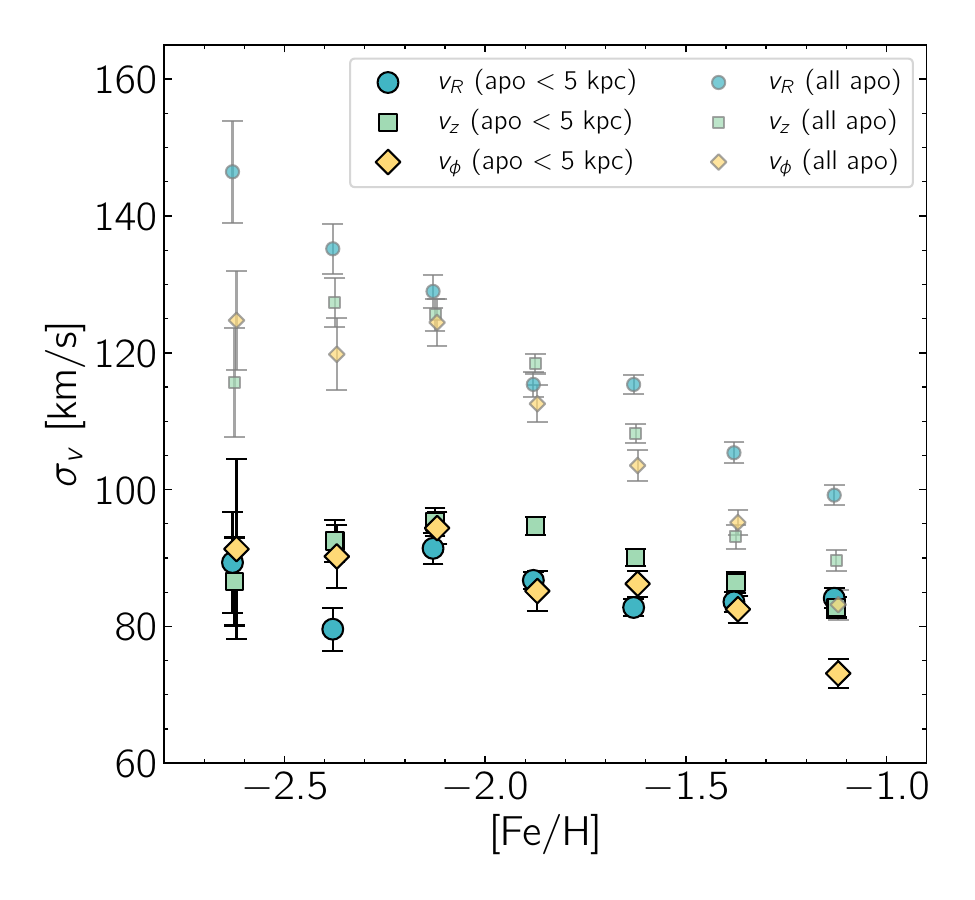}
\caption{Velocity dispersions in the three cylindrical velocity components as a function of metallicity, for stars with $1.5 < \RGC < 3.5$~kpc and apocentre $<5$~kpc (large solid symbols) and for stars with all apocentres (smaller, semi-transparent symbols). Small offsets in \feh have been applied for visual clarity. 
}
\label{fig:vsig} 
\end{figure}

To compare with azimuthal velocities for ``typical bulge'' stars, we use the reduced proper motion APOGEE bulge sample from \citet{queiroz21}, selecting metal-rich stars with [M/H]~$>0$ in our three $\RGC$ and apocentre bins (limiting to $|l| < 15^{\circ}$ and $|z| > 0.35$~kpc to make the footprint more similar to PIGS, see also Section~\ref{sec:apogee}). The median $v_\phi$ for these samples is shown by the horizontal lines. The rotational signature for the metal-rich APOGEE stars with $\RGC > 1.5$~kpc ($v_{\phi} = 108$~\kms) and with apocentres between 5 and 15 kpc ($v_{\phi} = 168$~\kms) is stronger than the signal in PIGS for all metallicities. For the inner $\RGC$ bin, the APOGEE stars have the same azimuthal velocity as the PIGS stars. The relative difference between metal-rich and metal-poor becomes larger the further away from the Galactic centre stars are located (and/or the larger their apocentres are), which is not surprising given e.g. a larger fraction of disc contamination in the metal-rich regime, and the disc becoming a more dominant component further away from the centre.

\subsubsection{Velocity dispersions}\label{sec:disp}

We also derive the velocity dispersion for our stars. These are corrected for the uncertainties because we used extreme deconvolution to fit the velocity distribution. 
If we do not split the sample by apocentre, we find that the velocity dispersion strongly and continuously increases for stars with decreasing metallicity, see Figure~\ref{fig:vsig} (only including stars with $1.5 < \RGC < 3.5$~kpc). This appears to be mostly due to the increased fraction of stars with larger apocentres at low metallicity (Figure~\ref{fig:interlopers}), which enhances the velocity dispersion. For this sample, the velocity dispersions are similar for each of the velocity components, and the anisotropy parameter $\beta = 1 - (\sigma_\theta^2 + \sigma_\phi^2)/(2\sigma_r^2)$ is therefore low ($\sim 0.2$ or below).  

After removing stars with apocentres $>5$~kpc, the trend with metallicity is much less strong. There is still a weak trend of increasing velocity dispersion with metallicity (for $-2.0 < \feh < -1.0$), most strongly for $v_\phi$ and only very weakly for $v_R$. The trends between velocity dispersions and metallicity are possibly still due to the different apocentre distributions at different metallicities. At every metallicity in our sample, we find that $v_\phi/\sigma_z < 1$ -- the population is pressure-supported. 

We find spurious results for stars with $\RGC < 1.5$~kpc, with lower velocity dispersions in $v_\phi$ and $v_R$ that are constant with metallicity (or even decreasing) at $\sim 75$~\kms, and higher velocity dispersions in $v_z$, rising from 90~\kms at $\feh = -1.0$ to 120~\kms at $\feh = -2.4$. For stars this close to the Galactic Centre, a small change in distance can place a star in front or behind the Galactic Centre, changing the direction of its velocity. This results in non-Gaussian velocity distributions for $v_\phi$ and $v_R$ (see Figure~\ref{fig:vphivr_draws}), and these are not well-represented by the ``1$\sigma$'' uncertainties we assigned. The $v_\phi$ and $v_R$ uncertainties for these stars are likely overestimated, allowing the XDGMM to therefore find lower ``intrinsic'' velocity dispersions for these velocity components. 
This issue should not affect $v_z$, because it does not flip sign depending on the distance to the star (although it does change in magnitude). We find that the velocity dispersion for $v_z$ is about $10-15$~\kms higher for the $\RGC < 1.5$~kpc sample compared to the higher $\RGC$ sample, down to $\feh = -2.2$. This suggests that the metal-poor inner Galaxy population has a (slightly) higher velocity dispersion closer to the Galactic centre.  

\subsection{No evidence for different kinematics for CEMP stars}

Carbon-enhanced metal-poor (CEMP) stars are chemically peculiar VMP stars with large over-abundances of carbon ($\cfe > 0.7$). A large fraction of VMP stars ($\sim$10$-$30$\%$, and larger towards lower metallicity) is thought to be carbon-rich \citep[e.g.][]{placco14,arentsen22}, so it is important to understand them better. They are thought to have either received a large amount of carbon from a former asymptotic branch star binary companion (CEMP-s, see e.g. \citealt{lucatello05,bisterzo10,abate15a}), or were born with their large carbon abundance (CEMP-no). The latter type is thought to be connected to the properties of the First Stars and their explosions \citep[see e.g.][]{chiappini06, meynet06, umedanomoto03, nomoto13, tominaga14}. 

\citet{arentsen21} found that the fraction of CEMP stars appears to be lower in PIGS than in other VMP surveys that target the halo. The authors suggested that this difference could indicate differences in the early chemical evolution and/or binary fraction between the building blocks of the inner Galaxy and the rest of the halo \citep[see also][for discussions around low CEMP fractions in the inner Milky Way]{howes16, pagnini23}. Other explanations for the apparently low CEMP fraction in PIGS could be photometric selection effects in the \Pristine survey \citep{arentsen21} and/or systematic differences in the analysis and sample selection biases of comparison halo samples \citep{arentsen22}. However, it is likely that these cannot fully explain the low fraction of CEMP stars in PIGS (as argued by those authors), and we might therefore expect to see a difference in the orbital properties between the CEMP stars and carbon-normal stars with similar metallicities in PIGS. If the fraction of CEMP stars is lower in the large early Galactic building blocks, we would expect the CEMP stars that are currently in the inner Galaxy to come from smaller building blocks, and hence to be less confined.  

An important caveat when looking at the orbital properties for CEMP stars in PIGS is that there might be systematic effects in the derived {\tt StarHorse} distances of CEMP stars. The first reason for this is that {\tt StarHorse} includes photometry in the distance derivation, and the photometry for CEMP stars can be different from that of normal VMP stars due to large molecular carbon-bands being present in the spectrum \citep[see e.g.][]{dacosta19, chiti20}. This effect is very sensitive to the effective temperatures and carbon abundances of the CEMP stars, being worse for cooler and more carbon-rich stars. It typically makes stars look fainter in filters that contain strong carbon features (e.g. $g_\mathrm{PS}$), but the effects have not yet been quantified systematically. The second reason is that the PIGS/FERRE stellar parameters for very carbon-rich objects are sometimes less accurate -- especially the surface gravities, which are found to be systematically lower for CEMP stars than for carbon-normal VMP stars \citep{arentsen21}. These issues (biased photometry and spectroscopic stellar parameters) might be expected to be worse for CEMP-s stars, which typically have higher \feh and \cfe and therefore stronger carbon features than CEMP-no stars.  

\begin{figure}
\centering
\includegraphics[width=1.0\hsize,trim={1.0cm 1.0cm 0.5cm 1.0cm}]{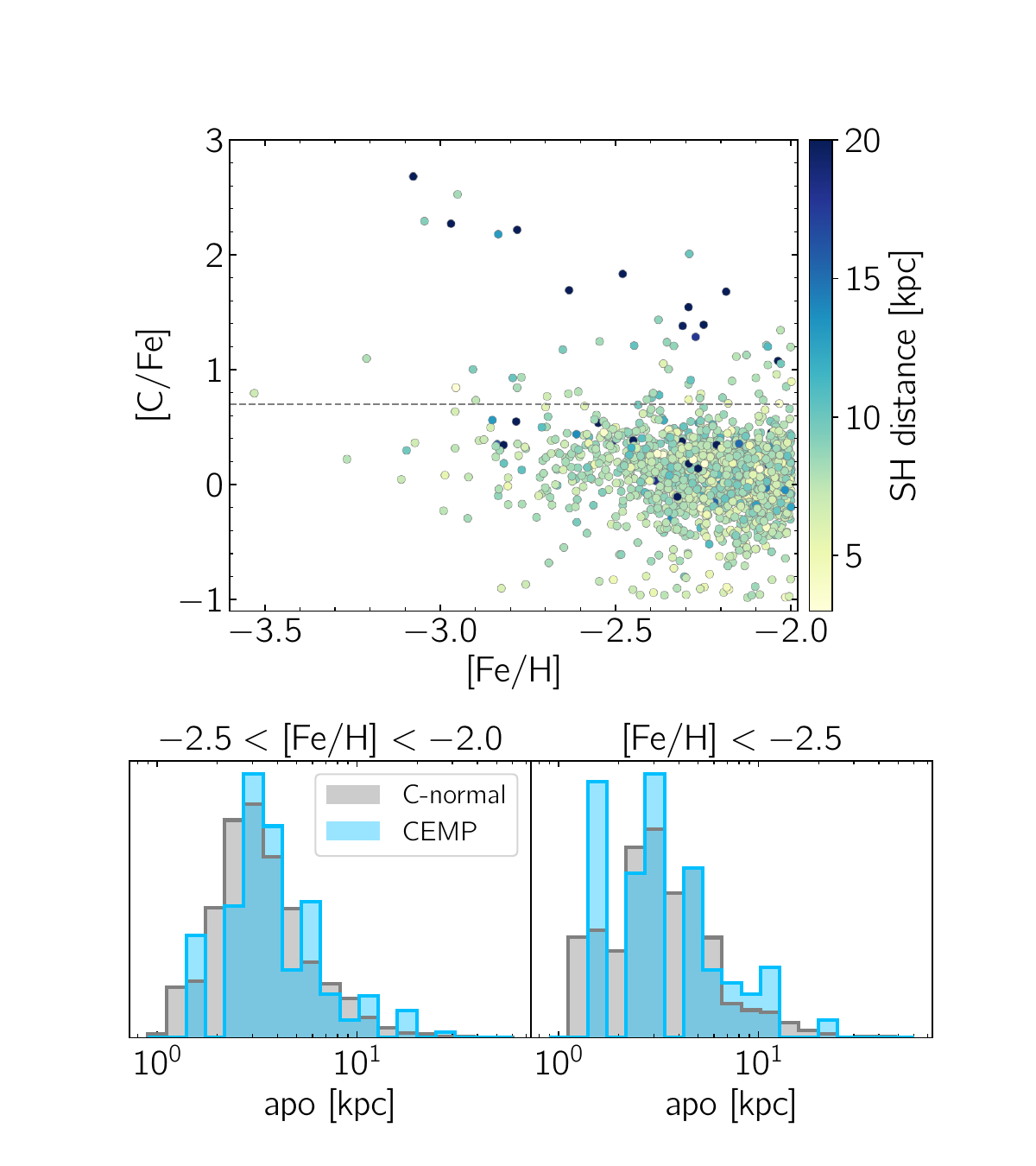}
\caption{Orbital properties of carbon-normal and CEMP stars. Top: \feh -- \cfe diagram colour-coded by StarHorse distance. CEMP stars have $\cfe > 0.7$ (above the dashed grey line). Bottom: apocentre distributions for carbon-normal and CEMP stars in two metallicity ranges. }
\label{fig:cemp} 
\end{figure}

The top panel of Figure~\ref{fig:cemp} presents the carbon abundances as a function of metallicity for the VMP stars in the sample used in this paper, colour-coded by {\tt StarHorse} distances. We find that most of the very carbon-rich stars ($\cfe > 1.5$) have large distances from the Sun ($d > 20$~kpc). This is likely due to the above-described caveats, and not necessarily because they truly are that far away. The moderately carbon-enhanced ($0.7 < \cfe < 1.5$) CEMP stars appear to be less drastically affected, although there could still be subtle systematic issues with their distances. 

Keeping these caveats in the back of our minds, we select all stars in our sample with Galactocentric distances $< 5$~kpc (which removes the drastic outliers above), and present the apocentre distributions for carbon-normal stars ($\cfe < 0.5$) and CEMP stars ($\cfe > 0.7$) in the bottom panels of Figure~\ref{fig:cemp}. We split it into two metallicity bins, since we previously found that the number of halo interlopers varies with metallicity (see Figure~\ref{fig:interlopers}). The apocentre distributions for CEMP and carbon-normal stars look very similar, possibly with a slight offset to larger apocentres for the CEMP stars. A Kolmogorov-Smirnov test suggests that there is no evidence for the C-normal and CEMP stars coming from different underlying populations (KS statistics of 0.15 and 0.20, and p-values of 0.49 and 0.68, for $-2.5 < \feh < -2.0$ and $\feh < -2.5$, respectively, meaning that there is no statistically significant difference). Given the limitations of the data and our approach, we conclude that further work is necessary to investigate the origin of CEMP stars in the inner Galaxy.

\section{Discussion}\label{sec:discussion}

We first briefly discuss some of the limitations of our data and our approach. We then dive into a comparison with the literature and possible interpretations of our results, and finally discuss some directions for future work. 

\subsection{Possible limitations}

As discussed earlier while describing the results, there are some limitations as to what is possible to derive from the data alone given the large uncertainties on some of the parameters (distance, in particular). Systematic biases might also affect the results. 

For the distances, systematic biases could, for example, come from limitations in the \texttt{StarHorse} analysis. We used one set of isochrones (PARSEC), but it is well-known that there are some variations between different isochrone sets (and there may also be systematics between models and data). These variations tend to be larger for metal-poor stars, where the assumptions on physical processes are less well-constrained. Testing \texttt{StarHorse} with different isochrone sets is beyond the scope of this work, but this might be an interesting exercise for future work. Furthermore, the extinction law has been fixed in the \texttt{StarHorse} analysis, although it has been shown that the extinction law may vary across the sky as well as in the inner Milky Way region \citep[e.g.][]{schlafly16, nataf16}. Finally, the density priors assumed in \texttt{StarHorse} may drive the derived distances for some stars, but \citet{Queiroz18} show that this is only the case for stars with large uncertainties (e.g. very distant stars, such as Sgr stars, see Appendix~\ref{sec:sgrappendix}). 

We tested the effect of a systematic shift in distances and whether it could potentially produce a spurious $v_\phi$ signal (especially closer to the Galactic centre) or significantly impact the estimates of the confined fraction. We tested the effect of possible biased distances by artificially reducing or increasing the distances by 15\%, the median uncertainty for the distances of MP to VMP stars (see Figure~\ref{fig:disterrhist}), and rerunning our key analyses. We find that the main conclusions/results of this paper do not change if the distances would be biased by this amount, although some of the details would be different -- see Appendix~\ref{sec:biastest} for further discussion. 

Additionally, the LSR assumption could also affect the mean azimuthal velocity. For stars in the bulge region (i.e., to the first order in Galactic longitude $l$, and in the Galactic plane, $b=0$), the tangential velocity $v_\phi$ is:
\begin{equation}
    v_\phi \approx -\left[\frac{sl(U+U_\odot)-(R_\odot-s)(V+v_\odot)}{|R_\odot-s|}\right],
\end{equation}

\noindent where $v_\odot \equiv v_0+V_\odot$, $s$ is the distance, and $U$ and $V$ are the stars Cartesian velocities (in the $x$ and $y$ direction respectively) with the respect of the Sun. Hence, for $U_\odot$ and $v_\odot$ off from their true values by $\Delta U_\odot$ and $\Delta v_\odot$,
\begin{equation}
    \Delta v_\phi \approx -\left[\frac{sl\Delta U_\odot-(R_\odot-s)\Delta v_\odot}{|R_\odot-s|}\right].
\end{equation}

\noindent However, especially in the VMP range, we have stars on both sides of the Galactic centre in terms of distance (changing the sign of $R_\odot-s$) as well as in longitude (changing the sign of $l$). Calculating $\Delta v_\phi$ for VMP stars in PIGS with $1.5 < \RGC < 3.5$~kpc, taking combinations of $\Delta U_\odot=\pm 10~\kms$ and $\Delta v_\odot=\pm 10~\kms$ (which are extreme values), $|\Delta v_\phi|$ is always less than 15~\kms. Any reasonable uncertainty in the LSR can therefore not remove the net rotational signal that we find.  

It is also possible that \Gaia systematics towards the bulge have an effect on our results. \citet{luna23} found that the proper motion uncertainties in \Gaia DR3 are underestimated in crowded bulge regions, up to a factor of 4 in fields with stellar densities larger than 300 sources per arcmin$^2$. However, most of our sample is not in the extremely crowded regions of the bulge but in its outskirts, so it is unlikely to have a large effect.   

Another limitation of part of our analysis, is that we must necessarily assume a potential to represent the gravitational field of the Milky Way and integrate the orbits of stellar particles for each star in PIGS. Although the S22 potential is one of the most up-to-date potentials for explaining the kinematics and distribution of stars in the Galactic centre, it is, as with all current Milky Way models, a simplification of the complexity of the Milky Way. In our analysis we compared the results with the axisymmetric version of the S22 potential, and with two other axisymmetric potentials -- which showed no significant changes to our main results. However, remaining among the non-axisymmetric models, even the S22 potential would require extensions, first of all the inclusion of spiral arms. These could change the orbital properties of PIGS stars, and for example induce radial migration for some of them, both in the case of recurring spirals with changing pattern speed \citep{sellwood02} or in the case of overlapping bar/spiral arms resonances \citep{minchev2010}. However, at the moment there are no detailed models of the spiral arms of the Milky Way, both because the parameter space is extremely large and because the data are very difficult to interpret unambiguously. Another possibility, is that the bar's pattern speed changed with time \citep[e.g.][]{chiba2021}. A changing bar's pattern speed can facilitate radial migration as the position of resonances (especially corotation) changes over time. The bar may bring out a fraction of the stars in the Galactic centre, but the variation in pattern speed over time must be quite high \citep{yuan23, li23}. In any case, exploring all these (still rather vague) possibilities is outside the scope of this paper.

As discussed in previous PIGS papers \citep{arentsen20_II, arentsen21}, the selection function of PIGS is complex. The targets were selected based on their brightness and their location in the metallicity-sensitive \textit{Pristine} colour-colour diagram, with the most metal-poor candidates having the highest priority and the rest of the fibres being filled with stars of increasing metallicity. Our main goal was to get as many of the most metal-poor stars as possible, not to have a homogeneous selection function. This main goal drove our observational strategy, which we optimised during the course of the survey based on our earlier observations (e.g. changing the source of broad-band photometry, updating the \CaHK photometric calibration, using different extinction maps, changing colour and magnitude ranges, etc.). Throughout the PIGS follow-up effort, we did not yet have a homogeneous and finalised photometric catalogue, nor calibrated photometric metallicities, and we selected the best candidates in each field relative to the other stars in that field only. As a result, the selection function will be slightly different across our 38 different AAT pointings, which leads to differences in the distributions of the metallicities and distances. We also do not have a complete understanding of the sources of contamination (metal-rich stars) in our sample, which also depend on the extinction. In summary, other kinds of samples (like those based on \Gaia, e.g. \citealt{rix22}) might be better suited for analyses that require knowledge of the selection function, such as estimates of the mass and/or the metallicity distribution function of the metal-poor inner Galaxy population.

\subsection{Comparison with the literature and interpretation}

\subsubsection{Fraction of confined stars/halo interlopers}
Previous work studying the confinement of metal-poor stars to the inner Galaxy found a range of different fractions. The first orbital properties for a set of very metal-poor stars ($\feh < -2.5$) were published by \citet{howes15}; this was pre-\Gaia so the authors used OGLE-IV \citep{ogle} proper motions and distances based on absolute magnitudes. They found that 3 out of 10 stars had apocentres less than 3.5~kpc, and 7 out of 10 stars had apocentres less than 10~kpc. These fractions are in good agreement with ours for stars in the same metallicity range ($30-40\%$ and $\sim80\%$ for 3.5 and 10~kpc, respectively). 

\citet{kunder20} find that 75\% of their RR Lyrae sample was confined to within 3.5~kpc. The metallicity distribution function of their stars peaks around $\feh = -1.4$ \citep{savino20} -- our fraction of stars confined to within 3.5~kpc at this metallicity is around 60\%. {One possible explanation for the difference is the fact that their observations are located closer to the Galactic plane than ours. For stars in our sample with $0.5 < R_{GC} < 1.5$~kpc at this metallicity the confined fraction is over 80\% (see Figure~\ref{fig:interlopers_sub}). Given the differences between the surveys (e.g. spatial coverage, target types, Galactic potential, definition of a confined star), our results are in good agreement. \citet{kunder20} refer to the stars with apocentres larger than 3.5~kpc as ``halo interlopers'' and suggest they are part of a different Galactic component compared to the more centrally concentrated RR Lyrae, but most of these still have apocentres $<8$~kpc. 

\citet{lucey21} studied the fraction of confined stars in their sample of metal-poor inner Galaxy stars from the COMBS survey, and also find a decreasing fraction with decreasing metallicity. Their absolute values, however, are roughly a factor of two lower than ours: 30\% around $\feh = -1.0$ and $\sim 20$\% for VMP stars, for stars confined to within 3.5~kpc for 75\% of Monte Carlo samples. They concluded most of the metal-poor stars in the inner Galaxy are therefore ``halo interlopers'' based on their apocentres. Their sample size is significantly smaller than ours ($\sim 35$ stars for $\feh < -1.5$), and we also find when comparing to their $R-z$ distribution for $\feh < -1.5$ that, compared to PIGS, their stars are typically more distant from the Galactic centre and at higher latitude -- these are less likely to be closely confined. 

\citet{rix22} find that most of the metal-poor ([M/H]~$<-1.2$ in this particular analysis) stars in their inner Galaxy sample have apocentres less than 5~kpc, but they do not provide a specific fraction. They also find that more metal-poor stars have a larger tail towards higher apocentres and eccentricities. Our results are in qualitative agreement, and we are likely probing the same population. 

In the PIGS high-resolution spectroscopic follow-up sample of \citet{sestito23}, 7/17 ($41\%$) VMP stars have apocentres less than 5~kpc, and only 3/17 ($19\%$) stay within 3.5~kpc. These fractions are slightly lower compared to the results in this work, which may partly be due to the high-resolution follow-up being performed for brighter stars that are closer to us and therefore further away from the Galactic centre. 

The number of halo interlopers in the inner Galaxy has also been studied indirectly by \citet{yang22}, using LAMOST stars currently beyond 5~kpc whose orbits bring them into the inner Galaxy (within 5~kpc). They estimate the total luminosity of the halo interloper population and compare to the results by \citet{lucey21} to derive the interloper fraction in the inner Galaxy: 100\% for $\feh < -1.5$ and 23\% for $-1.5 < \feh < -1.0$. This does not match at all with our results, likely because it has been calibrated against the incomplete and sparse sample of \citet{lucey21}. They also use the mass estimate of the metal-poor inner Galaxy ([M/H]~$<-1.5$) by \citet{rix22}, which is $> 10^8$~M$_\odot$, to redo the calculation and find the fraction of halo interlopers with apocentres $<5$~kpc to be more minor, less than 50\%. 

Overall, we find that our estimates of the fraction of confined metal-poor inner Galaxy stars are consistent with previous results in the literature. Thanks to the large sample size, range in metallicity and spatial coverage of PIGS, our estimates are likely the most representative of the underlying population. 

\subsubsection{Origin of the central metal-poor population}

The consensus of these works as well as from PIGS is that there is a population of metal-poor stars confined to the inner Galaxy, which would have largely been missed by previous ``typical halo'' surveys that do not probe within $\sim5$~kpc of the Galactic centre (although it was included in several RR Lyrae studies). 
As discussed in the introduction, galaxy simulations typically show signatures of ancient, metal-poor, centrally concentrated, spheroidal populations in Milky Way-like galaxies. Such a concentration was found in the Milky Way by \citet{rix22}, who refer to it as the proto-Galaxy (or the ``poor old heart'' of the Milky Way), as well as in \citet{kunder20} for RR Lyrae stars, who refer to it as a classical bulge component. The bulk of the stars in this population would have been accreted early on in a galaxy's history, when it was much less massive, or formed inside the main progenitor halo. However, the distinction between the main halo and an accreted galaxy for these early building blocks (in spatial distribution, dynamics and chemistry) may not be very clear if they are of similar mass, see e.g. the analyses of Auriga and FIRE simulations by \citet{orkney22} and \citet{horta23b}, respectively. We next discuss various literature works regarding the origin of the metal-poor inner Galaxy population(s), and how our results fit in. 

\begin{figure}
\centering
\includegraphics[width=0.95\hsize,trim={0.0cm 0.0cm 0.0cm 0.0cm}]{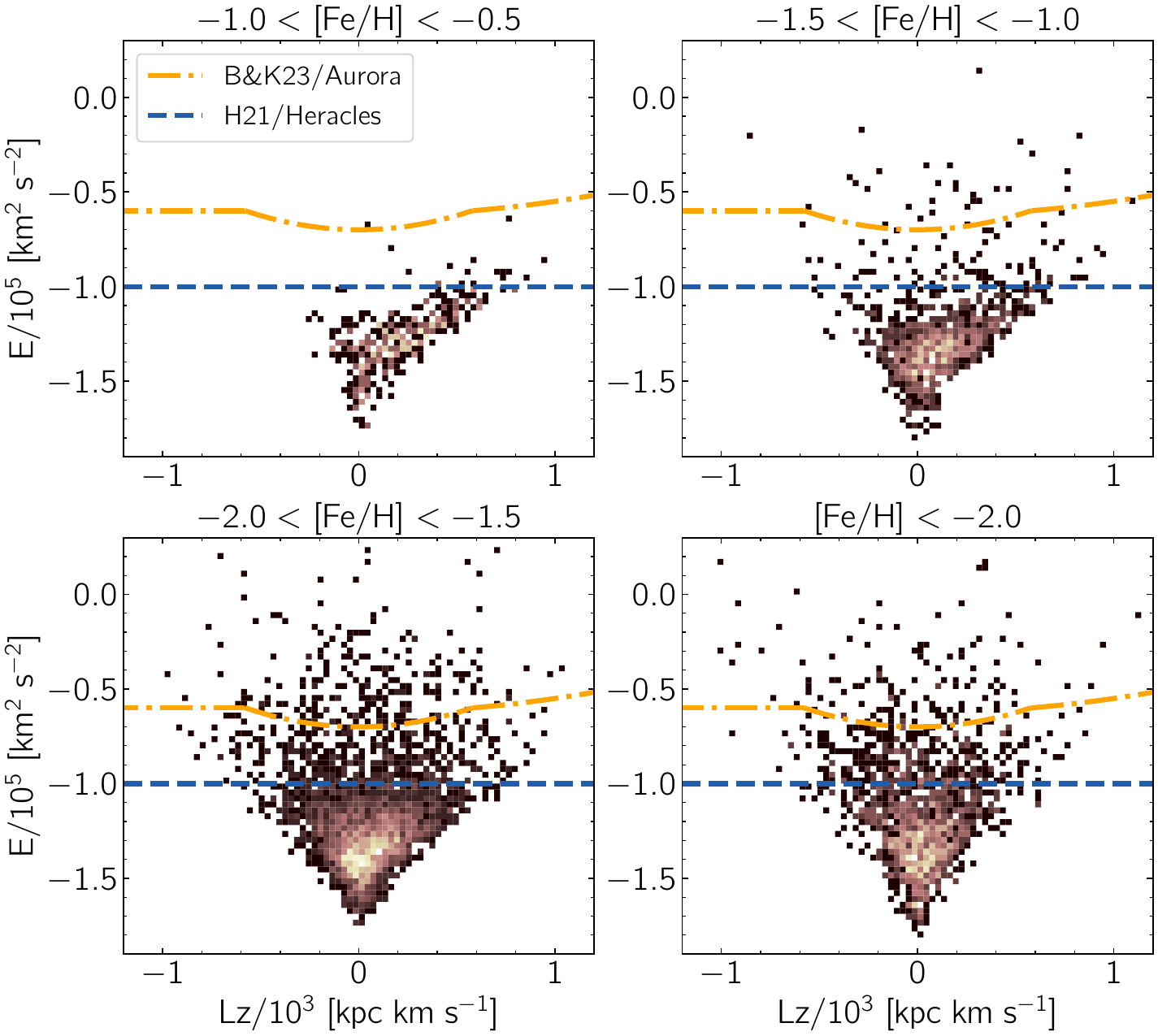}
\caption{Energy vs. angular momentum for PIGS stars with $\RGC < 3.5$~kpc in our four different metallicity ranges, colour-coded by density. Energy levels used by \citet{horta21a} and \citet{belokurovkravtsov23} are indicated (after correcting for differences in the employed potentials, see text). The Sun has an energy of $-0.57 \times 10^5$ km$^2$ s$^{-2}$ in our employed potential (S22).}
\label{fig:ELz} 
\end{figure}

\textbf{In-situ population --} \citet{belokurovkravtsov22} gave the name Aurora to the metal-poor pre-disc in-situ population of the Milky Way --  those born in the main progenitor. It has been identified among Solar neighbourhood stars with $\feh \gtrsim -1.7$, and the connection between Aurora and the (very) low-metallicity inner Galaxy can be inferred, but has not yet been established. However, \citet{belokurovkravtsov23b} show that the Aurora stars must be strongly centrally concentrated, following the very steep density profile to globular cluster-like stars, and \citet{belokurovkravtsov23} use the [Al/Fe] abundances of APOGEE stars to suggest a cut in energy and angular momentum below which most stars are expected to have been born in-situ. This cut is constant for $L_z/10^3 < -0.58$ at the E$_\mathrm{aMW14}/10^5 = -1.3$~km$^2$ s$^{-2}$ level and slightly dependent on $L_z$ for higher $L_z$, see their Equation~1. 

We present the E-$L_z$ diagram for stars in the PIGS sample in Figure~\ref{fig:ELz}, split by metallicity bins. We find that the Aurora level corresponds to E$_\mathrm{S22}/10^5 = -0.6$~km$^2$ s$^{-2}$ in our potential and shift Equation~1 from \citet{belokurovkravtsov23} to the appropriate level for our potential -- it is plotted as the orange dashed-dotted line in Figure~\ref{fig:ELz} and goes through the low-density energy tail of the PIGS population. This cut is roughly similar to a cut in median apocentre of $<10$~kpc in our PIGS sample -- according to this definition of ``halo interlopers'' there would only be $<20\%$ of these across the entire PIGS metallicity range, down to at least $\feh = -2.5$ (see Figure~\ref{fig:interlopers}). 

\citet{belokurovkravtsov23} find that around $20-30$\% of stars below the Aurora boundary are chemically classified as accreted, which may partly be contamination due to uncertainties in the abundances, and/or these may be stars from the very early evolutionary phase of Aurora before [Al/Fe] became high (see e.g. the discussions in \citealt{horta21a} and \citealt{myeong22}), and likely there are some truly accreted stars in this region as well. The Aurora boundary has been determined for stars with $\feh \gtrsim -1.5$ (and mostly for stars in the solar Neighbourhood), due to a lack of more metal-poor stars with abundances in APOGEE and due to the difficulty to chemically separate in-situ and accreted stars for lower metallicities. The fraction of accreted stars below the Aurora line may be higher at lower metallicity. 

\textbf{Large building blocks --} The possibility of identifying the larger building blocks to the primordial Milky Way is under discussion, but not entirely excluded. \citet{kruijssen20} use the globular cluster system of the Milky Way (with their ages and metallicities) to infer a large accretion event having taken place early on in the Galaxy's history (which they name Kraken) -- this population would be centrally concentrated. \citet{horta21a} use chemistry ([Mg/Mn] versus [Al/Fe]) to identify accreted stars in APOGEE (with some contamination fraction due to in-situ stars born in the ``unevolved'' phase of the early Milky Way), and find that there is a large population of these stars centrally concentrated in the bulge region, which they name Heracles and which peaks around $\feh = -1.3$. 

The energy level within which \citet{horta21a} find most of these stars to be present is E$_\mathrm{MM17} = -2.0/10^5$~km$^2$ s$^{-2}$, corresponding to E$_\mathrm{S22} = -1.0/10^5$~km$^2$ s$^{-2}$ in our potential, which is shown as a blue dashed line in Figure~\ref{fig:ELz}. This cut roughly corresponds to a limit of $\sim5$~kpc in the apocentre distribution of our PIGS stars, and closely follows the edge of our high-density distribution in E-$L_z$ for $\feh < -1.5$. The rough extent of the proto-Galaxy as determined by \citet{rix22} was also $\sim5$~kpc. Note that \citet{lanebovy22} suggested that the energy cut used by \citet{horta21a} was not necessarily physical and was the result of the APOGEE selection function, but we find that it matches well with our concentrated population of ancient inner Galaxy stars. So it does appear to be physical -- there is a metal-poor stellar population that drops off steeply beyond this energy level -- but it cannot be used to separate accreted from in-situ stars. For example, \citet{horta23b} show that, in simulations, stars from large early building blocks (of not much lower mass than the Milky Way main progenitor) have very similar present-day spatial distributions compared to the in-situ stars. 

According to the E-$L_z$ definition of \citet{belokurovkravtsov23, belokurovkravtsov23b}, both the Kraken globular clusters and the Heracles stars are considered part of Aurora, hence born in-situ -- this was also argued by \citet{myeong22} based on similarity between Heracles and Aurora in various chemical spaces. \citet{horta23a} show that there is a statistical difference between the detailed chemistry of the Heracles building block and Aurora stars -- this might partly be due to the Heracles stars being selected to have specific chemistry (low [Al/Fe] and high [Mg/Mn]), but the authors argue that their results are still an indication that the Heracles stars are of distinct origin from the in-situ stars. 

Furthermore, the contribution from Gaia-Sausage/Enceladus \citep[GS/E,][]{belokurov18,helmi18}, which was accreted more recently ($\sim 10$~Gyr ago), is expected to be small in the inner few kpc of our Galaxy. This is especially true among confined stars, since GS/E stars typically have low energies (above the orange line in Figure~\ref{fig:ELz}) and very radial orbits with large apocentres. But if any GS/E debris is present/confined in the inner Galaxy, it is more likely to be metal-rich since those stars are expected to have been most tightly bound to the GS/E progenitor and therefore sunk deepest in the MW potential \citep[e.g.][]{amarante22, orkney23}. GS/E is therefore unlikely a significant contributor to the metal-poor inner Galaxy.

\textbf{Small building blocks --} The expectation is that smaller dwarf galaxies should have contributed to the primordial Milky Way as well, although their total mass contribution is expected to be minor according to e.g. the analysis of the FIRE simulations by \citet{horta23b} -- they find that a proto-Milky Way is typically dominated by one or two dominant systems and a few smaller ones. Small (and/or very small) building blocks may contribute more in the lower metallicity tail of the population; see e.g. \citet{orkney23}, who investigated the fraction of \textit{ex-situ}/accreted stars in the Auriga simulations. Focusing on $R_\mathrm{GC} < 5$~kpc, they find that the fraction of accreted stars (from a range of progenitors) steadily rises with decreasing metallicity. In their main simulation, the accreted fraction is $\sim35$\% for $\feh = -1.5$ and rises to $60/70$\% for $\feh < -2.0$ and $<-2.5$, respectively. The accreted fraction of $\feh < -2.5$ stars is between $60-80$\% for nine out of their ten Milky Way simulations, only one simulation has a somewhat lower fraction of 40\%. These numbers do not significantly change when considering stars that have apocentres within 5~kpc (instead of just their present-day location). The decrease in fraction of tightly confined stars with decreasing metallicity we observe in PIGS (Figure~\ref{fig:interlopers}) might be connected to the increase in accreted stellar populations. It is worth noting that our metallicities are not necessarily on the same scale as those in the simulations, but the trends remain. 

Finally, as discussed in the introduction, it has also been shown that the fraction of stars with globular cluster-like chemistry (based on their high nitrogen abundances) is larger in the inner Galaxy \citep{schiavon17,horta21b, belokurovkravtsov23}. The latter authors argue that these mostly come from disrupted in-situ globular clusters -- and are therefore part of the same population as Aurora. All these studies are based on APOGEE data, typically for metallicities of $\feh \gtrsim -1.7$, and it is still an open question what the contribution of disrupted globular clusters might be for very metal-poor stars. Some VMP stars in the inner Galaxy do appear to come from globular clusters; \citet{sestito23} found two stars with $\feh < -2.5$ that show the typical globular cluster signature of enhanced [Na/Mg]. \citet{arentsen21} discussed the possibility that the apparently low fraction of carbon-enhanced very metal-poor s-process enhanced (CEMP-s) stars in the inner Galaxy could be due to a lower VMP binary fraction, which in turn could be due to a larger contribution from disrupted globular clusters (that have low binary fractions). However, further work on the chemistry of inner Galaxy VMP stars is necessary to investigate the role of disrupted VMP globular clusters in this region. 

\subsubsection{Rotational properties of the metal-poor inner Galaxy}

Simulations show that proto-galactic populations typically show weak but systematic net rotation up to a few tens of \kms \citep{belokurovkravtsov22, horta23b, mccluskey23, chandra23}. We indeed find rotation for PIGS metal-poor inner Galaxy stars (Figure~\ref{fig:vphitrend}). Our results are consistent with previous work that also found some azimuthal velocity among metal-poor inner Galaxy stars and/or local likely in-situ stars \citep{lucey21, wegg19, kunder20, rix22, belokurovkravtsov22, conroy22}, although \citet{rix22} claim that there is no net rotation among inner Galaxy stars with [M/H]~$<-2.0$. A net $v_\phi$ of $\sim +70$~\kms has also been found for metal-poor ($\feh < -1.0$) in-situ globular clusters with $\RGC < 5$~kpc \citep{belokurovkravtsov23b}, where in-situ was defined using the E-L$_Z$ Aurora boundary. 

A proto-galactic population could have been spinning since its formation (e.g. due to a net angular momentum from the combination of building blocks), but this does not have to be the case -- in their simulations, \citet{mccluskey23} show that the oldest in-situ stars were \textit{born} with low velocity dispersion and no net rotation, but currently this population has the highest dispersion and has been spun up (and flattened) over time \citep[see also][]{chandra23}. The mechanism(s) for the spinning up of the primordial Milky Way are still under discussion -- the growing disc likely plays a role, as well as the Galactic bar, which can severely affect the orbits of stars, moving them onto preferentially prograde orbits \citep[see e.g.][]{perezvillegas17, dillamore23}. Preferentially prograde merger events could also add to the spin of the ancient inner Galaxy population. 

In our PIGS sample, we find that the azimuthal velocity decreases as a function of metallicity for stars with $1.5 < \RGC < 3.5$~kpc between $-2.0 < \feh < -1.0$ (Figure~\ref{fig:vphitrend}), even after removing halo interlopers with large apocentres. It remains roughly constant for $\feh < -2.0$, with $v_\phi \sim 40$~\kms. We also find that $v_\phi$ stays constant with metallicity at a similar $v_\phi$ closer to the Galactic centre ($\RGC < 1.5$~kpc), across all metallicities (possibly with a slight drop for $\feh < -2.0$). One effect to keep in mind is that the more metal-rich stars in our sample are typically slightly further away from the Galactic centre than the more metal-poor stars (see Figure~\ref{fig:xyz-feh}) even within our selected rings of $\RGC$, which might allow them to have slightly higher net azimuthal velocities. A rise in rotation is also seen in the metal-rich APOGEE data from \citet{queiroz21}, which has higher $v_\phi$ further away from the Galactic centre -- this might be related to the rise in the rotation curve of the Galaxy (although in APOGEE it might also be connected to a rising contribution from the disc/bar). 

In previous work with the PIGS data \citep{arentsen20_I}, the rotation curve was studied as a function of metallicity by projecting the radial velocities with respect to the Galactic centre. They observed that there was no significant rotational signature for stars with $\feh < -2.0$. This difference with the results in this work may be due to several factors: the previous work did not do a cut on apocentre and therefore included many ``halo interlopers'' (which have small azimuthal velocities), the sample of VMP stars was only half the size (and less well-distributed in $l$ and $b$), and there is limited sensitivity when only using projected radial velocities. 

Additionally, \citet{rix22} claim to find no net rotation among inner Galaxy VMP stars, looking at $\overline{J_\phi/J_\mathrm{tot}}$ (see their Figure~6). However, compared to PIGS, their analysis is limited to stars with slightly larger apocentres ($3-7$~kpc, partly because their kinematics sample has larger distances from the Galactic centre), and their Figure does not actually show a drop to completely zero rotational support for [M/H]~$<-2.0$. Their results are therefore not in conflict with ours. 

\subsubsection{Interpretation}

There are currently no predictions for velocity trends with metallicity \textit{within} proto-galactic populations to compare to. We can, however, speculate about the origin for the trends in the PIGS data. Previously, \citet{arentsen20_I} interpreted the trend of decreasing inner Galactic rotation and increasing velocity dispersion as a function of metallicity in PIGS (see also the small symbols in Figure~\ref{fig:vsig}) as the result of a density transition between two overlapping populations: the metal-rich, cylindrically rotating bulge that formed from the disc and a hotter metal-poor halo component with no net rotation. However, the cylindrically rotating bulge is shown to be predominantly made up of stars with $\feh > -1.0$ with no strong tail towards lower metallicities, and stars with $\feh < -1.0$ are usually interpreted as coming from the halo or a spheroidal bulge component \citep[e.g.][]{wylie21, queiroz21, razera22}. This is consistent with recent works suggesting the MW thick disc does not start to form until $\feh \sim -1.3$ and only becomes prominent above $\feh = -1.0$ \citep[e.g.][]{belokurovkravtsov22, chandra23}. It therefore appears unlikely that the $v_\phi$ trend with \feh is (entirely) due to a transition between metal-poor halo and metal-rich discy bulge.

Recently, \citet{zhang23} used the \citet{andrae23} metallicities and kinematics from \Gaia to fit Gaussian Mixture Models to the velocity distributions of Solar neighbourhood stars. They showed that there appear to be two different halo components (separate from GS/E) in the metal-poor and very metal-poor regime -- a stationary halo with $\overline{v}_\phi \approx 0 \, \kms$ and $\sigma_{v_\phi} \approx 130 \, \kms$ and a prograde halo with $\overline{v}_\phi \approx +70 \, \kms$ and $\sigma_{v_\phi} \approx 90 \, \kms$ (which they suggest might be related to Aurora). They find the stationary halo to be dominant for [M/H]~$< -2.0$ (although the prograde halo is still present), and for $-2.0 < \mathrm{[M/H]} < -1.0$ the prograde halo takes over while the contribution from the stationary halo drops strongly (and the disc becomes a relevant component for [M/H]~$> -1.3$).

If these two halo components are also present in the inner Galaxy, as suggested by our detection of two components in the IMP regime with similar mean and dispersion in $v_\phi$ (Section~\ref{sec:rotation} and Table~\ref{table:gmm}) compared to the two inner halo components in \citet{zhang23}, we might interpret the trends we see between $v_\phi$, $\sigma_{v_\phi}$ and \feh for $1.5 < \RGC < 3.5$~kpc as a change in contributions from the stationary and prograde halos. 
The fraction of the cooler and faster component in our analysis varies from 0.83 to 0.62 to 0.40 for the MP, IMP and VMP regimes, respectively. 
Connecting it to the rest of the discussion here as well as that in \citet{zhang23}, this suggests a transition from predominantly ``accreted'' stars (or smaller building blocks) at lower metallicity to predominantly ``in-situ'' stars (or larger building blocks) at higher metallicities. 
We find that our slower halo component still has a slight rotation ($v_\phi = 26$~\kms, as opposed to the ``stationary'' component by \citet{zhang23}. Our estimate is for stars with $1.5 < R_{GC} < 3.5$~kpc, while \citet{zhang23} focus on stars closer to the Solar neighbourhood -- it is possible that this component rotates faster closer to the Galactic centre.
Along the same line of reasoning, the population of stars even closer to the Galactic centre ($0.5 < R_{GC} < 1.5$~kpc, for which $\overline{v}_\phi \approx +45$~\kms independent of metallicity, may be dominated by the in-situ population at all metallicities.

\subsection{Future work}

There is limited information currently available for the metal-poor inner Galaxy. In this work we employed spectroscopic metallicities and dynamical properties for thousands of metal-poor inner Galaxy stars, which have been observed with low/medium-resolution spectroscopy. Improvements in both the chemistry and dynamics could be made with higher resolution spectroscopic observations. 

The dynamics will be improved thanks to the improved proper motions in the next \Gaia data release, but in particular the dynamics would improve if it were possible to get more precise distances. One limitation is the parallax quality, which will only get slightly better in the next \Gaia data release. On the other hand, higher resolution spectra would allow for more precise stellar parameter estimates, which would improve the \texttt{StarHorse} distance estimates. In addition, more work could go into investigating how appropriate the adopted isochrones are for very metal-poor stars (as well as e.g. HB stars). Finally, there is still much to learn about (inhomogeneous) extinction towards the Galactic bulge, which also impacts distance estimates that include photometry. 

Higher resolution spectroscopic follow-up also allows for the measurement of elemental abundances other than just ``metallicity'' (and carbon). The current samples of high-resolution observations in the inner Galaxy are relatively small and/or inhomogeneous, although we have already learned a lot from them (see the Introduction). Large, homogeneous samples of metal-poor inner Galaxy stars will be observed with the upcoming Southern hemisphere 4-metre Multi-Object Spectroscopic Telescope \citep[4MOST,][]{4most}. Over 25\,000  metal-poor candidates pre-selected from PIGS photometry will be observed in the VMP\_PIGS sub-survey (more details in a future paper, Ardern-Arentsen et al. in prep.) within the 4MIDABLE-LR consortium survey \citep{chiappini19}, as well as VMP candidates based on metallicities from \texttt{StarHorse} + \Gaia XP spectra. The low-resolution (LR) mode in 4MOST is significantly higher resolution ($R\sim6500$) than what is currently available in terms of ``low-resolution'' surveys (which are more typically $R \lesssim 2000$). A variety of chemical abundances is expected to be derived from these spectra, among them carbon abundances, individual alpha and iron-peak elements, as well as some s-process elements. This will be the first large sample of thousands of very metal-poor inner Galaxy stars with chemical abundances beyond simple metallicity. There will also be some dedicated follow-up of VMP candidates in the high resolution 4MIDABLE-HR survey \citep{bensby19}, for which detailed chemical abundances from all nucleosynthetic channels are expected to be derived. Other ongoing and upcoming surveys will also target the inner Galaxy (e.g. SDSS-V, \citealt{sdssV}, and MOONS, \citealt{moons}), but they are not necessarily targeting VMP stars specifically, and/or focusing on the infrared, which has fewer lines for VMP stars.

What are the questions that can be probed with these large samples of homogeneous chemical abundances? Hopefully these large samples of stars with homogeneous chemical abundance will shed light on the origin of inner Galaxy metal-poor stars, e.g. are they more likely to have been born in the main progenitor of the Milky Way, or in a small dwarf galaxy? What were the properties of the First Stars in different environments, and how did they affect the next stellar generations? What kind of CEMP stars are present in the inner Galaxy? What is the contribution of disrupted globular cluster stars in the VMP regime? Additionally, in such a large homogeneous dataset, there is also ample opportunity for unexpected discoveries. 

Finally, to be able to compare our dynamics results to simulations, the right approach would be to forward model those to take into account the effect of large parameter uncertainties in the observations. This is beyond the scope of the current work, but is certainly worth revisiting in the future. 

\section{Summary and conclusions}\label{sec:summary}

In this work, we investigated the dynamical properties of (very) metal-poor stars in the inner Galaxy, using spectroscopic data from the Pristine Inner Galaxy Survey (PIGS, \citealt{arentsen20_II}). Metal-poor stars in the central regions of the Milky Way are expected to be among the oldest in our Galaxy. We derived spectro-photometric distances for the stars in PIGS using {\tt StarHorse} \citep[][]{Santiago16, Queiroz18}, which have typical uncertainties between $10-20\%$ (Figure~\ref{fig:disterrhist}). Combining these with radial velocities from PIGS and proper motions from \Gaia, we derive orbital properties by integrating the stars in the \citet{portail2017}/\citet{sormani22} potential, which includes a realistic representation of the Galactic bar. After spectroscopic and photometric quality cuts, there are $\sim 7500$ stars in the final sample used for our dynamical analysis, of which $\sim 1700$ have $\feh < -2.0$ (VMP stars). This is the largest sample of VMP stars with detailed dynamical properties in the inner Galaxy. 

With this paper, we release all the PIGS spectroscopic data products (see Appendix~\ref{sec:datarelease}). This includes the stellar parameters from two independent methods and the radial velocities, for which the methods are described in \citet{arentsen20_II}, and the dynamical properties inferred in this work, if available. The stars in PIGS are mostly inner Galaxy stars, but there is also a significant number ($\sim 800$) of Sagittarius dwarf galaxy stars in the release \citep[see][]{vitali22} -- although the distances and dynamics in this work should not be used for Sagittarius stars. The data release of the PIGS photometry will be part of a future paper (Ardern-Arentsen et al. in prep.). 

\vspace{0.3cm}

Our main findings in this paper are as follows: 

\begin{itemize}
    \item Almost all stars in our sample are currently located within 3.5~kpc of the Galactic centre (Figure~\ref{fig:xyz-feh}). We find that VMP stars are found more symmetrically distributed around the Galactic centre than less metal-poor stars ($-2.0 < \feh < -1.0$), which are systematically closer to us. This might be influenced by selection effects in our sample (e.g. VMP stars are intrinsically brighter).
    
    \item We compare trends in the dynamical properties ($v_\phi$, velocity dispersion, $z_{max}$, apocentres, eccentricities) of metal-poor stars in PIGS ($\feh < -1.5$) to metal-rich inner Galaxy stars from APOGEE ($\feh > 0$, from \citealt{queiroz21}), see Figure~\ref{fig:apogeecomp}. 
    The dynamics of metal-rich stars are dominated by the bar, whereas the metal-poor stars do not clearly follow the same structure, and behave more like a spheroidal, pressure-supported population.
    
    \item We find that the majority of metal-poor stars in PIGS have orbits that are confined to within $\sim 5$~kpc of the Galactic centre (Figure~\ref{fig:interlopers}). The fraction of confined stars decreases with decreasing metallicity, but even VMP stars have a significant fraction of stars that is confined to the inner Galaxy ($\sim 40\%$ within 3.5~kpc, $\sim 60\%$ within 5~kpc and $\sim 80\%$ within 10~kpc). Given the different selection functions, sample sizes and adoption of different potentials, definitions and distance estimates, previous estimates in the literature are in reasonable agreement with our results. We conclude that the Milky Way is hosting a significant, centrally concentrated, (very) metal-poor Galactic component (in agreement with e.g. \citealt{rix22}). 
    
    \item We find that metal-poor stars in the inner Galaxy show coherent rotational motion around the Galactic centre (Figure~\ref{fig:vphitrend}), down to the very metal-poor regime. For a ring in Galactocentric radius between 1.5 and 3.5~kpc, we find that $\overline{v}_\phi$ changes roughly linearly from $\sim +80$~\kms to $\sim +40$~\kms between $\feh = -1.0$ and $-2.2$, respectively. At lower metallicity, $\overline{v}_\phi$ remains roughly constant. For stars within a Galactocentric radius of 1.5~kpc, we find $\overline{v}_\phi \approx 45$~\kms, independent of metallicity, with a  slight decrease below $\feh < -2.0$. We suggest these trends may be the result of a transition between two spheroidal populations, a more metal-rich one that rotates faster (possibly connected to stars formed in-situ/in large building blocks) and a more metal-poor one that rotates less fast or not at all (possibly connected to stars that were accreted/formed in smaller building blocks). This is supported by a preferred two-component fit to the velocity distribution for stars with $-2.0 < \feh < -1.5$, with parameters similar to the two halo components recently found by \citet{zhang23}.
    
    \item The high velocity dispersion of metal-poor inner Galaxy stars is mostly driven by ``halo interlopers'', and previous PIGS results of a strongly rising velocity dispersion with decreasing metallicity \citep{arentsen20_I} can be explained by this (see Figure~\ref{fig:vsig}, consistent with previous findings by \citealt{lucey21} and \citealt{kunder20}). After removing stars with apocentres larger than 5~kpc, the velocity dispersion is still rising with decreasing metallicity, but only gently: from $\sim 80$~\kms for $\feh = -1$ to $\sim 95$~\kms for $\feh < -2.0$. This is likely due to the VMP inner Galaxy stars having a slightly more extended distribution compared to less metal-poor stars.  
    
    \item We investigated the orbital properties of PIGS carbon-enhanced metal-poor (CEMP) stars in the inner Galaxy \citep{arentsen21}, and found no significant difference between the apocentre distributions of carbon-rich and carbon-normal VMP stars (Figure~\ref{fig:cemp}). More work is necessary to further investigate whether the occurrence of CEMP stars is different between ``true inner Galaxy stars'' and ``halo interlopers''. 

\end{itemize}

The past years have seen significant progress in the study of the metal-poor inner Galaxy, but there are still many open questions about the ancient stellar population(s) in this region. The next years will see many new results from several large spectroscopic surveys targeting the Galactic bulge (e.g. SDSS-V, MOONS, 4MOST), with the 4MOST surveys 4MIDABLE-LR and -HR \citep{chiappini19,bensby19} being particularly promising for the investigation of the metal-poor inner Milky Way. Large, homogeneous samples of metal-poor stars with chemical abundances will allow us to probe the chemical evolution in the earliest phase of our Galaxy. 

\section*{Acknowledgements}

We thank the reviewer for their helpful suggestions. AAA thanks all members of the Pristine collaboration and the Cambridge Streams group for valuable discussions. 
We thank Sarah Martell and Jeffrey Simpson for observing for the PIGS/AAT program.
AAA acknowledges support from the Herchel Smith Fellowship at the University of Cambridge and a Fitzwilliam College research fellowship supported by the Isaac Newton Trust. AAA, GM, NFM and ZY gratefully acknowledge funding from the European Research Council (ERC) under the European Unions Horizon 2020 research and innovation programme (grant agreement No. 834148). GM, NFM and ZY gratefully acknowledge support from the French National Research Agency (ANR) funded projects ``MWdisc'' (grant N211483) and ``Pristine'' (ANR-18-CE31-0017). ES acknowledges funding through VIDI grant "Pushing Galactic Archaeology to its limits" (with project number VI.Vidi.193.093) which is funded by the Dutch Research Council (NWO). This research has been partially funded from a Spinoza award by NWO (SPI 78-411). DA acknowledges financial support from the Spanish Ministry of Science and Innovation (MICINN) under the 2021 Ramón y Cajal program MICINN RYC2021-032609. GCM thanks the Leverhulme Trust and the Isaac Newton Trust for their support on his research. HZ thanks the Science and Technology Facilities Council (STFC) for a PhD studentship. WHO acknowledges financial support from the Carl Zeiss Stiftung and the Paulette Isabel Jones PhD Completion Scholarship at the University of Sydney.

This research was supported by the International Space Science Institute (ISSI) in Bern, through ISSI International Team project 540 (The Early Milky Way).

We thank the Australian Astronomical Observatory, which have made the PIGS spectroscopic follow-up observations used in this work possible. We acknowledge the traditional owners of the land on which the AAT stands, the Gamilaraay people, and pay our respects to elders past and present. Based on data obtained at Siding Spring Observatory (via programs S/2017B/01, A/2018A/01, OPTICON 2018B/029 and OPTICON 2019A/045, PI: A. Arentsen and A/2020A/11, PI: D. B. Zucker). 

The spectroscopic follow-up used in this work was based on selection from observations obtained with MegaPrime/MegaCam, a joint project of CFHT and CEA/DAPNIA, at the Canada-France-Hawaii Telescope (CFHT) which is operated by the National Research Council (NRC) of Canada, the Institut National des Science de l'Univers of the Centre National de la Recherche Scientifique (CNRS) of France, and the University of Hawaii.

This work has made use of data from the European Space Agency (ESA) mission {\it Gaia} (\url{https://www.cosmos.esa.int/gaia}), processed by the {\it Gaia} Data Processing and Analysis Consortium (DPAC, \url{https://www.cosmos.esa.int/web/gaia/dpac/consortium}). Funding for the DPAC has been provided by national institutions, in particular the institutions participating in the {\it Gaia} Multilateral Agreement.

The Pan-STARRS1 Surveys (PS1) and the PS1 public science archive have been made possible through contributions by the Institute for Astronomy, the University of Hawaii, the Pan-STARRS Project Office, the Max-Planck Society and its participating institutes, the Max Planck Institute for Astronomy, Heidelberg and the Max Planck Institute for Extraterrestrial Physics, Garching, The Johns Hopkins University, Durham University, the University of Edinburgh, the Queen's University Belfast, the Harvard-Smithsonian Center for Astrophysics, the Las Cumbres Observatory Global Telescope Network Incorporated, the National Central University of Taiwan, the Space Telescope Science Institute, the National Aeronautics and Space Administration under Grant No. NNX08AR22G issued through the Planetary Science Division of the NASA Science Mission Directorate, the National Science Foundation Grant No. AST-1238877, the University of Maryland, Eotvos Lorand University (ELTE), the Los Alamos National Laboratory, and the Gordon and Betty Moore Foundation.
	
This publication makes use of data products from the Two Micron All Sky Survey, which is a joint project of the University of Massachusetts and the Infrared Processing and Analysis Center/California Institute of Technology, funded by the National Aeronautics and Space Administration and the National Science Foundation.

This publication makes use of data products from the Wide-field Infrared Survey Explorer, which is a joint project of the University of California, Los Angeles, and the Jet Propulsion Laboratory/California Institute of Technology, funded by the National Aeronautics and Space Administration.

This research made extensive use of the \textsc{matplotlib} \citep{matplotlib}, \textsc{pandas} \citep{pandas} and \textsc{astropy} \citep{astropy13, astropy18} Python packages, and of \textsc{Topcat} \citep{topcat}. We thank Sergey Koposov for sharing his column-density normalised 2D histogram Python code. 

\section*{Data Availability}

We release the PIGS spectroscopic parameters (methodology described in \citealt{arentsen20_II}) with this paper, as well as the distances and orbital properties derived in this work (summarised by the median and 16th + 84th percentiles). The {\tt StarHorse} Gaussian Mixture model for the distance solutions and the complete samples for the derived PIGS orbits are available on request to A. Ardern-Arentsen. 

\section*{Author Contributions Statement}

AAA is the PI of PIGS and initiated and designed the study, performed the analyses, made all the figures and led the writing of the manuscript. GM derived the orbital properties for the sample. AQ derived the \texttt{StarHorse} distances for the sample. Both wrote the respective method sections in the manuscript. ES, NFM and CC contributed through numerous discussions and assisted in the development of the project and the manuscript. DA derived the PIGS/AAT spectroscopic parameters with \texttt{FERRE}. AAA, NFM, SB, GFL, WHO, ZW and DBZ performed the observations at the AAT. DBZ also assisted in preparing for the AAT observations. All authors provided feedback and helped shape the manuscript.



\bibliographystyle{mnras}
\bibliography{pigs-orbits} %


\newpage


\appendix

\section{Distances for Sgr stars}\label{sec:sgrappendix}

Here we check the distances for stars in the Sagittarius (Sgr) dwarf galaxy. The distance to this galaxy is well-known at $\sim 26.5$~kpc \citep[e.g][]{ferguson20}, but this is very far away to get good distances. A Sgr density prior is included in \texttt{StarHorse}, but this is only for stars close to the core \citep{Anders22}. Additionally, the alpha abundances of Sgr stars are lower than those in the Milky Way, which might affect the \texttt{StarHorse} distances. 

We select Sgr stars as follows: $\sqrt{(\mu_\alpha + 2.69)^2 + (\mu_\delta + 1.35)^2} < 0.65$, RV $> 100$~\kms, $b < 0$ and $\feh < -1.0$. This is expected to give a clean selection of Sgr stars \citep[see][]{vitali22}. Figure~\ref{fig:sgr} shows their distance distribution (top) and on-sky distribution colour-coded by distance (bottom). These are not all the Sgr stars in the PIGS sample, only those with converged distances (which partly depends on the PanSTARRS footprint). Many stars are placed at the correct distance (dark blue), but there is also a large fraction of stars placed much closer (lighter colour). For the stars with RA~$> 285$~deg, this may partly be due to limited availability and quality of PanSTARRS photometry since it is reaching the edge of the PanSTARRS footprint (Dec $\sim -31$~deg, but variable with RA). For the stars with RA~$< 283$~deg, it might be possible that the bulge prior starts to dominate over the Sgr prior. The Galactic centre is located towards the left and slightly up.

This might be a hint that some distant halo stars behind the bulge could be placed inside the bulge by \texttt{StarHorse}, although they do not belong there. However, we expect stars \textit{in} the inner Galaxy to dominate our sample, given the much higher density as well as our cut in apparent magnitude.

\begin{figure}
\centering
\includegraphics[width=1.0\hsize,trim={0.0cm 0.5cm 0.0cm 0.0cm}]{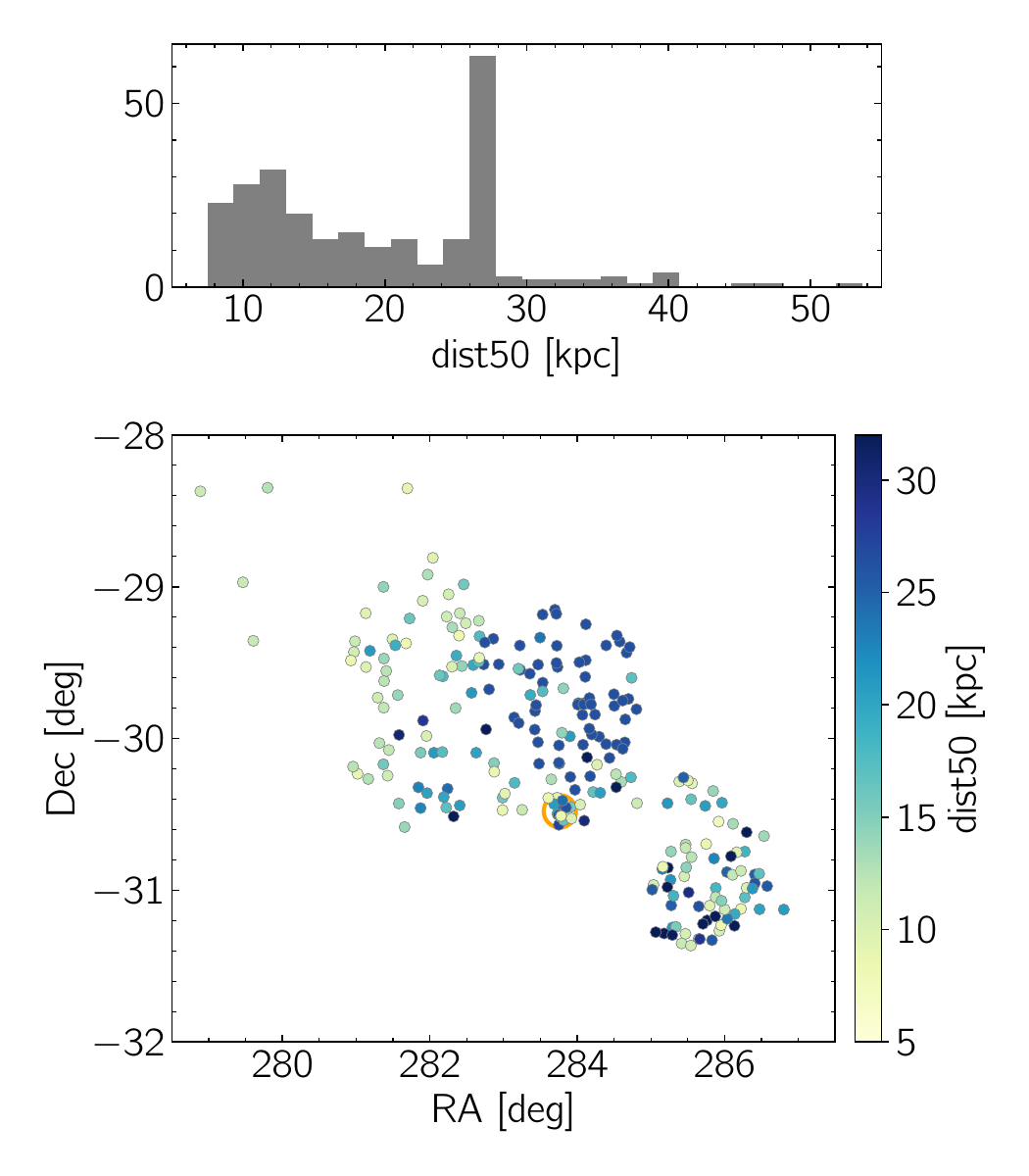}
\caption{PIGS-\texttt{StarHorse} distances for Sagittarius dwarf galaxy stars. The top panel shows the 1D distribution of the distances, the bottom panel the on-sky distribution colour-coded by distance. The orange circle is centred on the location of M54, the central globular cluster in Sgr.}
\label{fig:sgr} 
\end{figure}

\section{Horizontal Branch stars}\label{sec:hb}

As discussed in the main body of this article, there is a large number of horizontal branch (HB) stars present in the PIGS sample. 
It is not yet straightforward to interpret the HBs of systems -- the diversity of HB morphologies in globular clusters, for example, is still not entirely explained, although the main parameters appear to be age, metallicity and helium abundance \citep[see e.g.][]{torelli19}. 
In the metallicity range $-1.5 < \feh < -1.0$, the PIGS sample consists roughly half/half of normal RGB stars and HB stars. We noticed that there was a difference in the x-y distributions between RGB and HB stars for this metallicity range, which is shown in Figure~\ref{fig:xy-feh_hb}, with the HB stars appearing to be more distant than the RGB stars. It is unclear whether the spatial difference is real or caused by systematics in the distance estimation. It is beyond the scope of this paper to investigate the HB stars in detail, but here we briefly present one analysis supporting our decision to remove HB stars from our sample. 

The top panels of Figure~\ref{fig:kiel} present the Kiel diagram of the RGB and HB stars in the metallicity range $-2.0 < \feh < -1.0$ with the FERRE parameters on the left and StarHorse parameters on the right. The StarHorse parameters look ``cleaner'' because they are strongly influenced by the adopted isochrones. The red box in the StarHorse panel is what we use in this work to select and remove HB stars from our sample. The bottom panels show the stellar parameter differences for these two populations, highlighting that the differences are stronger for the HB sample, especially in [M/H] (the FERRE \feh has been converted to [M/H] as described in the main text). These can result in different distance biases for RGB and HB stars. We trust the HB {\tt StarHorse} distances less than the RGB distances, we therefore removed the HB stars.

\begin{figure*}
\centering
\includegraphics[width=0.7\hsize,trim={0.0cm 0.0cm 0.0cm 0.0cm}]{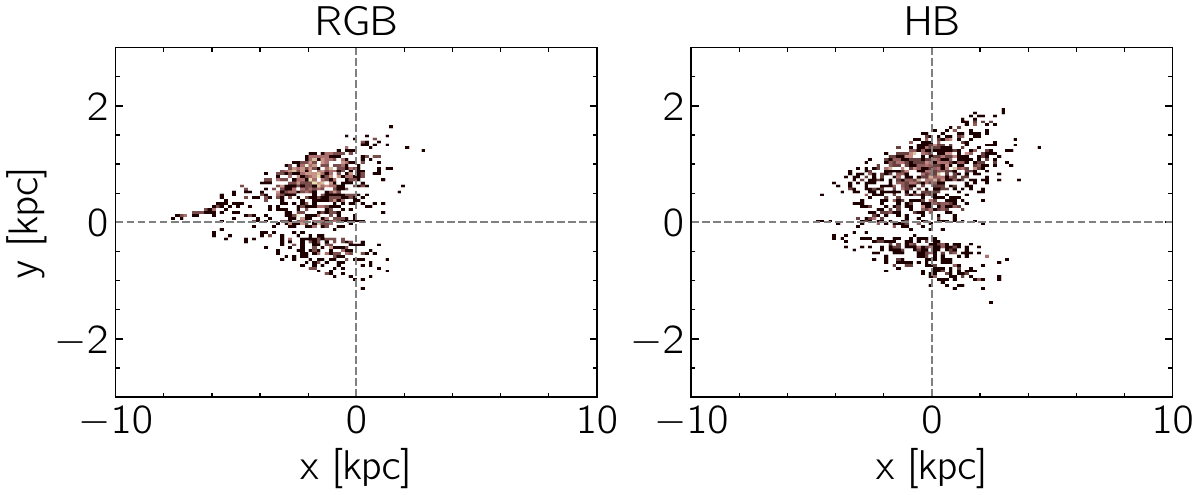}
\caption{Spatial distributions of RGB and HB stars, in the metallicity range $-1.5 < \feh < -1.0$.}
\label{fig:xy-feh_hb} 
\end{figure*}

\begin{figure*}
\centering
\includegraphics[width=0.7\hsize,trim={0.0cm 0.0cm 0.0cm 0.0cm}]{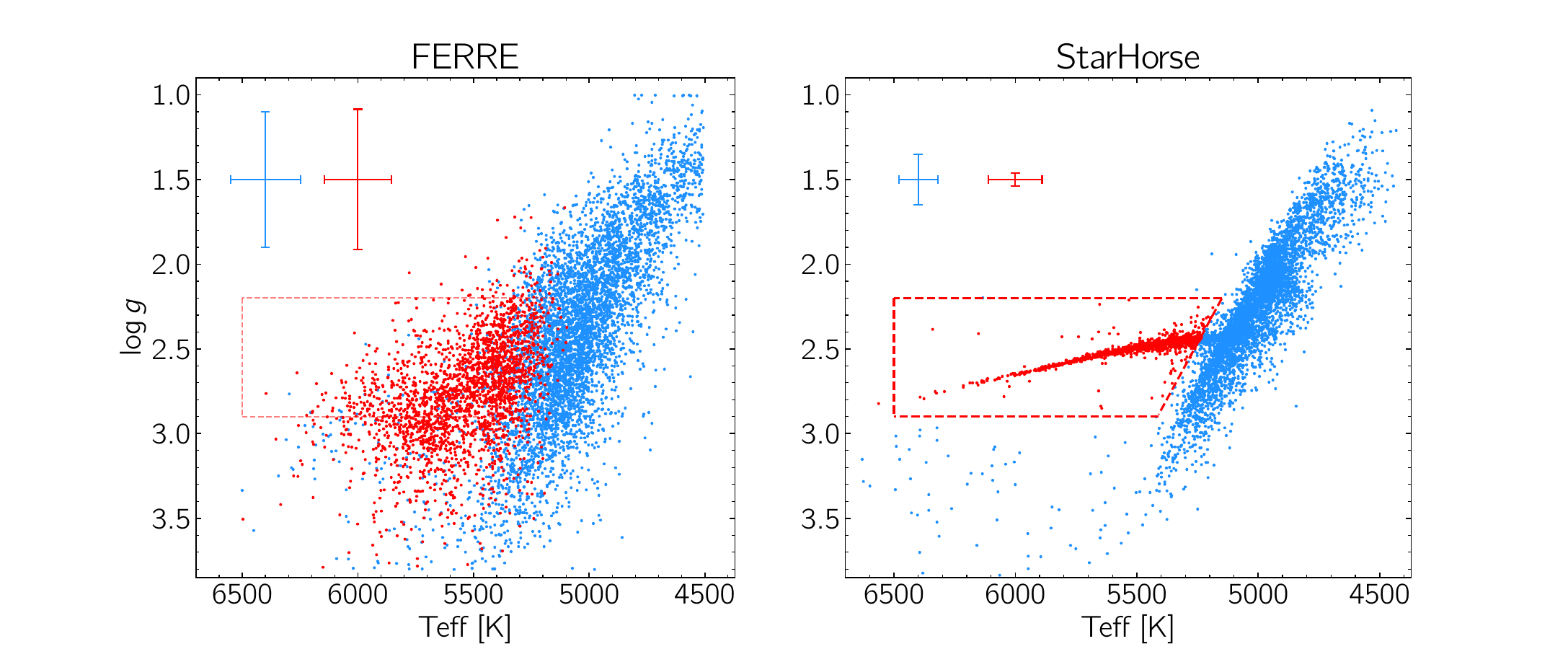}
\includegraphics[width=0.9\hsize,trim={0.0cm 0.0cm 0.0cm 0.0cm}]{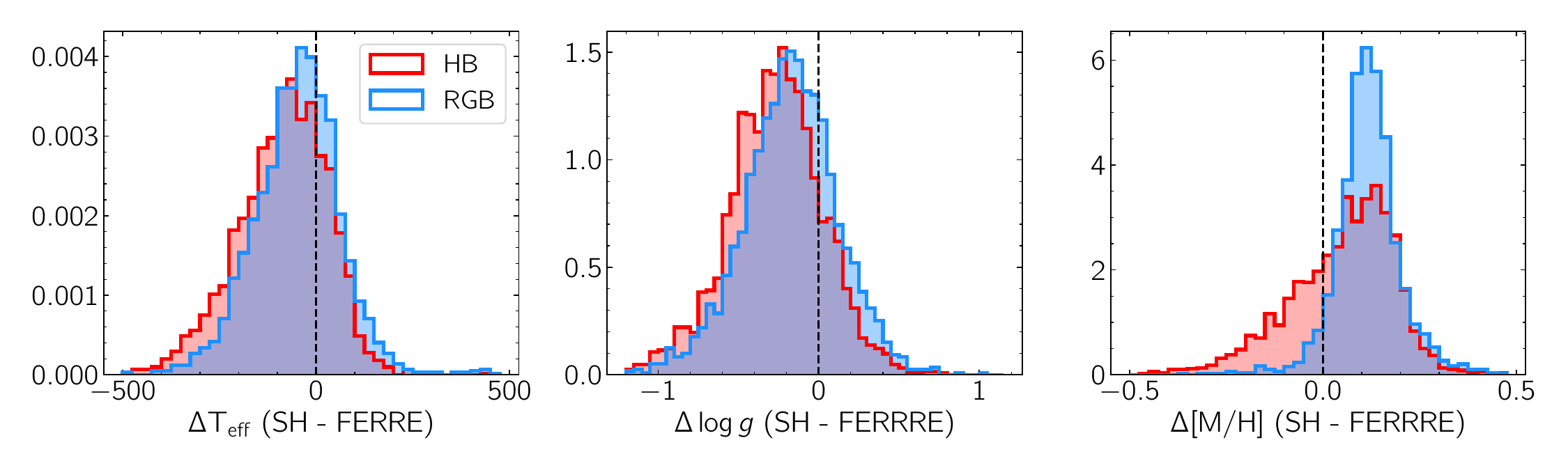}

\caption{Selection of and differences between RGB stars and HB stars, in the metallicity range $-2.0 < \feh < -1.0$. Top row: Kiel diagram colour-coded by selection -- the red stars are our horizontal branch selection, and have been selected inside the red box in the StarHorse panel, the blue stars (RGB) are the rest. Bottom row: Difference between FERRE and SH \teff and \logg, for the same groups. 
}
\label{fig:kiel} 
\end{figure*}

\section{Distance uncertainties/biases and their effect on the velocities}\label{sec:biastest}

Here we investigate the effect of distance uncertainties and biases. Figure~\ref{fig:vphivr_draws} shows the effect of the input parameter uncertainties on the derived velocities, which are typically dominated by the distance uncertainties. The left-hand panel focuses on IMP stars, which are systematically in front of the Galactic centre, and the right-hand panel focuses on VMP stars, which are more symmetrically located around the Galactic centre (see Figure~\ref{fig:xyz-feh}). A few examples are highlighted (coloured by star). There are often strong correlations between $v_\phi$ and $v_R$, because a change in distance can place a star either in front or behind the Galactic centre. This can be recognised as the largely ``circular'' pattern, most prominently in the VMP panel. The median properties (the larger symbols) are not always the best representation of a given set of draws -- in this parameter plane and also in others. We try to avoid using the median/mean/dispersion of the samples for the velocities and other parameters where possible.

\begin{figure*}
\centering
\includegraphics[width=0.45\hsize,trim={0.0cm 0.0cm 0.0cm 0.0cm}]{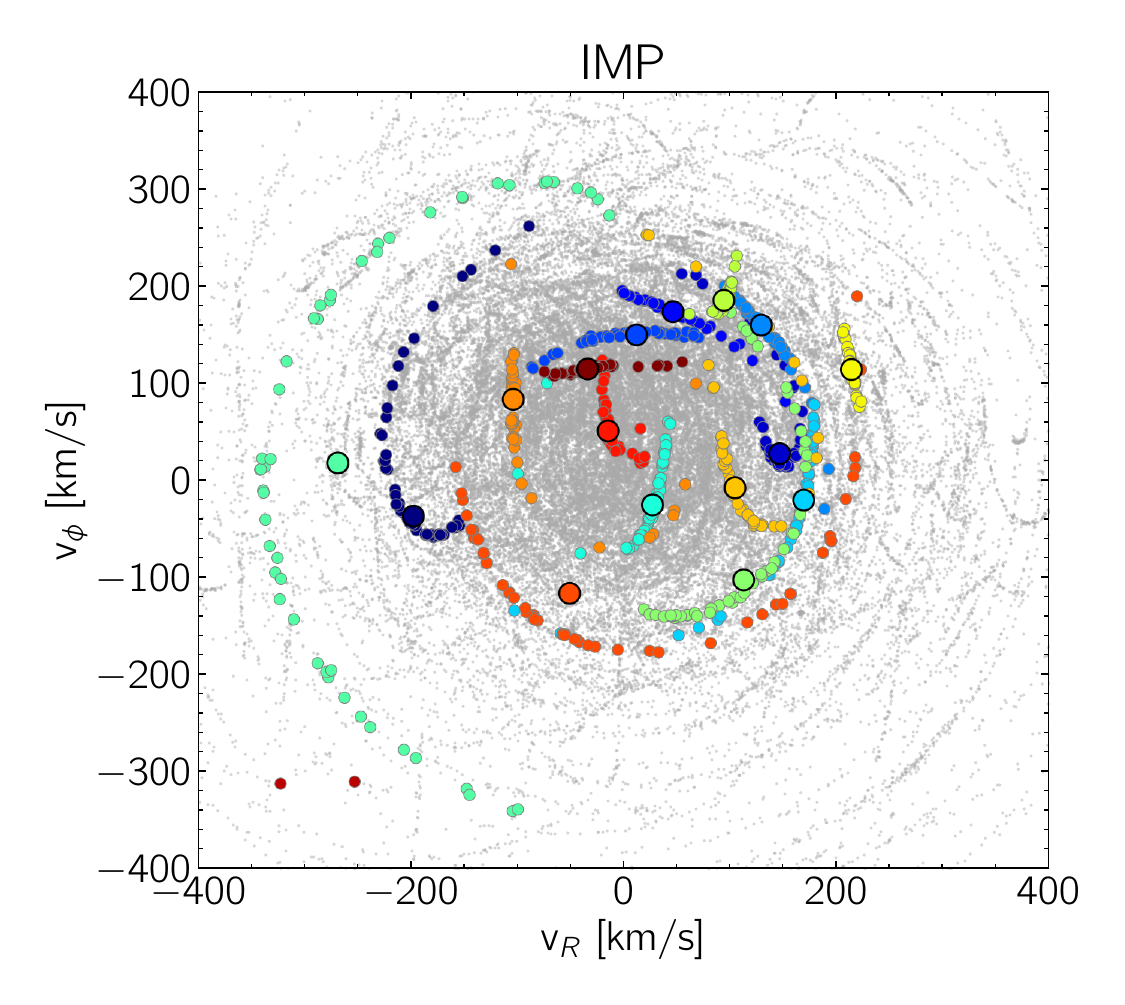}
\includegraphics[width=0.45\hsize,trim={0.0cm 0.0cm 0.0cm 0.0cm}]{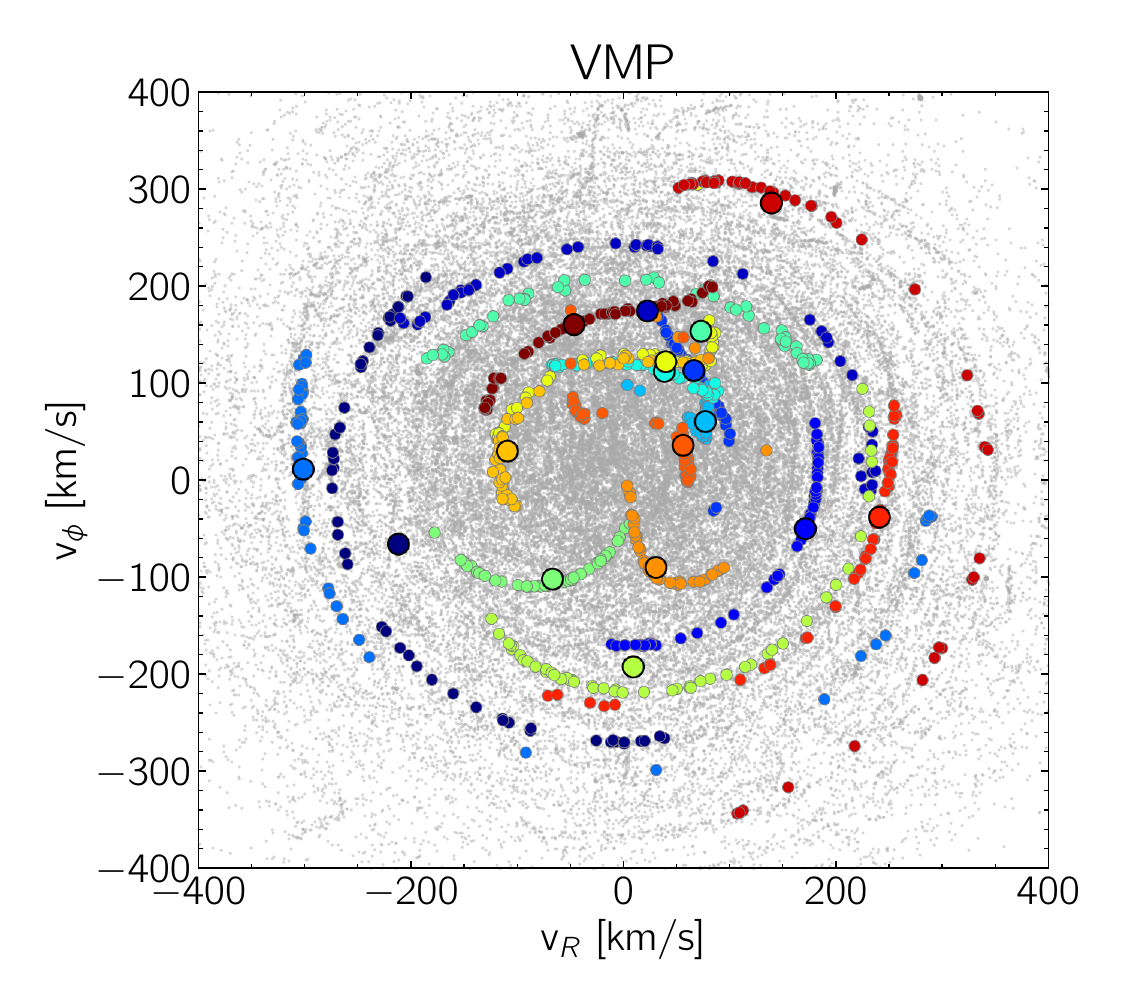}
\caption{Distributions of $v_\phi$ and $v_R$ for all Monte Carlo draws of IMP stars ($-1.6 < \feh < -1.4$, left) and VMP stars ($\feh < -2.0$, right). A few examples have been highlighted in colour, coloured by \textit{Star\_ID} number. The median value for each star is highlighted with big round symbols with black borders and filled by the same colour.
}
\label{fig:vphivr_draws} 
\end{figure*}

To test the effect of a possible distance bias, we artificially bias the distances by $\pm 15\%$, and discuss the effect on the confined fraction and the mean azimuthal velocity as a function of metallicity. The results are summarised in Figure~\ref{fig:comp_apofrac}, comparing the biased distances ($-15\%$ left and $+15\%$ right) to the fiducial results presented in the main body of the paper (middle). 

For the fraction of confined stars as a function of metallicity, if the distances are reduced by 15\%, the main change is that the fraction of stars confined to within 3.5~kpc becomes constant between $-2.7 < \feh < -1.4$, at around $40-45$\%. The trends for the confinement within 5 and 10~kpc remain very similar to before. If the distances are increased by 15\%, the trends become steeper for all three confinement regimes (meaning fewer confined stars at lower metallicities), with a clear break at $\feh = -2.0$, below which the trend flattens. In this scenario, still $\sim 25\%, 45\%, 75\%$ of the VMP stars are confined to within 3.5, 5 and 10~kpc, respectively. The stronger differences for the more metal-rich stars can be explained by the fact that they have their (fiducial) distances peaking in front of the Galactic centre, while the VMP stars are more symmetrically distributed around $x=0$ (see Figure~\ref{fig:xyz-feh}).

For the analysis of $v_\phi$ as a function of metallicity, we find that the mean $v_\phi$ slightly increases for the reduced distances (with stronger changes for the stars more distant from the Galactic centre), while the mean $v_\phi$ becomes slightly lower for the increased distances. Changes are of the order of maximum $\sim 20 \, \kms$. We find that the trends stay the same in any of these cases, and the mean azimuthal velocity is still (significantly) positive for all metallicities as well. Our distances would have to be severely underestimated for this signal to go away, which is unlikely to be the case.

\begin{figure*}
\centering
\includegraphics[width=0.33\hsize,trim={0.1cm 0.0cm 0.35cm 0.0cm}]{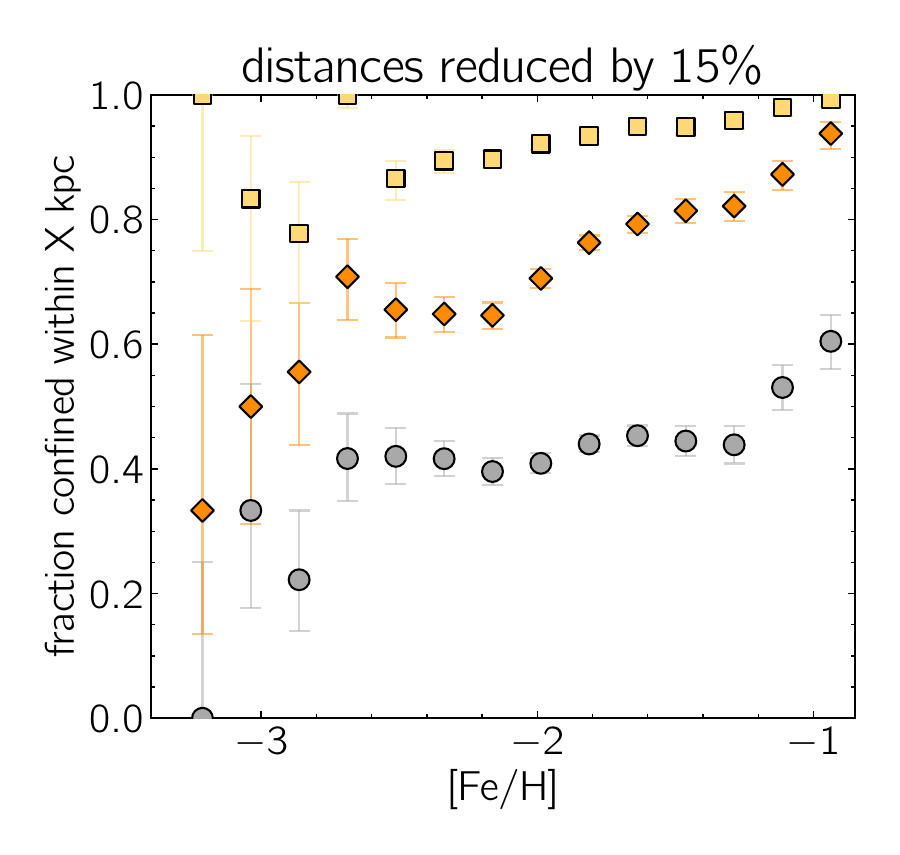}
\includegraphics[width=0.33\hsize,trim={0.1cm 0.0cm 0.35cm 0.0cm}]{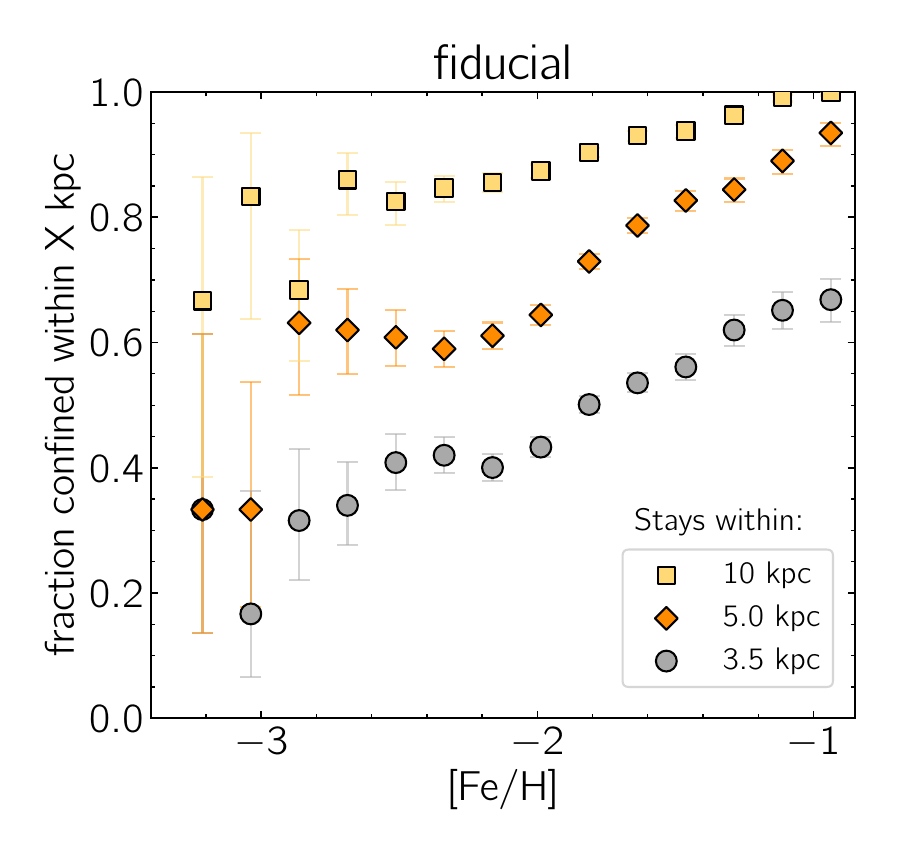}
\includegraphics[width=0.33\hsize,trim={0.1cm 0.0cm 0.35cm 0.0cm}]{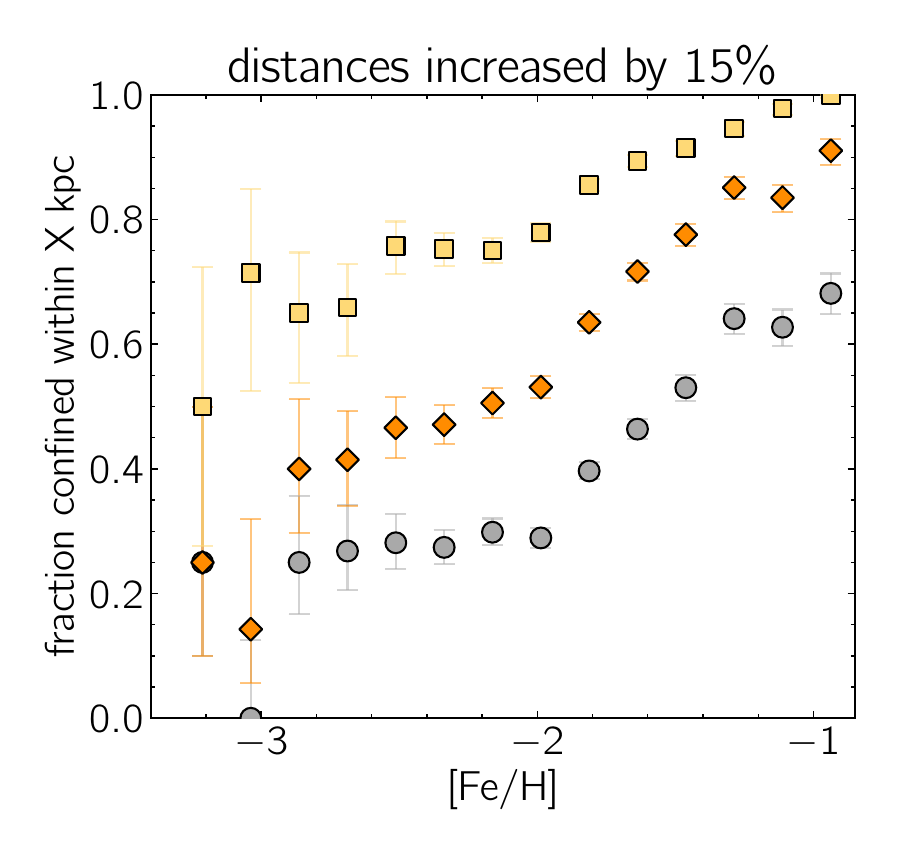}

\includegraphics[width=0.32\hsize,trim={0.0cm 0.0cm 0.0cm 0.0cm}]{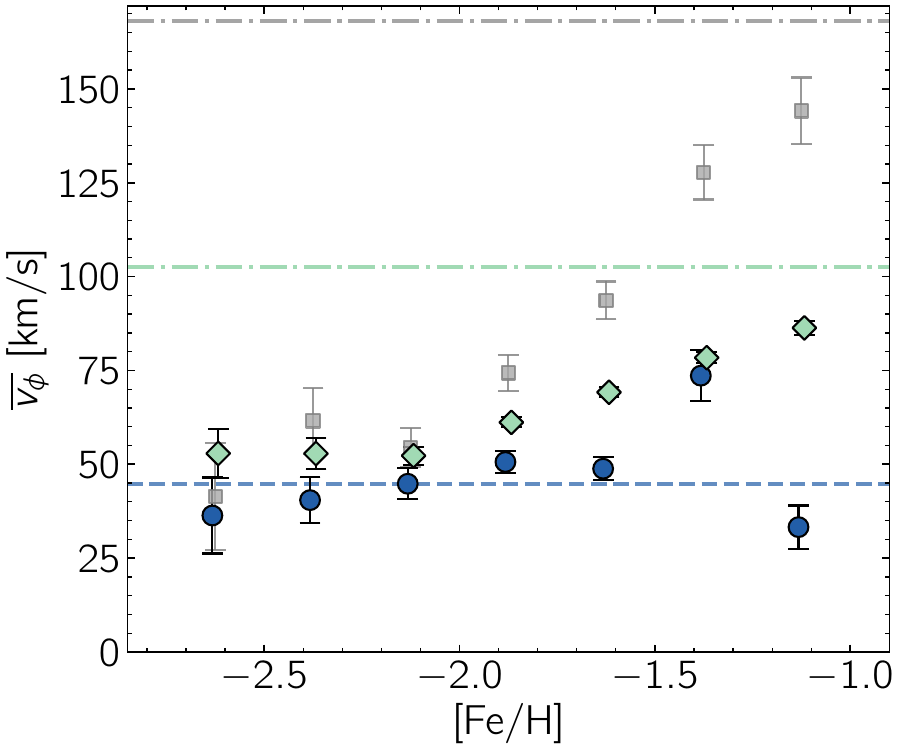}
\includegraphics[width=0.32\hsize,trim={0.0cm 0.0cm 0.0cm 0.0cm}]{figures/vmean_vpi-feh_50draws_xdgmm_35kpc_cyl_apo5_Sormani.pdf}
\includegraphics[width=0.32\hsize,trim={0.0cm 0.0cm 0.0cm 0.0cm}]{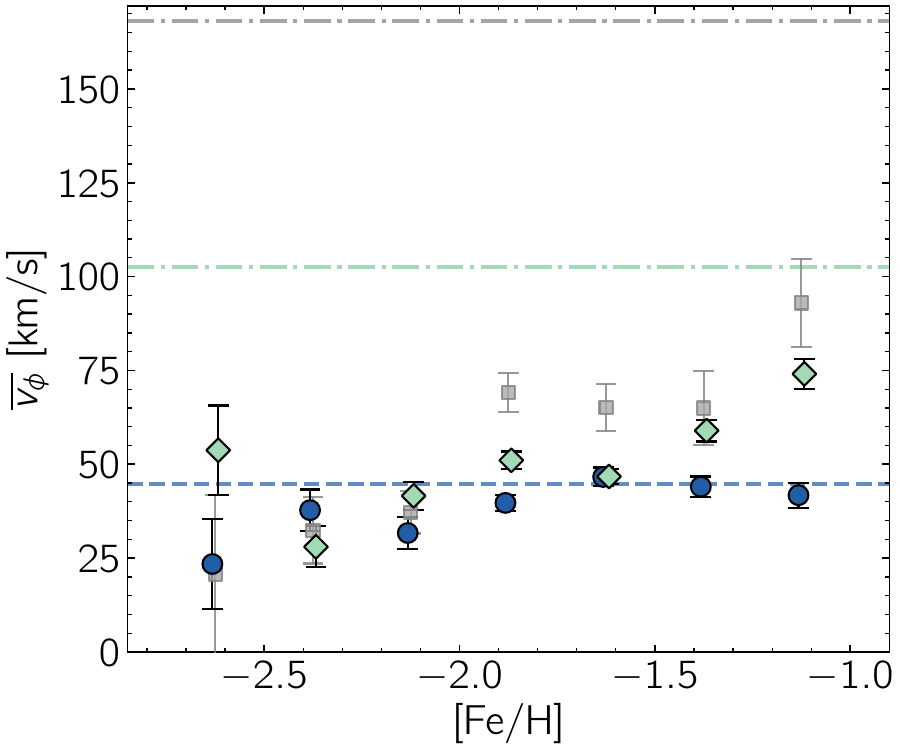}

\caption{Testing the impact of potential biased distances for the confined fraction (top row) and the azimuthal velocity (bottom row). The middle column shows the reference figures presented in the main body of the paper (bottom panel of Figure~\ref{fig:interlopers} and Figure~\ref{fig:vphitrend}), whereas the left and right columns present distances biased low and high by 15\%, respectively. 
}
\label{fig:comp_apofrac} 
\end{figure*}

\section{PIGS/AAT data release}\label{sec:datarelease}

\begin{table*}
\centering
\caption{\label{tab:log}Observed AAT fields. Field names (except for the pilot program, where the numbers do not mean anything anymore, and two Sgr fields, which have an additional suffix) follow the convention FieldRADec, with RA and Dec the central right ascension and declination in degrees.}
\begin{tabular}{lccl}
 \hline
 Fieldname & Total exposure time [s] & date(s) [DD/MM/YYYY] & Note \\
 \hline
Field 108153 & 10800 &  23/8/2017, 6/6/2018 & pilot program  \\
Field 217218 & 7200 &  24/8/2017 & pilot program  \\
Field251.2-29.7 & 5400 &  2/4/2019 &   shifted RV (see Section~\ref{sec:pigs}) \\
Field253.2-27.3 & 7200 &  4/4/2019 &   \\
Field253.7-29.5 & 7200 &  3/4/2019 &   \\
Field254.2-25.1 & 7200 &  5/4/2019 &   \\
Field255.4-27.9 & 7200 &  27/2/2020, 28/2/2020 &   \\
Field256.57-26.42 & 12600 &  31/3/2019, 1/4/2019 & overlaps with CAPOS field \\
Field256.9-29.5 & 7200 &  29/2/2020, 1/3/2020 &   \\
Field258.0-24.0 & 7200 &  17/6/2020 &   \\
Field258.3-21.8 & 7200 &  18/6/2020 &   \\
Field259.0-29.0 & 9000 &  8/8/2018, 1/4/2019 & overlaps with CAPOS field (two slightly different pointings combined) \\
Field259.0-29.1 & 5400 &  8/8/2018 &   \\
Field259.5-20.2 & 5400 &  19/6/2020 &   \\
Field260.7-25.4 & 7200 &  16/6/2020 &   \\
Field261.1-18.8 & 7200 &  15/6/2020 &   \\
Field262.8-16.3 & 7200 &  16/6/2020 &   \\
Field263.3-18.4 & 9000 &  18/6/2020, 19/6/2020 &   \\
Field265.5-17.9 & 9000 &  18/6/2020, 19/6/2020 &   \\
Field271.0-29.4 & 9000 &  24/6/2018 &   \\
Field271.5-27.8 & 9000 &  5/8/2018, 6/8/2018 &   \\
Field273.0-29.4 & 9000 &  4/8/2018, 5/8/2018 &   \\
Field273.3-26.5 & 7200 &  8/8/2018 &   \\
Field273.7-27.1 & 7200 &  7/6/2018, 4/8/2018 &   \\
Field275.5-25.4 & 7200 &  6/8/2018 &   \\
Field275.7-23.3 & 7200 &  15/6/2020 &   \\
Field275.85-27.6 & 7200 &  7/6/2018 &   \\
Field275.9-29.5 & 7200 &  5/8/2018 &   \\
Field277.3-26.5 & 7200 &  7/8/2018, 8/8/2018 &   \\
Field277.95-28.05 & 7200 &  7/6/2018 &   \\
Field278.1-24.5 & 7200 &  16/6/2020 &   \\
Field278.6-26.4 & 7200 &  6/4/2019 &   \\
Field281.0-26.2 & 7200 &  7/8/2018 &   \\
Field281.6-24.9 & 5400 &  7/6/2018 &   \\
Field282.0-29.8\_Sag & 10800 &  6/8/2018, 7/8/2018 & partly dedicated Sgr targets   \\
Field282.9-32.1 & 9000 &  16/6/2020 &  partly dedicated Sgr targets \\
Field284.0-30.0\_Sag & 10800 &  4/8/2018, 5/8/2018 & partly dedicated Sgr targets \\
Field286.0-31.1 & 5400 &  15/6/2020 & partly dedicated Sgr targets  \\
 \hline
\end{tabular}
\end{table*}

\begin{table*}
\centering
\caption{\label{tab:pigscat}Contents of the AAT/PIGS catalogue, containing 13,235 stars (see \citealt{arentsen20_II} for further details regarding the stellar parameters, and \citealt{arentsen21} for carbon-related details)} 
\begin{tabular}{lcl}
 \hline
 Column & units & description \\
 \hline
 starname &  & unique Pristine target name \\
 fieldname & & AAT field name (see Table~\ref{tab:log})\\
 source\_id & & Gaia DR3 source ID \\
 ra & degrees & right ascension from Gaia DR3 \\
 dec & degrees & declination from Gaia DR3 \\
 l & degrees & Galactic longitude \\
 b & degrees & Galactic latitude \\
 ebv\_green19 &   &  Reddening E(B-V) from \citet{green19} \\
 SNR\_4000\_4100 &   &  Signal-to-noise ratio between $4000-4100$~\AA \\
 SNR\_5000\_5100 &   &  Signal-to-noise ratio between $5000-5100$~\AA \\
 SNR\_CaT &  &   Signal-to-noise ratio in the calcium triplet (red arm) \\
 rv & \kms &  {\tt FXCOR} radial velocity  \\
 rv\_err & \kms & {\tt FXCOR} radial velocity uncertainty (incl. 2~\kms floor) \\
 teff\_ferre & K & {\tt FERRE} effective temperature \\
 eteff\_ferre\_tot & K & {\tt FERRE} effective temperature uncertainty (incl. 107~K floor)   \\
 logg\_ferre &  &  {\tt FERRE} surface gravity   \\
 elogg\_ferre\_tot &  & {\tt FERRE} surface gravity uncertainty (incl. 0.24 floor) \\
 feh\_ferre &  &  {\tt FERRE} metallicity   \\
 efeh\_ferre\_tot &  &  {\tt FERRE} metallicity uncertainty (incl. 0.13 floor)  \\
 cfe\_ferre &  &  {\tt FERRE} carbon over iron abundance   \\
 ecfe\_ferre\_tot & & {\tt FERRE} carbon over iron abundance uncertainty (incl. 0.20 floor)    \\
 mccor100 &  &  median \citet{placco14} evolutionary carbon correction \\
 sccor100 &  &  standard deviation of  \citet{placco14} evolutionary carbon correction \\
 chi2\_ferre & & {\tt FERRE}  $\log{\chi^2}$ \\ 
 teff\_ulyss & K & {\tt ULySS} effective temperature \\
 eteff\_ulyss\_tot & K & {\tt ULySS} effective temperature uncertainty (incl. 107~K floor)   \\
 logg\_ulyss &  &  {\tt ULySS} surface gravity   \\
 elogg\_ulyss\_tot &  & {\tt ULySS} surface gravity uncertainty (incl. 0.24 floor) \\
 feh\_ulyss &  &  {\tt ULySS} metallicity   \\
 efeh\_ulyss\_tot &  &  {\tt ULySS} metallicity uncertainty (incl. 0.13 floor)  \\
 srr\_ulyss &  &   {\tt ULySS} signal-to-residual ratio  \\
 sig\_ulyss &  &   {\tt ULySS} broadening width  \\
 f\_badpix\_C2a\_ulyss &  & {\tt ULySS}  fraction of rejected pixels in the fit between $4600-4800$~\AA \\
 f\_badpix\_C2b\_ulyss &  & {\tt ULySS}  fraction of rejected pixels in the fit between $5000-5200$~\AA \\
 double & [True or False]  &   if True, the CaT spectrum is clearly double-lined (most likely chance-alignment) \\
 good\_ferre & [True or False] & if True, star passes all \citet{arentsen20_II} quality flags (FERRE case) \\ 
 good\_ulyss & [True or False] & if True, star passes all \citet{arentsen20_II} quality flags (ULySS case) \\ 
 \hline
\end{tabular}
\end{table*}

\begin{table*}
\centering
\caption{\label{tab:pigscat_orbits}PIGS distances and orbital properties, where each parameter ``par'' has an entry for the median (par50) and the 16th and 84th percentiles (par16 and par84, respectively). For the apocentre, the 75th percentile has also been included. } 
\begin{tabular}{lcl}
 \hline
 Column & units & description \\
 \hline
 starname &  & unique Pristine target name \\
 feh &   &   {\tt FERRE} metallicity (repeated from Table~\ref{tab:pigscat} for convenience) \\
 dist & kpc  & {\tt StarHorse} distance \\
 x & kpc  &  \\
 y & kpc  &  \\
 z & kpc  &  \\
 vR & \kms  &  \\
 vphi & \kms   &  \\
 vz & \kms   &  \\
 apo & kpc  &   \\
 peri & kpc  &   \\
 zmax & kpc  &    \\
 ecc &    &    \\
 JR & kpc \kms   &    \\
 Jphi & kpc \kms   &    \\
 Jz  & kpc \kms   &    \\
 Energy & \ensuremath{\rm{km^2}\,s^{-2}}\xspace  &   \\
 \hline
\end{tabular}
\end{table*}


\bsp	
\label{lastpage}
\end{document}